\documentclass{article}
\usepackage[utf8]{inputenc}
\usepackage{DejaVuSansMono}
\usepackage{authblk}
\usepackage{setspace}
\usepackage[margin=1.25in]{geometry}
\usepackage{graphicx}
\graphicspath{ {./figures/} }
\usepackage{subcaption}
\usepackage{amsmath}
\usepackage{amssymb}
\usepackage{mathtools, cuted}
\usepackage{fancyvrb}
\usepackage{tcolorbox}

\usepackage{bbm}
\usepackage{courier}
\usepackage{hyperref}

\newcommand{\dataset}{{\cal D}}

\newcommand{\indep}{\rotatebox[origin=c]{90}{$\models$}}

\usepackage[style=ieee, 
citestyle=numeric-comp,backref=true,hyperref=true, indexing=true,
sorting=nyt]{biblatex}
\title{A Causal Research Pipeline and Tutorial for Psychologists and Social Scientists}

\author[]{Matthew J. Vowels}

\affil[]{Institute of Psychology, University of Lausanne (UNIL), Lausanne, Switzerland.}

\date{}

\begin{document}

\maketitle


\begin{abstract}
Causality is a fundamental part of the scientific endeavour to understand the world. Unfortunately, causality is still taboo in much of psychology and social science. Motivated by a growing number of recommendations for the importance of adopting causal approaches to research, we reformulate the typical approach to research in psychology to harmonize inevitably causal theories with the rest of the research pipeline. We present a new process which begins with the incorporation of techniques from the confluence of causal discovery and machine learning for the development, validation, and transparent formal specification of theories. We then present methods for reducing the complexity of the fully specified theoretical model into the fundamental submodel relevant to a given target hypothesis. From here, we establish whether or not the quantity of interest is estimable from the data, and if so, propose the use of semi-parametric machine learning methods for the estimation of causal effects. The overall goal is the presentation of a new research pipeline which can (a) facilitate scientific inquiry compatible with the desire to test causal theories (b) encourage transparent representation of our theories as unambiguous mathematical objects, (c) to tie our statistical models to specific attributes of the theory, thus reducing under-specification problems frequently resulting from the theory-to-model gap, and (d) to yield results and estimates which are causally meaningful and reproducible. The process is demonstrated through didactic examples with real-world data, and we conclude with a summary and discussion of limitations.
\end{abstract}

\begin{figure}[!ht]
\centering
\includegraphics[width=1\linewidth]{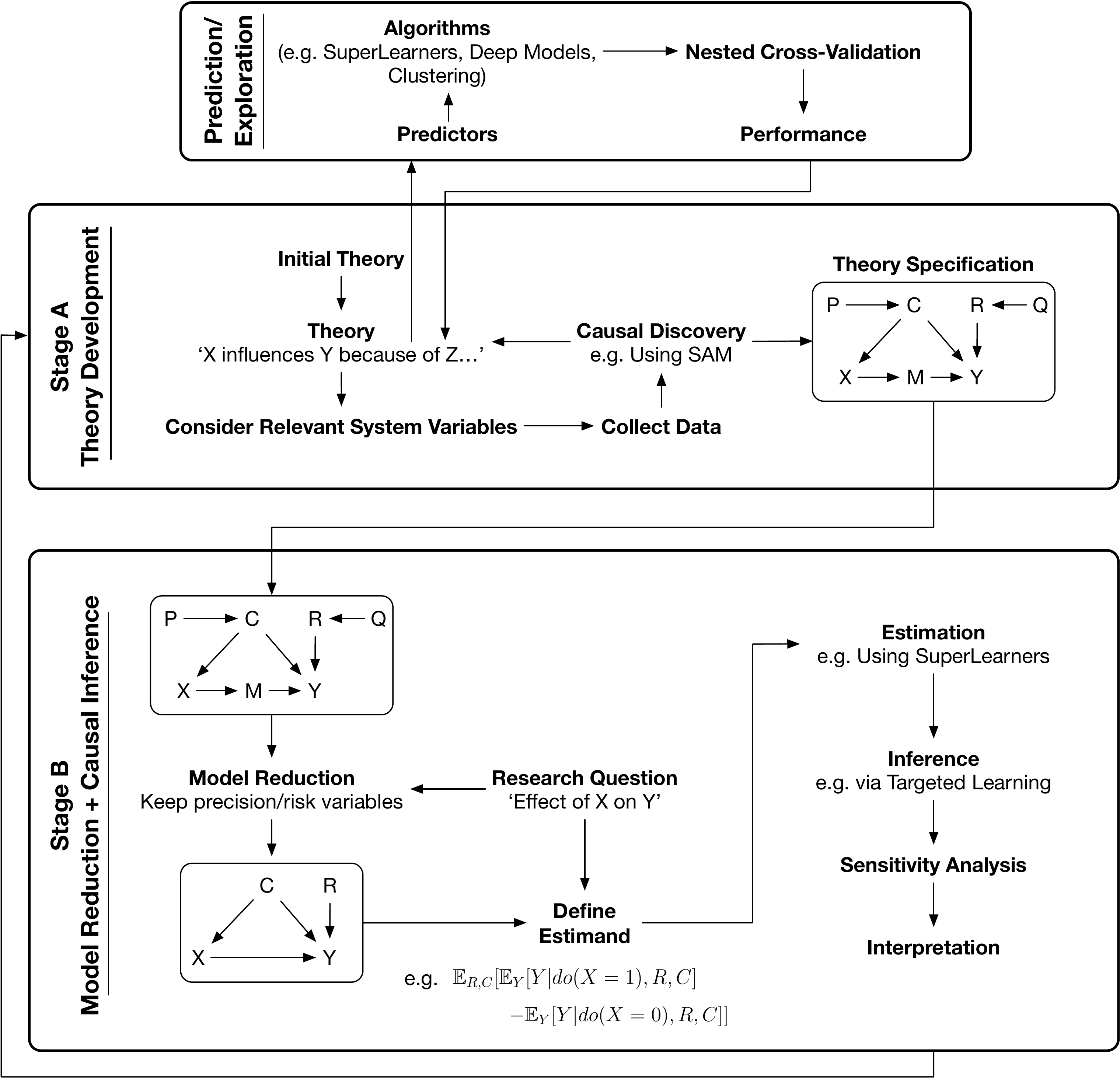}
\caption{Complete Research Pipeline. Stage A begins with an initial theory, deriving (for instance) from qualitative interviews, and which is used to start iterative theory development process. This theory can be used to inform a predictive/exploratory stage, which in turn can either help to validate the theory itself \parencite{Yarkoni2017}, or be used in place of a causal approach in cases where the phenomenon is not easily represented in mathematical form (in this work, we only briefly discuss the predictive approach). The theory can be used in tandem with data-driven causal discovery approaches to yield (\textit{e.g.}) a causal graph, which can be used for causal inference (Stage B). Stage B involves specification of a research question and an estimand expressed as a function of the graph and the observed data, for subsequent estimation, inference, sensitivity analysis, and finally, interpretation. The pipeline is iterative insofar as new discoveries which occur at any stage can influence the overall process.} 
\label{fig:pipeline}
\end{figure}

\section{Introduction}
\label{sec:intro}
As scientists, we are fundamentally interested in understanding the cause-effect relationships in the world, such that we may design effective interventions to manipulate and improve the world around us \parencite{Machamer2000causalphil, Pearl2009, Glymour2014, Mackie1974causalcement} . Unfortunately, meta-researchers have highlighted that in spite of our underlying pursuit of answers to causal questions, causality remains taboo in psychology and social science \parencite{Grosz2020, Hernan2018}. The consequence is that there exists a distinct mismatch between the inherently causal nature of our theories, and the nature of all downstream elements of the research pipeline. This precludes smooth transitions between (causal) theories, and (correlational) hypotheses, data collection methodologies, modeling practices, and interpretations. Furthermore, theories in psychology and social science have been criticized for being underspecified and untestable \parencite{Scheel2020}. As such, it is often unclear what exactly about the causal relationships described by the theories we are actually testing. Thus, in addition to the mismatch between the causal nature of the theory and the statistical/correlational nature of the tests, there is also often a large gap between the specification of the theory itself and how this specification is linked to the model. The consequence is that the model is likely to be causally misspecified, yielding arbitrary results which are impossible link transparently with the theory and thus impossible to meaningfully interpret \parencite{Vowels2021}.

In this work, we present a `causally homogeneous' research pipeline which encourages theoretic specificity to facilitate increased reliability of testing. We split the pipeline into two main stages A and B, with an optional predictive/exploratory stage, and present an overview in Figure~\ref{fig:pipeline}. In stage A we are concerned with theory development, and propose and iterated development and validation of an initial theory (one which may, for instance, have been informed by qualitative interviews) by accommodating techniques from the domain of \textit{causal discovery}. This stage uses causal discovery techniques \parencite{Vowels2021DAGs} to inform and augment our theories, by testing putative causal relationships in data. The methodology for data collection itself is informed by our developing theory, such that we can be confident \textit{e.g.} that sampling occurs at sufficiently frequent and/or regular intervals \parencite{VowelsSpectralTutorial}, that we collect all causally relevant variables, that we have a sufficient sample size, etc. Once we have sufficient evidence that our theory is reflected by the data, we will have derived a representative graph - a mathematical and unambiguous object which captures our theory and enables us to hypothesise clearly about certain theoretical relationships. In parallel, or instead, one may also use the initial theory to specify a set of variables to be used for predictive and/or exploratory modeling. Such an approach is non-causal, but may nonetheless help researchers establish the predictive validity of their initial theory, and/or to develop tools for automating decision processes (such as automated assessment or behavioral coding).

In stage B, we use the graphical representation of our theory to facilitate \textit{causal inference} in order to test a hypothesis which is specified in terms of the graph. Causal inference \parencite{Pearl2009, Imbens2015} involves the specification of a target effect of interest (for instance, the cause-effect relationship between two variables or constructs of interest). By virtue of the fact that this effect can be tied directly to the causal graph, it can therefore also be tied directly to our theory, thus resolving the issue of the indeterminate links between our statistical results and the aspects of the theory with which they are intended to correspond. Tools from the domain of causal inference, notably \textit{do}-calculus \parencite{Pearl2009} and the rules of \textit{d}-separation \parencite{Koller2009} enable us to understand under what assumptions the quantity we are interested in is estimable from the data. It is possible, for instance, that the theory precludes non-experimental estimation. The choice is then left open to the researcher to continue anyway (all the while maintaining transparency about the concomitant assumptions), or to rethink their approach.

In summary, the research pipeline is a proposal to (a) harmonise our research pipeline with the causal nature of our intentions and goals, (b) encourage transparent representation of our theories as unambiguous mathematical objects, (c) to tie our statistical models to specific attributes of the theory, thus reducing the under-specification problems frequently resulting from the theory-to-model gap, and (d) to yield causally meaningful estimates and results. Whilst both parametric and non-parametric approaches to our proposal exist, we prefer to take a general approach, and thus demonstrate our proposal using non-parametric machine learning and semi-parametric machine learning approaches to causal discovery and causal inference, respectively. This reduces the number of assumptions we impose upon our problem considerably, because the assumption of parametric form is, in many cases, both unnecessary and problematic \parencite{vanderLaan2014}.

\begin{tcolorbox}
\textbf{Application/Code Note :}
Accompanying code is provided at \url{https://github.com/matthewvowels1/causalPipeline}  which includes a tutorial notebook \texttt{causal\_pipeline\_tutorial.ipynb} to aid in implementation.
\end{tcolorbox}

We begin with a review of the literature pertinent to motivating the current pipeline in Section~\ref{sec:motivation}. We then present a short summary overview of the proposed pipeline in Section~\ref{sec:overview}, and in Section~\ref{sec:background} we review relevant background theory. Two sections follow, one for Stage A (Section~\ref{sec:stageA}) on theory specification, development, causal discovery, and structural validation, and the next for Stage B (Section~\ref{sec:stageB}) for estimation of causal effects, statistical inference, and robustness checks. In each of these two sections, we review relevant concepts and theory, present the proposal for a shift in research methodology, and present example implementations of the proposed changes. Next, in Section~\ref{sec:prediction} we briefly review the possibilities for predictive/exploratory approach. In Section~\ref{sec:assumptions} we discuss a number of key assumptions as well as discussing the handling of missing data. Finally, in Section~\ref{sec:discussion} we close with a summary of the proposed pipeline and its limitations.


\section{Motivation for Causal Approaches to Research}
\label{sec:motivation}
Research in psychology and social science has come under fierce criticism for a number of years, and the domains have, in general, been slow to react \parencite{aarts, Sedlmeier1989, Spellman2015, Hullman2022, Cassidy2019, Vowels2021, Meehl1990, Scheel2022}. These critiques concern a wide variety of aspects relating to all aspects of the research pipeline, from theory development \parencite{Oberauer2019, Scheel2022, Borsboom2021}, measurement \parencite{Flake2020, Barry2014}, data collection \parencite{Button2013, Henrich2010, Vankov2014}, model specification and analysis \parencite{Vowels2021, Scheel2022, Hullman2022}, and reliability of interpretations \parencite{Rohrer2018, Grosz2020, Szucs2017, Yarkoni2019, Hernan2018}. Notwithstanding problems of (non)specific and untestable theories \parencite{Scheel2020}, there also exists a distinct mismatch between the inherently causal nature of the theories, and the nature of all downstream elements of the research pipeline. Indeed, we are not aware of a theory in psychology or social science which does not invoke cause-effect terminology. And yet, it seems that in psychology and social science there exists a taboo around causality \parencite{Grosz2020, Hernan2018}, and a conflation of causal intent and correlational language \parencite{Shmueli2010}, which leads to a dilution of already indistinct causal theories into arbitrarily specified tests of statistical association. Sometimes, the tests themselves involve ostensibly causal models. For example, Structural Equation Models (SEMs) are a popular method in psychology today \parencite{Blanca2018}, and can be traced back to work on path models by \textcite{Wright1921}. They provide a means to directly encode an experts causal knowledge in a statistical model and are equivalent to Structural Causal Models with an assumed linear functional form \parencite[pp.135]{Pearl2009}. And yet, even with such inherently causal models, the concomitant interpretations are often loose and ambiguous, reflecting an underlying desire for causal interpretation disguised in vague `associational' terminology \parencite{Shmueli2010,Grosz2020, Rohrer2018, Pearl2009}.

Of course, one direct route to the estimation of causal quantities involves the use of experimental, randomized controlled trial (RCT) data. RCT studies facilitate comparison between groups in such a way that the differences are not immediately confounded by unobserved differences. However, RCTs have their own limitations; they may be expensive, unethical, and/or impractical \parencite{Deaton2018}. For instance, one cannot force people to smoke or not smoke to investigate the role of smoking on cancer. In contrast, observational data is much more abundant, less expensive, and in general, has less ethical implications. One of the principal challenge with observational data, of course, is that causal effects are less readily derivable. Fortunately, in spite of the general reluctance to adopt them, there exist a wide variety of tools which facilitate the identification and estimation of causal quantities from observational data \parencite{Imbens2015, Pearl2009, Wright1921, Wright1923, Vowels2021DAGs, Guo2020, Glymour2019}, and such tools are comparatively commonplace in other fields such as epidemiology \parencite{Venderweele2006} and economics \parencite{Chernozhukov2017, Fisher1970sem}. Causal inference tools enable us to derive mathematical expressions for desired causal quantities in terms of observational quantities. Similarly, causal discovery tools enable us to uncover causal structures from unobserved data, and do so by testing for a number of features indicative of cause-effect relations. 

Such tools are not without their own risks, and have, in turn, be faced with some notable rebuttals. \textcite{Dawid2008} encourages researchers to `beware of the DAG' - Directed Acyclic Graphs (DAGs) being a popular tool for causal inference. Other researchers refer to the quest for the identification of causal structures as ``a glorious perversion'', and liken it to the "search for the philosopher’s stone" \parencite[pp.551]{Korb1997}. These cautions are not all that surprising, particularly given that the validity of causal inference and discovery rests on a number of \textit{untestable} assumptions. For instance, causal discovery techniques assume \textit{faithfulness} \parencite{Pearl2009}, which holds that the causal process does not include effects which cancel each other out, which would otherwise make the distribution of the data generated by such a process equivalent to one without such a causal link. Or, alternatively, there are similar assumptions known as \textit{sufficiency} in causal discovery, and  \textit{ignorability} (also known as conditional exchangeability) for causal inference, which hold that there are no unobserved confounders which otherwise render the goal of the respective analyses significantly more challenging \parencite{Pearl2009, Imbens2015}. However, even in the face of such strong assumptions, the question would remain as to what the alternative would or should be. Either researchers in psychology and social science should stick to purely associational research (in the absence of experimental data), and resign themselves to never addressing fundamentally causal problems, at least not explicitly \parencite{Hernan2018, Grosz2020, Shmueli2010}; or, they can start assimilating causal techniques in a structured and rigorous manner, stating any pivotal assumptions and limitations clearly and transparently in order to push their estimates closer to the (causal) target. Indeed, causal approaches are being recommended and applied in other fields such as epidemiology \parencite{vanderLaan2011, Tennant2021}, biology \parencite{Triantafillou2010}, medicine \parencite{Yazdani2015, Castro2019}, advertisement \parencite{Bottou2013},and economics \parencite{Hunermund2021, Imbens2015}. 

Even if it is challenging or even impossible to make causal claims without caveats, the techniques enable researchers to engage in this worthwhile pursuit with more precision and rigor, thereby moving us towards reliable interpretations and repeatable science \parencite{Tennant2021, Grosz2020}. Indeed, in psychology, meta-researchers have also been recommending the adoption of causal approaches for similar reasons \parencite{Vowels2021, Grosz2020, Rohrer2018}. They explain how, without the tools from the fields of causal discovery and causal inference, we struggle to elucidate theory in a formal manner and to test it meaningfully. For instance, without a careful consideration for causal structure, we are more likely to misspecify a statistical model, rending the results uninterpretable \parencite{Vowels2021}. The tools from causal inference also help us understand whether our data are sufficient for testing a particular hypothesis. If we know there exists a confounder which jointly influences (\textit{i.e.}, causes) treatment choice as well as recovery rates, but we have not collected data for this variable, we may not be able to estimate the effect of treatment. For instance, consider that younger patients may be more likely to opt for surgery, and older patients more likely to prefer medication. Older and younger patients may also have different recovery rates. As such, age represents a crucial confounder in this causal system, and it may be impossible to derive meaningful estimates without having data for it. Note that certain confounders may not be so obvious/easy to identify, as age was in the previous example. The tools we discuss in this work help us identify such variables, and to understand how to integrate them into statistical models.

We also note the limitations associated with popular \parencite{Blanca2018} but nonetheless restrictive parametric approaches to analysis in psychology and social science. Indeed, \textcite{vanderLaan2014} argue that the choice of linear, parametric models represents a poor choice from the outset, dooming the family defined by a choice of linear parametric models to ever be able to contain within it the true distribution. Furthermore, such impositions are rarely necessary. Whilst such approaches have the deceptive appeal of being easy to explain and interpret, they are more likely to be misspecified both in terms of the functional form (\textit{i.e.}, they are linear and parametric), and in terms of their structural form (\textit{i.e.}, if a causal structure is not well specified, they will not yield meaningful estimates regardless of the functional form). As such, meta-researchers have argued that both functional and structural misspecification must be taken seriously, and that in general neither are addressed seriously in practice \parencite{Vowels2021, Hullman2022}. As such, we do not only advocate for a transparent causal approach to research, but also for the use of machine learning algorithms with flexible, data-adaptive functional form. There exist a range of techniques for deriving popular statistics such as standard errors and \textit{p}-values, regardless of whether non-parametric statistical models are used \parencite{Efron1981}. Indeed, in this paper we demonstrate our proposal using non-parametric techniques for causal discovery \parencite{Kalainathan2020}, and semi-parametric techniques for causal inference \parencite{vanderLaan2011}.

The approach presented here is, of course, only one possible approach which we believe improves upon current \textit{status quo}. Note that, by way of example, Stage A involves the development and validation of an initial theory (which itself might come as the result of qualitative interviews) using expert-guided causal discovery. In contrast, \textcite[pp.709]{Navarro2021perspective} argues that `theory building is not a statistical problem' and uses Shepard's law of generalization for psychological science \parencite{Shepard1987} as an example demonstrating that theory building can also be done \textit{post hoc}, by identifying patterns in existing data. \textcite{Navarro2021perspective} identifies a potential (but nonetheless moot) contradiction between such an approach and the generally received wisdom that one cannot develop theory without preregistered confirmatory tests. Our approach falls somewhere between the two insofar as it is iterative, thereby integrating prior knowledge with conclusions derived from data:  Stage A involves an iterative development of methodology (and possibly multiple data-collection rounds) to converge on a mathematical representation of a theory derived according to a combination of expert knowledge (which itself may be influenced by qualitative reports) and data-driven causal discovery.

Similarly, \textcite{Borsboom2021} recently proposed their own `Theory Construction Methodology' (TCM) as a means to develop theory, and recommended a shift away from relatively isolated hypothesis testing, which does not directly contribute towards theory generation and development. Their framework involves an iterative process, beginning with an initial theory (what they refer to as a \textit{prototheory}), the formal specification of this theory as a mathematical model, simulations to establish whether the model can adequately represent the phenomenon of interest, and finally the empirical evaluation of the model. These steps form an iterative loop not dissimilar to the one we present in Figure~\ref{fig:pipeline}. Interestingly, they only briefly allude to causality once, despite an overarching causal motivation ``[Theories] allow us to predict and control our environment through strategic interventions...'' \parencite[p.756]{Borsboom2021}. This notwithstanding, our proposal shares a number of parallels with TCM. In particular, both are iterative - if the researcher encounters new/unexpected information, this information can be integrated into the theory and the process can be repeated until some form of convergence (of course, the time frame for this convergence is unknown). Furthermore, in a similar way to how \textcite{Borsboom2021} argue that null-hypothesis tests are not sufficient in themselves for the development of theory, we make a clear separation between the theory development and specification of Stage A, and the hypothesis testing in Stage B, even though the information from Stage B can, and should, subsequently be used to inform the theory. 

 In our view, whilst discussions such as those by \textcite{Navarro2021perspective, Borsboom2021} are essential and highly valuable, their work does not explicate exactly what is involved in the process, are are primarily conceptual. As such the work does not provide researchers with a tangible means or demonstration for how to put these conceptual ideas into practice. In contrast, we provide tangible examples of theory development, specification, and testing, using a wide array of tools from the causality, statistics, and machine learning literature. Indeed, we take a causal emphasis precisely because we do not see how a theory can be developed without one. Given the tendency for causal intentions to be gradually diluted as we progress through the \textit{typical} research pipeline, it is not surprising to hear many of the critiques leveraged against the fields of psychology and social science. The techniques and recommendations in this paper provide researchers with a means to answer causal queries without needing access to experimental data and without needing to impose unnecessary linear or parametric functional restrictions. These techniques encourage specificity of theory and hypothesis, thus improving reproducibility and transparency. They inform the requirements for data collection, and enable us to translate our theories into unambiguous and testable causal models. Finally, they move us towards a homogeneous, causal pipeline, thereby reflecting the inherently causal nature of our theories.

\section{Pipeline Overview}
\label{sec:overview}
A summary of the proposed pipeline is given in Figure~\ref{fig:pipeline} and may be broken down into the following steps:

\begin{itemize}
    \item Stage A: Theory development, validation, and formalisation.
    \begin{itemize}
    \item Input: An initial theory and subject of research (potentially informed by qualitative interviews).
    \item Primary Output: A developed and validated theory specified formally in terms of a graphical and/or structural and mathematical object.
    \item Secondary Output: top-level theory informing a set of variables amenable to use as predictors in a predictive model.
        \item Tools: Causal discovery techniques (\textit{e.g.}, Structural Agnostic Model, \cite{Kalainathan2020}) for discovering and validating putative cause-effect relationships.
    \end{itemize}
    \item Stage B: Model Reduction, Causal Inference, and Hypothesis Testing.
    \begin{itemize}
        \item Input: A model or graphical specification of the theory, as output from Stage A, and a hypothesis or target quantity of interest.
        \item Output: Estimates and statistics for inference (standard error etc.)
        \item Tools: Reduction algorithm for reducing the complexity of the model and providing the minimal representation necessary for a given quantity of interest; probabilistic ADMGs for further specification of unknown paths and expert plausibility ratings; an algorithm for checking the target quantity of interest is estimable (at least in principle) from the data; a means to specify this estimand mathematically; a means estimate the quantity (\textit{e.g.}, targeted learning with Super Learners \cite{vanderLaan2011, Polley2007}).
    \end{itemize}
    \item Prediction/Exploration: Predictive theory validation, exploration, and utility.
    \begin{itemize}
        \item Input: A top-level theory determining a set of predictor variables.
        \item Output: A predictive or latent/cluster model and a quantification of the associated performance.
        \item Tools: Predictive algorithms such as Super Learners \cite{Polley2007}, deep computer vision models \parencite{Himabindu2021compvision}, deep Natural Language Processing (NLP) models \parencite{Qiu2020NLPsurvey}, clustering and dimensionality reduction algorithms \parencite{mcinnes}.
    \end{itemize}
\end{itemize}

Starting with an initial theory, which may comprise many uncertainties and be lacking in specificity, Stage A of the pipeline is intended to aid in the development and specification of the theory in terms of a transparent and unambiguous mathematical object. There exist different options for this specification, such as Directed Acyclic Graphs (DAGs), Structural Causal Models (SCMs), and computational models. In this work, we focus on DAGs and SCMs and discuss these in detail in the following sections. In order to specify these models, we need some prior domain expertise to guide us and/or causal discovery techniques. Our recommendation would be to supplement domain expertise, using causal discovery techniques to validate this expertise and to generate suggestions for causal structure between certain variables about which the expert may have limited prior knowledge. The purpose of this stage is thus to derive a DAG and/or SCM which is (a) plausible in its reflection of our existing domain knowledge, (b) supported by the data insofar as it has been validated using causal discovery techniques, and (c) represents a combination of domain expertise and potentially novel, data-driven causal discovery.

\begin{tcolorbox}
\textbf{Application/Code Note :}
 It is worth noting that many approaches to causal discovery exist, and readers are encouraged to consult surveys by \textcite{Glymour2019, Vowels2021DAGs, Assaad2022} for a discussion of various options. In this paper, we provide a working example using the Structural Agnostic Modeling (SAM) method  by \textcite{Kalainathan2020}. SAM is a flexible, non-parametric, state of the art approach to structure learning, and has been demonstrated to perform well with as many as variables of mixed type (\textit{i.e.}, binary, categorical, continuous). The method also accepts a set of constraints, which preclude the discovery or specification of certain links which the expert may deem unrealistic or implausible (indeed, one can preclude putative causal effects which occur backwards in time). The output of SAM is a set of cause-effect link confidences (\textit{e.g.} the cause-effect relationships $X \rightarrow Y$ exists with a confidence of $0.6$) which have been averaged over multiple runs. At this stage, the researcher has two choices. Either they can threshold the confidences to directly yield a model of the underlying structure, or they can modify the graph and/or specify additional links and unobserved variables according to their domain knowledge. Regardless, the output of Stage A a mathematical object which represents the underlying theory of the phenomenon of interest.
\end{tcolorbox}

Once we have a model of our theory in the form of a DAG and/or SCM (again, these are introduced formally below), we can review our fundamental research questions and hypotheses. For instance, we may hypothesise that a treatment has an effect on an outcome, and the target quantity of interest in this case may be the average treatment effect. Note that we refer to this quantity as the \textit{target estimand}, or just the \textit{estimand} (\textit{i.e.}, the estimand is the quantity we which to estimate). We should be able to tie this estimand directly to attributes in the model. For example, it may be that within our model there exists a causal dependence between the treatment $A$, and the outcome $Y$, for which we are interested in estimating the causal effect. The link between these variables can be specified formally in terms of the model, and the estimand can thus be specified mathematically and unambiguously. Once we have specified our estimand in terms of the model, we can reduce the model to the minimal complexity necessary for this given estimand, in such a way that the estimation itself is not biased by this reduction. For instance, if we are interested in a particular causal relation, it may not be necessary to use all distal causes. Indeed, as we will discuss, there exist a number of techniques for identifying a submodel which captures the essence of the causal process for a given estimand \parencite{Koller2009, Vowels2022minsem}.

Then, we can choose from a variety of approaches for undertaking the estimation itself. For instance, if we lived in a linear world (which we do not), we might choose a linear regression algorithm, and use it to estimate our estimand as the coefficient on the treatment variable. Alternatively, and more realistically, we might use an ensemble of diverse machine learning algorithms, such as a Super Learner \parencite{Polley2007}. On their own, such an ensemble would not yield statistics such as standard errors, and we need a method to derive such quantities to facilitate testing. There exist a range of methods for this \parencite{Efron1981}, and we demonstrate this step using a version of the influence function method, known as targeted learning \parencite{vanderLaan2011}.

Before interpreting and discussing the implications of the estimation and inference, we recommend researchers perform some type of sensitivity analysis \parencite{Bonvini2020, Diaz2013, Thabane2013Sensitivity, Zhou2021sens}. Sensitivity analysis offers a means to quantify the robustness (\textit{i.e.}, sensitivity) of the results to various types of model misspecification and violations of assumptions. For example, one of the strongest assumptions associated with causal inference is that of ignorability, which is the assumption that there exist no unobserved confounders which render the estimand `unidentifiable' from the observed data. One heuristic means to simulate such violations is by assuming that the bias which has already been adjusted for by the inclusion of the \textit{observed} confounders is unlikely to exceed the magnitude of the bias associated with the \textit{unobserved} confounders (for which adjustment is not possible). One can estimate the impact of such bias by intentionally misspecifying certain aspects of the model and observing the resulting change in the estimate.

If researchers encounter difficulties in specifying their theory as a concrete mathematical object (such as a DAG) owing to the complexity of the phenomenon or problems with measurement (some attributes may be difficult to measure), one has the option to take a non-causal route and instead use predictive and/or exploratory models. This alternative approach can be used as a way to establish the predictive validity of their theory \parencite{Yarkoni2017}, as a means to engineer a solution to a particular automation task (\textit{e.g.}, automating assessment processes using machine learning decision models), or for identifying clusters or latent concepts for exploratory purposes. Indeed, if the researcher is not confident in the specification of their theory, they should not knowingly use misspecified models - the results are unlikely to be reliable or meaningful.

In summary, the proposal influences most if not all aspects of the research pipeline, from theory conception to the interpretation of statistical analyses. In what follows, we begin by introducing the necessary background theory before discussing each stage in more detail and providing demonstrations with real-world data.

\section{Preliminaries / Background}
\label{sec:background}
In this section, we review a number of important concepts relevant to the proposed pipeline. In particular, and in the following order, we review Directed Acyclic Graphs (DAGs) and Structural Causal Models (SCMs), pertinent concepts in causal discovery, and finally, relevant concepts relating to causal inference. Regarding notation, we use upper-case letters \textit{e.g.} $X,Y$ to denote random variables, and bold font, upper-case letters to denote sets of random variables \textit{e.g.} $\mathbf{X}, \mathbf{Y}$. The symbols $\indep$ and $\not\!\perp\!\!\!\perp$ denote statistical independence or statistical dependence, respectively. For example $X \indep Y$ means that random variable $X$ is statistically independent of $Y$. Notation such as $f(X)$ or $g(Y)$ denotes some arbitrary function $f$ or $g$ of $X$ or $Y$. We generally denote the existence of a dataset with $\dataset$ with sample size $N$, such that we may have access to $\dataset = \{x_i, y_i\}_{i=1}^N$, sampled from some true and inaccessible distribution $\mathcal{P}$, where lower-case $x$ and $y$ indicate specific realisations of random variables $X$ and $Y$. For a set of variables $\mathbf{X}$ we index individual variables with superscripts \textit{e.g.}, $X^j \in \mathbf{X}$.

\subsection{Causality, Directed Acyclic Graphs (DAGs), and Structural Causal Models (SCMs)}

We use causality in the sense defined by \textcite{LewisCausation1973} when he said: 

\begin{quote}
    ``We think of a cause as something that makes a difference, and the difference it makes
must be a difference from what would have happened without it. Had it been absent,
its effects – some of them, at least, and usually all – would have been absent as well.''
\end{quote}

This definition alludes to counterfactuals, because it alludes to realities which may never come to pass. For instance, if a patient chooses to engage in therapy, we can measure the consequences of this particular choice. In order to understand if therapy worked, we must reason about what \textit{would} have happened, had they \textit{not} chosen therapy. Information about what \textit{did} happen may help us in this endeavour. \textcite[pp.70]{Pearl2009} defines causality in a consistent, but more formal way (notation modified to align with ours):

\begin{quote}
    ``Given two disjoint sets of variables $X$ and $Y$, the causal effect of $X$ on $Y$, denoted as... $P(y|do(\mathbf{x}))$, is a function from $\mathbf{X}$ to the space of probability distributions on $Y$. For each realization of $\mathbf{x}$ of $\mathbf{X}$, $P(y|do(\mathbf{x}))$ gives the probability of $Y=y$ induced by deleting from the model [$x^j := f^j(pa^j,u^j), j=1...,J,$] all equations corresponding to variables in $\mathbf{X}$ and substituting $\mathbf{X}=\mathbf{x}$ in the remaining equations.''
\end{quote}

Here, $x^j := f^j(pa^j,u^j), j=1...,J,$ represents a SCM, where $j$ is the variable index in set of $J$ variables $\mathbf{X}$, $pa^j$ indicates parents of variable $j$, and $u^j$ indicates error or exogenous noise. The `walrus operator' $:=$ indicates that we are dealing with an asymmetrical relationship. We are \textit{assigning} the value of the right hand side to the variable on the left. To clarify this terminology, consider the following example SCM:

\begin{equation}
\begin{split}
    X^{j=1} := f^{X^1}(U^{X^1})\\
    X^{j=2} := f^{X^2}(X^{j=1}, U^{X^2})\\
    Y := f^Y(X^{j=1}, X^{j=2}, U^Y).
\end{split}
\label{eq:examplescm}
\end{equation}

\begin{figure}[h!]
\centering
\includegraphics[width=1\linewidth]{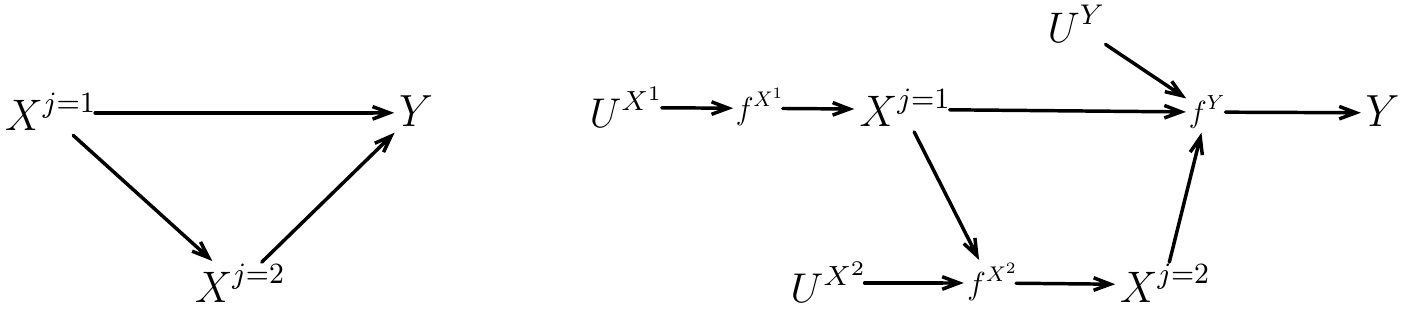}
\caption{\textbf{Left:} Directed Acyclic Graph (DAG) correspinding with the SCM in Eq.~\ref{eq:examplescm}. \textbf{Right:} An expansion of the DAG to show how the functions $f$ and unobserved, exogenous noise $U$ are encoded implicitly within the DAG.}
\label{fig:exampledag}
\end{figure}

From these equations, we see that variable $X^{j=1}$ is some function $f^{X^1}$ of noise $U^{X^1}$, $X^{j=2}$ is a function $f^{X^2}$ of $X^{j=1}$ and noise $U^{X^2}$, and that $Y$ is a function $f^Y$ of both $X^{j=1}$ and $X^{j=2}$, as well as noise $U^Y$. Thus an SCM can be used to specify a dependency structure, and this dependency structure can, in turn, be represented graphically, using the Directed Acyclic Graph in Figure~\ref{fig:exampledag}. As it can be seen from the figure, the unobserved, exogenous noise variables $U$ are not usually explicitly represented, and neither are the functions $f$. DAGs are nonparametric, meaning that the functions which describe the causal mechanisms linking variables are not restricted to belong to a certain family. Thus, only the directional, structural relationships between variables are indicated in the DAG, thus making them an efficient means to communicate causal relationships. The term $pa$ for parents thus becomes quite intuitive. If two variables $X$ and $Y$ are adjacent in the graph, that is, there exists a directed arrow between them $X$, then variable $X$ is \textit{parent} of \textit{child} $Y$ if the direction of the arrow is from $X$ to $Y$, \textit{i.e.} $X \rightarrow Y$. Extending this terminology, variables which are parents of $X$ are therefore \textit{ancestors} of \textit{descendent} $Y$. One can see from the DAG on the left in Figure~\ref{fig:exampledag} that $Y$ has two parents, such that $pa(Y) = \mathbf{X}$.

Thus returning to Pearl's statement about a causal effect, which holds that $p(y|do(\mathbf{x}))$ involves the replacement of all equations for $\mathbf{X}$ with $\mathbf{x}$:

\begin{equation}
\begin{split}
    X^{j=1} :=x^{j=1}\\
    X^{j=2} := x^{j=2}\\
    Y := f^Y(X^{j=1} =  x^{j=1}, X^{j=2} = x^{j=2}, U^Y).
\end{split}
\label{eq:examplescmint}
\end{equation}

Thus, part of the dependency structure is broken because we have intervened on the system of equations by setting $\mathbf{X} = \mathbf{x}$. The definition of causal holds that the consequence of such an action should involve a change to $Y$, if indeed some or all of $\mathbf{X}$ are causes of $Y$. In the real world, such an intervention could represent, for example, the assignment of a certain treatment. Indeed, this is one way in which RCTs differ from observational studies, because with observational studies we rarely have the opportunity to interfere with the causal system and intervene on a set of variables to observe any corresponding causal cascade.

DAGs are assumed to fulfil what is known as the Markov property, which is a key property on which much of the subsequent theory rests. The principal implication of the Markov property is that the graph factorises according to a set of implied \textit{conditional independencies}. To grasp the intuition behind this, is it worth remembering some basic rules of probability. The joint distributions of two independent and fair coin flips $X$ and $Y$, one after the other, can be written as $P(X,Y)$ (\textit{i.e.}, the probability of observing both $X$ and $Y$). The fact that both coins are fair and the flips are independent (\textit{i.e.}, $X \indep Y$) means that this joint distribution is equal to the product of the marginal distributions $P(X)P(Y)$. If the probability of a heads for coin $Y$ is 0.5, and the probability of a heads for coin $X$ is also 0.5, then the probability of flipping two heads in a row is $0.5\times 0.5 = 0.25$. The DAG for this scenario is simply two variables $X$ and $Y$, independent and unconnected. Consider now the case where $A \rightarrow B$ such that the second event $B$ is influenced by $A$. For example, drawing a card $A$ from a deck of cards without replacement influences the card drawn for event $B$. But if we know $A$, then we have all available information to describe the probability of a particular card at $B$. The graph $A \rightarrow B$ therefore factorises as $P(A,B) = P(A)P(B|A)$. These rules thus translate to DAGs very conveniently, such that the joint distribution of a graph comprising a set of $J$ variables $\mathbf{X}$ can be written as:

\begin{equation}
    P(\mathbf{X}) = \prod_j^JP(X^j |pa_j)
\end{equation}
 
 Thus, deriving the factorisation of the joint distribution for the DAG on the left of Figure~\ref{fig:exampledag} is as simple as tracing the dependencies in the graph, and is given as 
 \begin{equation}
     P(Y, X^{j=1}, X^{j=2}) = P(X^{j=1})P(X^{j=2}|X^{j=1})P(Y|X^{j=2}, X^{j=1}).
 \end{equation} 
 
 This decomposition implicates a set of \textit{d}-separation rules, which describes how different variables in the graph can be rendered independent given knowledge about others. Consider the following structures:
 
 \begin{equation}
     \begin{split}
         A \rightarrow B \rightarrow C\\
         A \leftarrow B \leftarrow C\\
         A \leftarrow B \rightarrow C.
         \end{split}
         \label{eq:chainfork}
 \end{equation}
 
 The first two structures are known as \textit{chains}, and the last one is known as a \textit{fork} or \textit{v-structure}. As a result of the structure in each of these examples, there exists statistical association/dependence between all variables. However, this dependence is `broken' if we condition on $B$. This can be stated as $A\indep C|B$, or equally, $C \indep A | B$. This is known as a statement of \textit{conditional independence}, and this concept will be used throughout this work. It is certainly intuitive in the graph $A \rightarrow B \rightarrow C$ that knowing $A$ tells us nothing about $C$ if we know $B$ already - there is no other way that $A$ can affect $C$ except via $B$. Hence, it is via the concept of \textit{d}-separation that we may also decompose the graph into a product of independent conditional or marginal distributions.
 
 There is one other key structure, besides, chains and forks, that should be described, and this known as the \textit{collider} structure: $A \rightarrow B \leftarrow C$. In this structure, $A \indep C$ without conditioning on $B$. In fact, conditioning on a collider, \textit{or a descendent of a collider} renders the two parents of the collider as dependent. \textit{i.e.}, for the collider structure, $A \not\!\perp\!\!\!\perp C | B$.  Note that all DAGs can be constructed as combinations of fork, chain, and collider structures.

\begin{figure}[h!]
\centering
\includegraphics[width=1\linewidth]{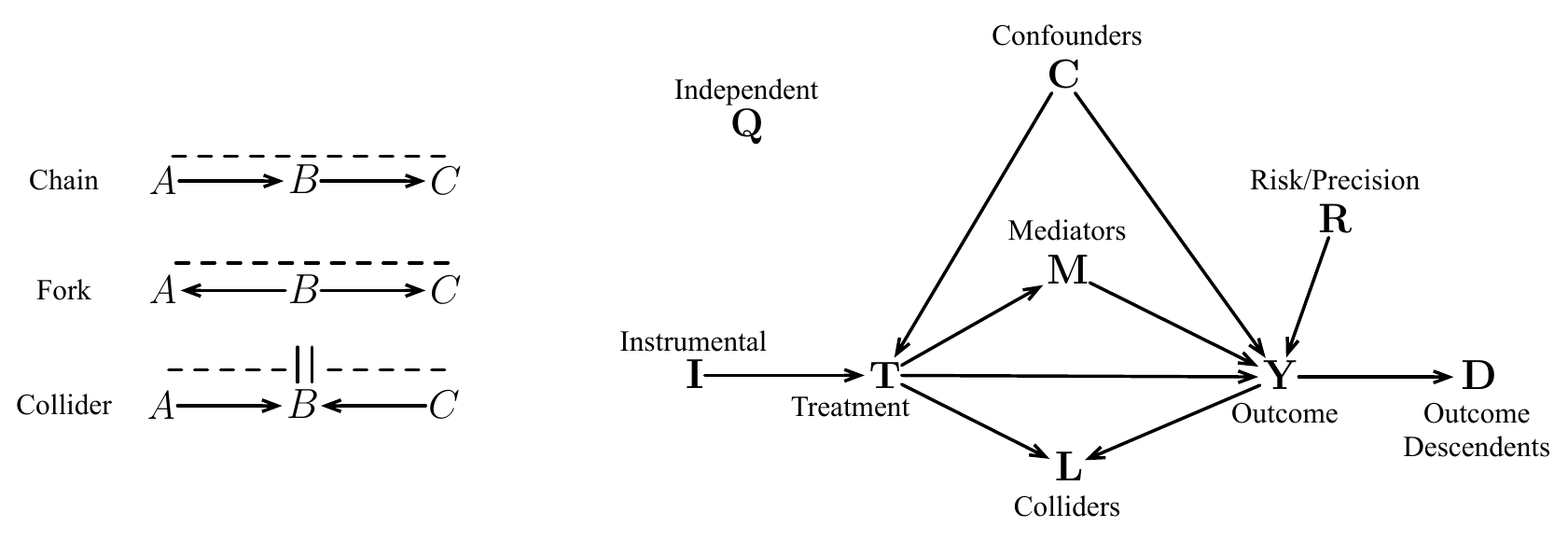}
\caption{\textbf{Left:} A summary of the three main graphical structures. Dashed line indicates statistical dependence. It can be seen that dependence exists between all three variables in the chain and the fork but is blocked by the collider. \textbf{Right:} A typical graphical structure with labelled variables based on their relationship to the causal effect between treatment variables $\mathbf{T}$ and outcome variables $\mathbf{Y}$. }
\label{fig:labelled}
\end{figure}

 More generally, there are a number of graph structures which are frequently used in practice and which are useful to define here. These structures are shown and labelled in Figure~\ref{fig:labelled}. In particular, \textit{confounders} represent variables (observed or unobserved) which are non-causal in terms of the path from treatments $\mathbf{T}$ to outcomes $\mathbf{Y}$, but for which there exists an induced statistical dependence through the path $\mathbf{T} \leftarrow \mathbf{C} \rightarrow \mathbf{Y}$. This is also known as a \textit{backdoor path} \parencite{Pearl1993backdoor, Pearl2009}, and usually, if we are interested in estimating the effect of treatment on outcome, we must condition or \textit{adjust} for the dependence between outcome on treatment via this non-causal path to be blocked (we will discuss this more later). \textit{Mediators} are causal variables which lie on an indirect path between treatment and outcome. \textit{Instrumental} variables are those which cause a particular choice of treatment but which does not directly affect descendants of treatment (for instance, proximity to a therapy centre). \textit{Risk} variables are those variables which explain variation in the outcome, but which are unrelated to the choice of treatment. These are also known as \textit{precision} variables \parencite{Cinelli2020} because they can be used to explain residual variation in the outcome thereby improving the precision of the estimation of the effect of treatment. We have indicated that these variables can be multi-dimensional (bold font notation) because in reality there may be multiple variables which fall into each category. 
 
\begin{figure}[h!]
\centering
\includegraphics[width=0.9\linewidth]{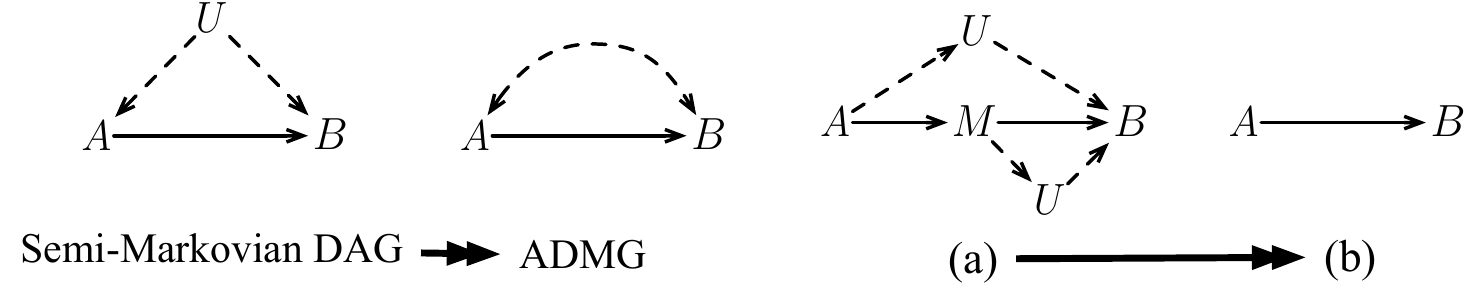}
\caption{\textbf{Left:} The graphical transition from a (semi-Markovian) DAG with unobserved confounder $U$, to the equivalent ADMG. \textbf{Right:} Showing a causal structure from $A$ to $B$ in graph (a) may comprise multiple indirect observed and unobserved paths, but which can nonetheless be \textit{projected} to graph (b).}
\label{fig:example_transitions}
\end{figure}

 If there exist unobserved confounders, the conditional independencies described above may not hold. Indeed, models containing unobserved confounders are known as \textit{semi-Markovian} models. One class of graph which facilitates the representation of unobserved confounders is known as Acyclic Directed Mixed Graphs (ADMGs). An example of an ADMG representation of a structure containing an unobserved confounder is given in Figure~\ref{fig:example_transitions} on the left. The unobserved confounding fork structure is simple replaced by the (dashed) bidirected arrow. One might wonder what happens if the unobserved structure is not a fork structure but a collider or chain structure. In such cases, and depending on the query of interest, it is possible to reduce the graph by a process known as projection \parencite{Richardson2017, Verma1990}, or simply as deleting intermediate variables \parencite{Glymour2001}. Consider the two graphs on the right hand side of Figure~\ref{fig:example_transitions}. In graph (a) there exists two sets of unobserved mediation paths, and one observed mediation path $A\rightarrow M \rightarrow B$. If one is, for instance, interested in the effect of $A$ on $B$, then all of these paths can be combined to graph (b). Indeed, it is conceptually difficult to imagine a circumstance where one could feasibly collect all intermediate mediators on a causal path, and in practice we deal with graphs representing the result of many (possibly an infinite number of) projection operations. The technique of projection will be useful later when we attempt to simplify our models such that they capture only the essential features necessary for a given research question or hypothesis.

\subsection{Causal Discovery}
\label{sec:causaldisc}
Now we have introduced relevant concepts relating to DAGs and SCMs, we turn our attention to the task of causal discovery, that is, the goal of deriving structure from data. Readers are encouraged to consult accessible introductions by \textcite{Glymour2019, Vowels2021DAGs, Heinze2018, Greenland1999, Forney2022causaledu}. In general, approaches to causal discovery are split into a number of categories: Constraint based approaches, which test for conditional independence relationships; asymmetry based methods, which test for distributional asymmetries under different cause-effect directions; score-based approaches, which derive a structure based on the fit to the data; and finally, interventional approaches, which involve some kind of experimental manipulation. We do not discuss interventional approaches in this work, but briefly review the others in turn, before summarising some limitations of causal discovery in general.

\textbf{Constraint-Based Conditional Independence Tests:} Given that the graph we intend to learn encodes a set of conditional independency relations, it should not be surprising that one of the most popular ways to learn structure is by testing for such conditional independencies. For instance, the structures on the left of Figure~\ref{fig:labelled} encode the following independencies: $A \indep C | B$ for the chain and fork, and $A \indep C |\emptyset$ or $A \not\!\perp\!\!\!\perp C |B$ for the collider. Thus, we can test for these relationships using statistical conditional independence tests \textit{e.g.} testing between the following hypotheses $h_0 : A \indep B | \mathbf{C}$ vs. $h_1 : A \not\!\perp\!\!\!\perp B | \mathbf{C}$. There are two principal challenges to such an approach, firstly, conditional independence tests are data hungry, in that they are sensitive to small sample sizes (this is particularly true when the conditioning set is large) \parencite{Shah2020}. Secondly, the direction of the arrows in the chain structure are not distinguishable, and the chain structure is also not distinguishable from the fork structure. In other words, these two (common) structures encode the same conditional independencies. This leads to what is known as a \textit{Markov Equivalence Class} (MEC). As such, in some cases it is only possible to identify the graph \textit{skeleton}, which has only undirected edges, resulting in a number of possibilities for the edge orientations which are not discernible via conditional independence tests.

 \begin{figure}[h!]
\centering
\includegraphics[width=1\linewidth]{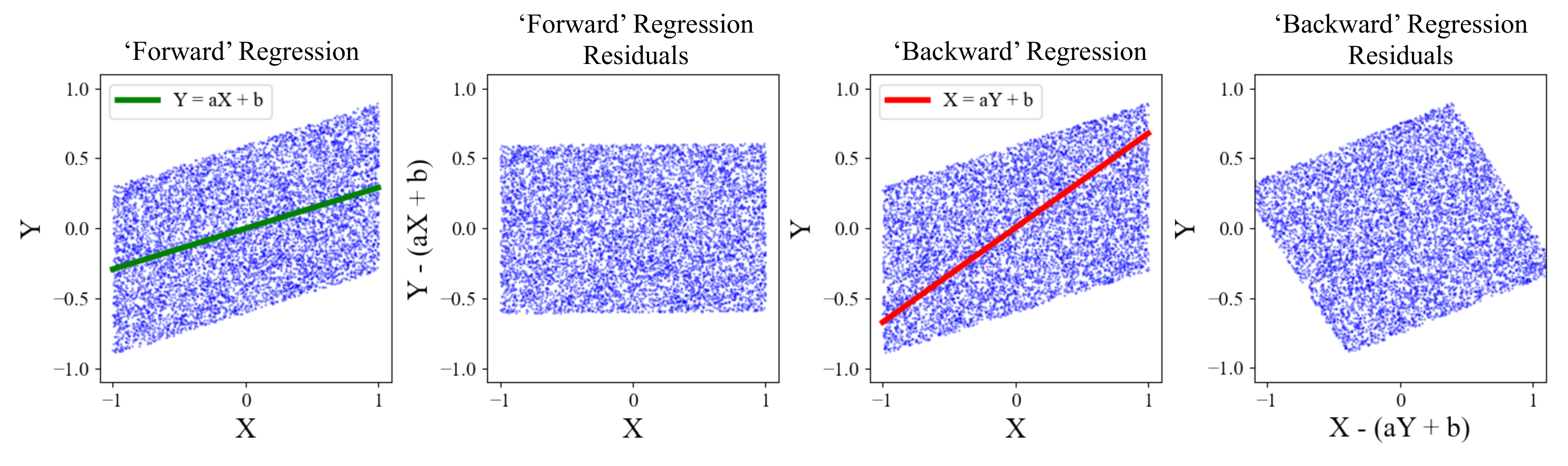}
\caption{Demonstrating the directional asymmetry of non-Gaussian residuals (adapted from \cite{Vowels2021DAGs,Peters2017})}
\label{fig:cd_regs}
\end{figure}

\begin{figure*}[t]
    \centering
    \begin{subfigure}[b]{0.48\textwidth}   
        \centering 
        \includegraphics[width=.6\textwidth]{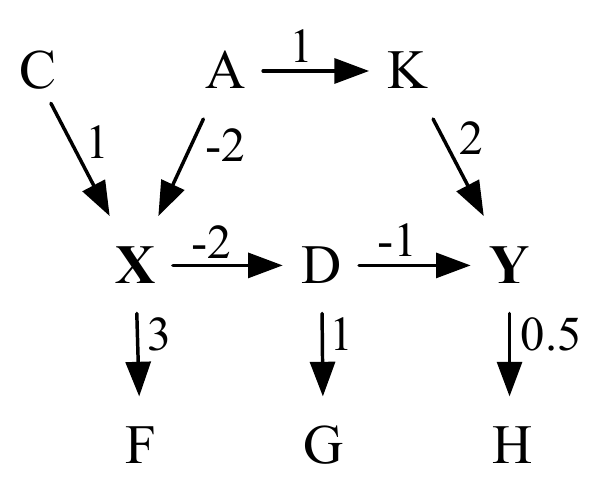}
        \caption[]%
        {{\small Test graph \parencite{Peters2017}.}}
        \label{fig:testgraph}
    \end{subfigure}
    \hfill
    \begin{subfigure}[b]{0.48\textwidth}
        \centering
        \includegraphics[width=.9\textwidth]{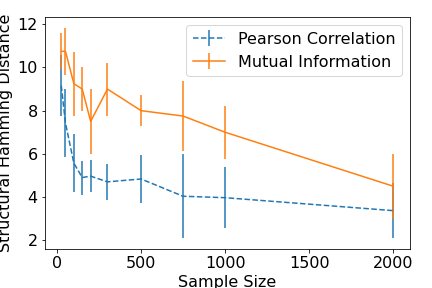}
        \caption[Network2]%
        {{\small Structural Hamming Distance (SHD).}}    
        \label{fig:shds}
    \end{subfigure}
    \hfill
    \begin{subfigure}[b]{0.48\textwidth}   
        \centering 
        \includegraphics[width=.9\textwidth]{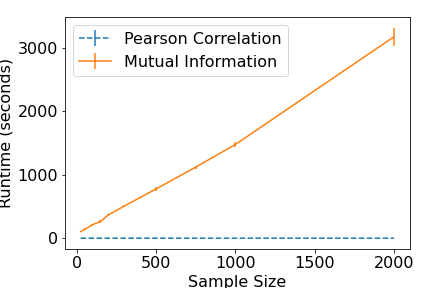}
        \caption[]%
        {{\small Runtime (seconds).}}
        \label{fig:runtime}
    \end{subfigure}
    \caption{Causal discovery results for the provided graph. Metrics: Structural Hamming Distance and runtime (seconds). Algorithms: the \texttt{pgmpy} \parencite{pgmpy} implementation of the PC \parencite{Spirtes2000} causal discovery algorithm using a linear correlation conditional-independence test and a non-parametric  conditional mutual independence based \parencite{Runge2018} conditional-independence test.}
    \label{fig:cdresults}
\end{figure*}

\textbf{Example Constraint-Based Causal Discovery:} In Figure~\ref{fig:cdresults} we show results for the well-known PC algorithm \parencite{Spirtes2000}, which uses conditional independence constraints to derive a putative structure. The results illustrate the differences in performance (across a range of sample sizes) between a linear, correlation-based conditional independence test, and a nonparametric, k-nearest neighbours conditional mutual information based conditional independence test \parencite{Runge2018}. In Figure~\ref{fig:testgraph} we show the true, 9-node, 9-edge graph for the underlying data generating structure. All exogenous (not shown) variables are Gaussian. In Figure~\ref{fig:shds} we show the Structural Hamming Distance, which is a measure of how successful the algorithm was at correctly inferring the graph (smaller the better). Finally, in Figure~\ref{fig:runtime} we show the algorithm runtime (in seconds). Firstly, it can be seen that the correlation based measure is notably better at inferring the structure than the mutual information based approach, particularly for small sample sizes. It can also be seen that the time taken to run the nonparametric version increases linearly with the sample size (over 50 minutes for a sample size of 2000) vastly exceeding the runtime of the correlation approach (which has a run of 0.26 seconds for a sample size of 2000). Thus the price paid for not having to make parametric assumptions is one of both computation time and, for a parametric data generating process, accuracy. Of course, one expects that in cases where the parametric assumption does not hold, the advantages of non-parametric conditional independence tests become self-evident.

\textbf{Distributional Asymmetries:} If one admits the possibility of either or both non-linearity and non-Gaussian exogenous noise in the structural equation models, it is possible to discern the direction of cause and effect. Consider the plots in Figure~\ref{fig:cd_regs}. The SCM used to generate the data is given by:

\begin{equation}
\begin{split}
        U^X \sim U(-0.5, 0.5)
        X := U^X\\
        Y := X + U^Y
\end{split}
\end{equation}

In the leftmost plot, we show the result of regression $Y$ onto $X$, which aligns with the structural direction $X\rightarrow Y$. In the second plot from the left, we shows the residuals from this regression which are, as expected, uniformly distributed between $[-0.5, 0.5]$, and there is no dependence between the residuals and $X$. Now, if we instead regress $X$ onto $Y$ (third plot from the left), which conflicts with the true structural directionality, the residuals (rightmost plot) exhibit dependence with $Y$. Thus one can test for this residual dependence and thereby identify the (in)correct directionality of the cause-effect relationship. Variations on these methods have been developed to account for variations where, for example, the noise is additive but the functions relating variables are non-linear, \textit{e.g.} $Y = f(X) + U^Y$. Interested readers are encouraged to consult work by \textcite{Hoyer2008b, Mooij2016, Peters2017}.

\textbf{Score Based:} It is also possible to derive a graph according to various fit/likelihood metrics. The idea is that the structure which best explains the data is the most likely to be correct. Unfortunately, these methods have been heavily criticised, particular because it is well known that `fitness to data is an insufficient criterion for validation causal theories' \parencite[pp.61]{Pearl2009}. Indeed, a number of methods which leverage model fit have been shown to exhibit problematic behaviour \parencite{Kaiser2021, Reisach2021}. Such a phenomenon is not hard to imagine if one considers that non-causally related but otherwise correlated variables could be useful for explaining relationships \parencite{Yarkoni2017}. Thus, the inclusion of non-causal relationships could also aid in explaining the data and thereby improving likelihood metrics. One of the appeals of likelihood driven methods is that they are amenable to continuous optimization methods which opens the door to a large range of approaches for structure learning \parencite{Vowels2021DAGs}. This, in turn, makes such methods appealing for structure learning in scenarios with a large number of variables (possibly thousands), because whilst constraint based approaches have to test an exponential number of conditioning statements (exponential in the number of nodes), score-based approaches can continuously optimize a fixed $J\times J$ adjacency matrix representing the graph structure with the highest log-likelihood (where $J$ is the number of nodes).

\textbf{Further Remarks:} As described above, causal discovery methods relying on conditional independence tests can identify a graph only up to the Markov Equivalence Class (MEC). Indeed, some related methods, including the popular Pairwise Markov Random Field (PMRF), use weaker causal heuristics and cannot even achieve this level of specificity (the edges are undirected), making the solution highly ambiguous \parencite{Epskamp2018PMRF, borkulo2014PMRF, Robinaugh2020, Ryan2022}. Furthermore, constraint based approaches such as the PC-algorithm \parencite{Spirtes2000}, which we demonstrated above, can be very computationally expensive when nonparametric conditional independence tests are used. Score based approaches provide a way around this, because unlike constraint based approaches, score based approaches do not need to undertake an exponential number of tests, but can rather optimize over a fixed graph size. If one uses neural-network based approaches, one can combine multiple heuristics simultaneously. One example, Structural Agnostic Modeling (SAM) \parencite{Kalainathan2020}, leverages a combination of conditional independence tests, asymmetries, \textit{and} likelihood to derive a putative structure, and therefore represents a strong candidate. Unlike the PMRF, and other parametric methods for causal discovery, SAM makes no assumptions about the underlying functional or parametric form of the data, making it a flexible approach (and one which we will demonstrate in later sections). However, as a neural network based method it is also subject to convergence and inconsistency issues \parencite{goodfellow}. An additional challenge relates to statistical power. Conditional independence tests, in particular non-parametric tests, are known to suffer in the presence of small sample sizes, where `small' depends on the number of nodes and underlying structure in the graph \parencite{Shah2020}.

\subsection{Causal Inference}
\label{sec:causalinf}
In the observational setting, the estimate of causal relationships and effects is challenging because of statistical dependence induced through confounding. One of the reasons why we adopt the graphical framework described above (DAGs, SCMs, etc.) is because this framework facilitates causal inference in clear and intuitive way \parencite{Greenland1999}. One alternative to the graphical causal framework is the Neyman-Rubin potential outcomes framework \parencite{Rubin2005}. There is some debate about the relative merits of each approach \parencite{Imbens2020}, but there is little doubt about the intuitive nature of the graphical representations. Indeed, one of the key components for deriving causal effects is to adjust for confounders, and confounders elude transparent definition within the potential outcomes framework. In contrast, one can identify confounders from a quick visual inspection of the graphical representation. Furthermore, the graphical representations makes it trivial to identify other classes of variables, such as those highlighted in Figure~\ref{fig:labelled}. As such, we assume the task of causal inference follows from the specification of a graph (\textit{e.g.}, a DAG).

 \begin{figure}[h!]
\centering
\includegraphics[width=0.4\linewidth]{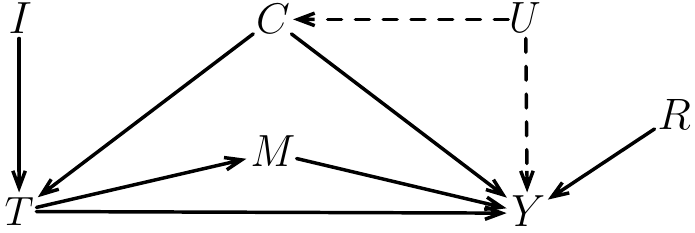}
\caption{Example graph.}
\label{fig:exampleinf}
\end{figure}

Once the graph is known, one could conceivably the Structural Equation Modelling (SEM) framework \parencite{Kline2005, Loehlin2017} to estimate the path coefficients on each cause-effect link. However, SEM assumes the functional form is known \textit{a priori} and thereby increases the chances for model misspecification. In particular, it is common to assume the causal relationships between variables in an SEM are \textit{linear}, and in many cases we can be quite sure this is more likely to be incorrect than it is to be correct \parencite{vanderLaan2014}. Furthermore, in order to reliably estimate all effects in an SEM, we must correctly specify the full structure. In contrast, and as we will discuss in later sections, it is often possible to consider a subgraph or substructure, thereby reducing the chances of structural misspecification by reducing the scope of our inquiries. Indeed, the current proposed method requires the specification of a target estimand, thereby narrowing/focusing our query to a particular attribute within our given structure. For instance, assume the graph in Figure~\ref{fig:exampleinf} represents the true causal process, and assume that we are interested in estimating the average causal effect (ACE) or average treatment effect (ATE) of binary treatment $T$ on binary outcome $Y$. The targeted \textit{estimand} (\textit{i.e.}, the quantity we wish to estimate) is given as:

\begin{equation}
    \Psi = P(Y=1|do(T=1)) - P(Y=1|do(T=0)).
    \label{eq:exinf}
\end{equation}

In words, we wish to estimate the difference in the probability of outcome $Y=1$ for when we `do' $T=1$ and for when we `do' $T=0$. This quantity specifically refers to the result of the causal links between $T$ and $Y$, and does not, therefore, concern the simultaneous estimation of all causal effects in the graph (as would be the case with SEM). Here, the $do$ operator signifies that we wish to estimate the effect of an \textit{intervention}, and this estimand is therefore to be clearly distinguished from a difference between conditional probabilities $P(Y=1|T=1) - P(Y=1|T=0)$ because it concerns counterfactual quantities. In practice, we will have data for the outcome given one of the treatments but not the other(s) (this can also be conceptualised as a missing data problem). Note that the potential outcomes framework \parencite{Rubin2005}, which is popular in economics and epidemiology, represents the difference between the counterfactual quantities (as in Eq.~\ref{eq:exinf}) as $Y(1) - Y(0)$, which is the difference between the outcome had the participant received treatment, and the outcome had the participant not received treatment. Again, in practice we will only have access to one of these two quantities

We can return to Pearl's definition of causality in the previous section, the effect of intervention is to cut the links between the parents of treatment $T$ and $T$ itself, and to assign the value specified. Of course, unless one is able to experiment in this manner, it will not be possible to literally achieve this. Therefore, one must establish whether this estimand is \textit{identifiable} from the data. That is to say, does there exist an expression for this estimand which is a function of the observed distribution and which can be estimated in principal and which is \textit{not} expressed in terms of $do$ operators but which is otherwise equivalent?\footnote{Certain difficulties or even impossibilities may still arise with respect to estimation, as discussed by \textcite{Maclaren2020}.} Note that if there exist no unobserved confounders, identification is always possible, and the goal of deriving an expression for the estimand in terms of functions of the observational distribution reduces to an application of what are known as the rules of \textit{do}-calculus \parencite{huang2012pearls}. However, rather than describing these rules, which would take some time, we instead focus on one popular and widely applicable method for establishing identification known as \textit{backdoor adjustment}. This method cannot be used to identify all estimands, but it is a staple technique that is widely applicable to common models.

In order to fulfil what is known as the backdoor criterion, one must identify a set of nodes which `block' all `backdoor paths' between the cause and the effect. To do so requires the utilization of the \textit{d}-separation rules introduced above. Once one has identified this set, which we will for now call $\mathbf{Z}$, we can use the following identification formula:

\begin{equation}
    P(Y=1|do(T=t)) = \sum_Z P(Y=1|T=t, \mathbf{Z}) P(\mathbf{Z})
\end{equation}

This tells us that the probability of the outcome (which might, for instance, be a state of recovery) for a given treatment is equal to the conditional probability of recovery for that treatment and for each value of $\mathbf{Z}$, averaged over all possible values for $\mathbf{Z}$. For binary outcome this is equivalent to finding $\mathbb{E}_{y\sim P(Y|do(T=t))}[y]$ which, for continuous variables is equivalent to:

\begin{equation}
    \mathbb{E}_{\mathbf{z} \sim P(\mathbf{Z})}[\mathbb{E}_{y \sim P(Y|T=t, \mathbf{Z}=\mathbf{z})}[y]] = \int\int y p(Y=y|T=t,\mathbf{Z}=\mathbf{z})p(\mathbf{Z}=\mathbf{z}) dyd\mathbf{z}
    \label{eq:backdoorcont}
\end{equation}.

To find which variables belong in $\mathbf{Z}$, one can begin by identifying all non-causal pathways between $T$ and $Y$. A non-causal pathway one in which there does not exist a directed route from cause to effect. For instance in the graph $A \rightarrow B \rightarrow C \rightarrow D$, there is a causal path from $A$ to $D$, but in $A \leftarrow B \rightarrow C \rightarrow D$, there is not. Then, one must establish whether, for each non-causal pathway between cause and effect, the cause and effect are already \textit{d}-separated either by a collider. For instance, for the non-causal sequence of links $A \rightarrow B \leftarrow C \rightarrow D$, there is no flow of statistical association between $A$ and $D$ because of the collider at $B$. As such, this non-causal path is already blocked. The non-causal paths which remain are therefore \textit{backdoor} paths, and result in a flow of statistical information between cause and effect. The graph $A \leftarrow B \rightarrow C \rightarrow D$ is an example of a graph with a backdoor path between $A$ and $D$.

Returning to the graph in Figure~\ref{fig:exampleinf}, we can identify two backdoor paths. The first is : $T \leftarrow C \rightarrow Y$, and the second is $T \leftarrow C \dashleftarrow U \dashrightarrow Y$. The other two paths between $T$ and $Y$ are both causal paths, in that one can follow the directed paths between the cause and the effect. In order to block the backdoor paths, one needs only to use the rules of \textit{d}-separation. In particular, for the fork $A \rightarrow B \leftarrow C$, one can \textit{d}-separate $A$ and $C$ by conditioning on $B$: \textit{i.e.}, $A \indep C | B$. Applying this to the graph, we can see that conditioning on $C$ blocks both backdoor paths, even those one of the paths involves an unobserved confounder. This would not be possible if, for instance, there was an unobserved confounding relationship $T \dashleftarrow U \dashrightarrow Y$, for which the effect would be un-identifiable (there would be no way to block the backdoor path through $U$). Thus in order to estimate the target quantity one needs to use:

\begin{equation}
    \Psi = \sum_C P(Y=1|T=1, C=c)P(C=c) - \sum_C P(Y=1|T=0, C=c)P(C=c).
\end{equation}

As above, for a binary outcome $Y$, $P(Y=1|T=1, C) = \mathbb{E}_{y\sim P(Y|T=1, C)}[y]$, and in this case the expression above is equivalent to:

\begin{equation}
    \Psi = \mathbb{E}_{c\sim P(C)}[\mathbb{E}_{y\sim P(Y|T=1, C=c)}[y] - \mathbb{E}_{y\sim P(Y|T=0, C=c)}[y]].
\end{equation}

Or, alternatively:

\begin{equation}
    \Psi = \int \int yp(y|t=1, c)p(c) dy dc - \int\int yp(y|t=0, c)p(c) dy dc.
\end{equation}

It is worth noting that each inner expectation \textit{e.g.} $\mathbb{E}_{y\sim P(Y|T=t,C=c)}[y]$ can be approximated via regression. The outer expectation can be achieve by averaging the predictions from the regression over all datapoints (thereby deriving an expectation over $C$). Such a process is known as \textit{plug-in} estimation, because it involves first fitting an estimator for the inner expectation, which we can call $\hat Q(T,C)$ (where the $\widehat{\mbox{hat}}$ notation indicates it is an empirical estimator), which then plug-in to the expression for the estimand (we will leverage this plug-in estimation approach for the practical demonstrations). The empirical approximation to the estimand thus becomes:

\begin{equation}
 \hat{\Psi} = N^{-1} \sum_i^N \left[\hat Q(T=1, C) - \hat Q(T=0, C) \right].
\end{equation}

In this example, adjusting for the confounder $C$ is thus a simple case of conditioning our regression on this variable for the inner expectation (which we can approximate with a regression model), and the marginalizing over this variable to compute the outer expectation (assuming our estimand is the Average Treatment Effect). The process of identifying the confounders was made straightforward by the graphical representation of the underlying causal theory. Of course, this assumes our graphically representation and/or our theory is correctly specified. In cases where the theory is structurally misspecified, it is possible to adjust for variables which are not confounders. For instance, imagine if the true causal process is given by Figure~\ref{fig:exampleinf} but that in our model, variables $M$ and $C$ were swapped. In such a case, we may mistakenly choose to condition on $M$ rather than $C$. This would, in fact, block the indirect effect from $T$ to $Y$ via $M$, thereby biasing our estimation of the causal effect. Thus there exists `good and bad controls', and one must apply the \textit{d}-separation rules to understand whether the inclusion of a particular variable hurts or helps estimation. For an excellent review of good and bad adjustment variables, readers are encouraged to consult the recent work by \textcite{Cinelli2020}.

Finally, note that there exist many target quantities that may be of interest other than the ATE. For example, one may be interested in the Probability of Necessity and Sufficiency (PNS) \parencite{Mueller2021ite, Pearl1999}, the effect among the treated, or risk and odds ratios \parencite{Rothman2008}. The PNS is a particularly interesting counterfactual quantity which helps us fractionate causal effects into the most meaningful components. Specifically, to paraphrase a statement by \textcite{Mueller2021ite}, PNS tells us the number of recovered patients that should credit therapy for their recovery are those who would recover if they had received therapy \textit{and} would not recover if they had not received therapy.


\section{Stage A: Theory Development and Causal Discovery}
\label{sec:stageA}
Having reviewed a number topics relevant to the presently proposed pipeline, we begin with a description and discussion of Stage A of the pipeline, which involves the iterative development and formal specification of a theory. As per the introduction and background sections, we assume the use of graphs (such as DAGs) for the formalisation of the theory, because they are visually intuitive and can be operationalized fairly quickly. Of course, there exist other tools which may be more suitable (complex systems theory, for example, \cite{Sayama2015, strogatz}), and the choice depends largely on the specifics of the phenomenon or theory being formalized.

As with much of current practice in psychology and social science, the current state and specification of theory has come under heavy critique.  Indeed, it has been argued that the replication crisis and problems of statistical practice have received a disproportionate amount of attention in comparison with the problems with theory development in psychology \parencite{Eronen2021, muthukrishna}, and that instead of just repeatedly urging researchers to preregister their studies \parencite{Claesen2019, Wagenmakers2012}, utilize more adaptive modeling techniques \parencite{Vowels2021, Yarkoni2017}, and collecting more reliable data \parencite{Vankov2014, Maxwell2004, Crutzen2017}, we should, in addition, be looking much more closely at our theories \parencite{Eronen2021, Oberauer2019, muthukrishna, Fiedler2017, Borsboom2021}. \textcite{muthukrishna} have argued that psychology exhibits a ``\textit{potpourri}'' of disconnected empirical findings which lack in an overarching or connecting framework. \textcite{Scheel2022} argues that theories and associated hypotheses are vague and underspecified, often presented as verbal statements with degrees of ambiguity typical for their lack of formalism. This makes it difficult to know what exactly is being tested, both in terms of tying the hypothesis to a specific attribute of the theory, and knowing what the theory itself represents. Similarly, \textcite{Oberauer2019} encourage theory specificity to facilitate direct links between theories and formally constructed hypothesis testing. In contrast, \textcite{Eronen2021} argue that in light of the complexity and nuance associated with psychological theory, more mathematical/statistical/computational formalism is not necessarily the answer; what is needed (at least first), is a broader range of consistent evidence for certain phenomena in order to constrain the possible space of compatible theories.  They also argue that `a key feature of good theories is that they should, in one way or another, track causal relationships', even if the discovery of such causes can be prohibitively difficult and impractical \parencite{Eronen2020, Eronen2021}. 

Whilst we agree with \textcite{Eronen2021} that reasoning about causality is immensely challenging, particularly in the complex domain of psychology, and whilst we also agree the field could strongly benefit from amassing a broader range of high-quality evidence, we nonetheless agree with \textcite{Oberauer2019, Scheel2020} and others that in order to test the theories reliably, repeatably, and quantitatively, we must be able to formalize them. Hence why we argue for the use of frameworks such as graphical models for this formalization. Furthermore, \textcite{Eronen2021} also argues that experiments in psychology are frequently `fat-handed', making it difficult to perform experimental interventions without simultaneously and unintentionally affecting a large number of variables simultaneously, thereby making causal inference in psychology especially challenging. However, in order to do our best in spite of these challenges, we must understand and have the tools to do so. Indeed, the goal of causal discovery remains central to psychology and science in general.

\subsection{Integrating Causal Discovery}
The principal tool used in Stage A which differs from most theory development processes is the use of causal discovery algorithms. The theory behind such algorithms has been introduced above, but note that there exist a number of decisions regarding the specific method used. For instance, should one use a constraint-based methods (which tests for conditional independencies), score-based methods (which test for model goodness of fit), or distributional asymmetries? If one chooses a method integrating conditional independence tests, should one assume parametric form (which implicates tests of partial correlations for continuous variables, and Chi-squared tests for discrete variables), or non-parametric form (which implicates tests for conditional mutual information). Furthermore, the choice of method has an impact on the nature of the putative structure. For instance, the seminal PC algorithm \parencite{Spirtes2000} yields a Markov Equivalence Class (MEC) and does not yield asymptotically correct results in the presence of unobserved confounding. The Structural Agnostic Model (SAM) \parencite{Kalainathan2020}, in contrast, is a continuously optimized neural network method, without formal guarantees, which provides a graph with possibly bidirected edges and cycles. It is possible to enforce and acyclicity constraint to encourage `DAG-ness', but some post-processing of the graph may be required. The point here is that there exist a large number of techniques, and their behaviour and performance varies accordingly. Reviews of such considerations are provided by \textcite{Glymour2019} and \parencite{Vowels2021DAGs}. 

In addition, we must emphasize that the ways in which researchers leverage the output is not intended to overrule theory or prior evidence. Robust causal discovery is challenged by finite sample variation, quality of measurement, various distributional assumptions, and unobserved confounding. In light of these challenges, we recommend that causal discovery methods be used to supplement and validate certain aspects of a theory, but not to replace the theory itself. Indeed, it may be the case that the structure underlying certain key processes relating to a particular theory already have some strong support, in which case we can use causal discovery to check this structure is reflected in the data using causal discovery. Alternatively, in the specification of the full theory, there may be a range of more distal causes for which the relationships with the core process (and the relationships with each other) are unknown. In such cases, causal discovery methods can yield a putative structure which can be reviewed and adjusted if necessary. The main incentive for the introduction of causal discovery techniques into the research pipeline for psychologists is to provide them with a means to construct a mathematical representation of their theories. The graphical framework and the associated causal discovery algorithms, in our opinion, provides the most accessible route for achieving this.

In the next subsections, we provide a step-by-step example of how to implement causal discovery in order to specify a theory.

\subsection{Step 1 - Initial Theory and Data Collection}
\label{sec:stageAstep1}
We assume that there exists some prior theory, even if this theory happens to be immature or vague. This initial theory might derive or be supported by qualitative studies which broadly captured the key components of a phenomenon from people with experiences relevant to this phenomenon. Whilst there exist numerous ways to specify this initial theory, we choose DAGs as a method growing in popularity and which are intuitive insofar as they visually represent key causal relationships. We therefore recommend starting by specifying the parts of the theory to which the research questions and hypotheses relate as a DAG, or a set of plausible DAGs. The DAGs(s) may be incomplete insofar as the researcher does not yet know or have sufficient prior knowledge to confidently specify all links, but note that the inclusion of a link is a much weaker assumption than the absence of a link (the causal effect in the presence of a link can still be zero, but the causal effect in absence of a link \textit{must} be zero).

In the current demonstration we use secondary data acquired with permission from \parencite{Umberson2015}.\footnote{Further details about these data can be accessed at: \url{https://www.icpsr.umich.edu/web/ICPSR/studies/37404/summary} } The data are dyadic and comprise responses from 403 couples for the following variables: the age and gender of both partners (Age R, Age S, Gender R, Gender S), the relationship type (Rel Type), number of adversities experienced before the age of 18 for both partners (Advrs R, Advrs S), retrospective level of distress in response to an emotionally distressing event experienced by each partner (Distress R, Distress S), restrospective level of experienced partner support in response to the emotionally distressing event (Support R, Support S),  length of cohabitation for the partners (Cohab Len), perception of the couple's current success at coping for each partner (DCI Dyd R, DCI Dyd S), perception of each respondent's own current supportive response to their partner's distress (R DCI R, S DCI S), perception of their partner's current supportive response to the respondents distress (S DCI R, R DCI S), relationship satisfaction for each partner (Rel Sat R, Rel Sat S), and levels of depressive symptoms (Dep R, Dep S). Given the didactic nature of the demonstration, we refrain from lengthy descriptions about the methodology for the data collection and the literature surrounding support and relationship satisfaction etc. In practice, of course, such reviews and methodologies may include information which is key to representing the theory graphically. The concomitant graphs may also be much more complex and contain many more variables. As such, the procedure we describe below is \textit{only used as an illustrative demonstration} of the general methodological process.

\begin{figure}[h!]
\centering
\includegraphics[width=0.7\linewidth]{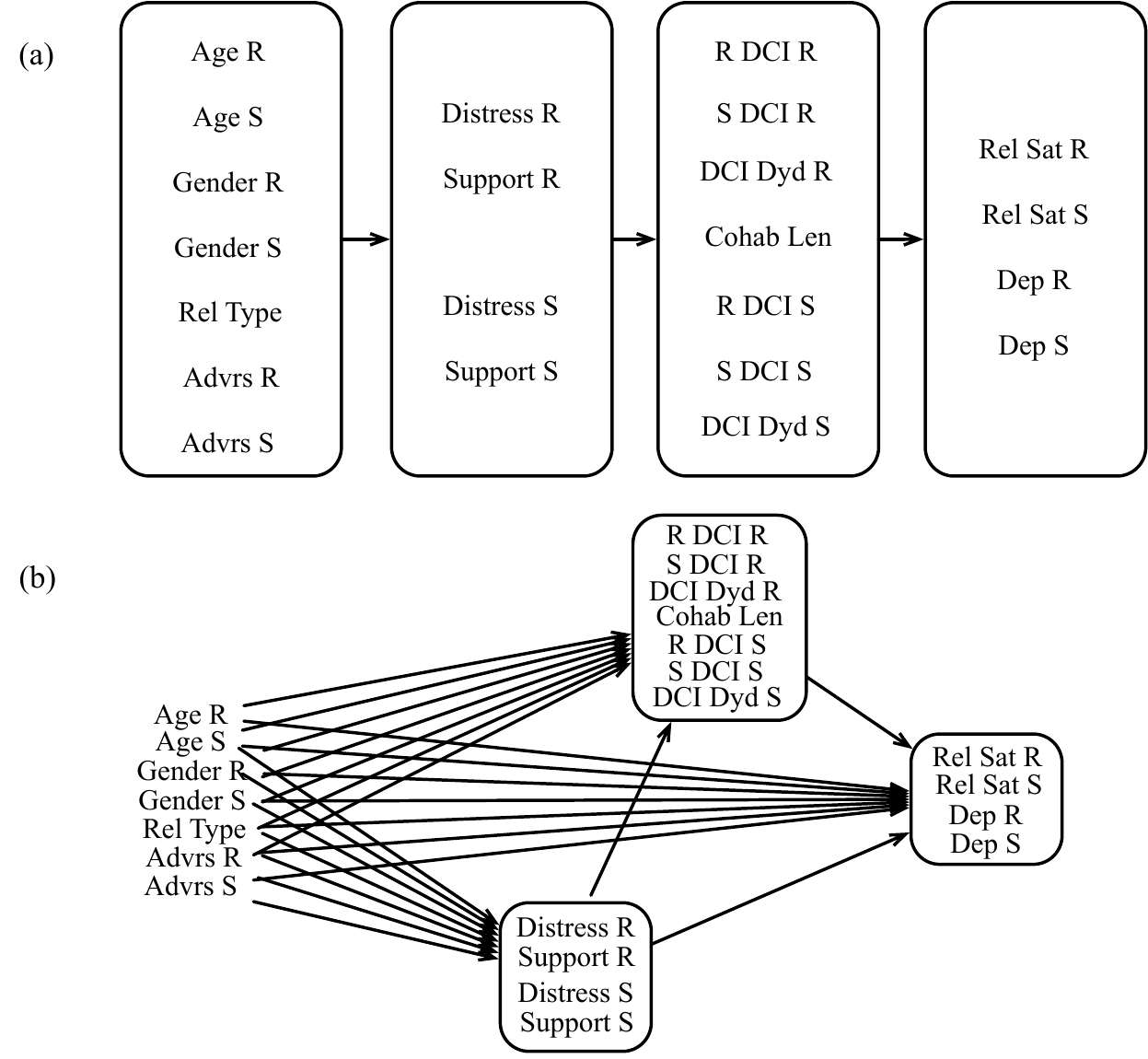}
\caption{Example graphs for initial theory.}
\label{fig:practical_ex_theory}
\end{figure}

Ideally, the graphical representation of the theory should be independent of the specifics of a particular dataset. However, measurements may take on a different meaning depending on the scales used and how the data were collected, and this can affect the structure in turn. This is particularly relevant for researchers using secondary data. For instance, in the secondary dataset being used for this demonstration, participants were asked to recall and report \textit{retrospective} levels of distress as opposed to \textit{current} levels of distress. Even if we are interested in levels of distress more generally, the specifics of the scale may, at least for this example, influence where in the temporal ordering the variable is placed, even though it may be collected at the same point in time as variables relating to current experiences. This is potentially problematic because, of course, current levels of distress could reasonably be expected to affect the perception of previous levels, leading to a `backwards in time' or, at least, a bidirectional effect. In contrast, one can plan a data collection methodology which purposefully avoids such difficulties by, for instance, collecting longitudinal data where the constructs only ever relate to current experiences (even if those current experiences relate to past events). It is worth noting that the degree to which temporal orderings are compromised depends also on how stable the constructs or attributes are. For instance, it may be reasonable to assume that the number of adverse childhood adversities experienced, as reported by the participant as an adult, will remain relatively constant regardless of when it is collected. Hence, these stable and/or `historic' variables are likely to precede the others in the system (and indeed they are placed as such in the graphs of Figure~\ref{fig:practical_ex_theory}). 

In cases where the researcher has access to secondary dataset (as we do for this example) the researcher should not feel constrained by what variables already exist this dataset. Indeed, before embarking on the analysis journey, the process of building a graphical representation may be an opportunity to note whether key variables are missing, and whether any initial research questions can actually be answered using these data. Similarly, depending on the research question, variables which are relevant for a general theory may not be required for analysis itself - this depends on the model structure and (in the graphical modeling framework) the consequences of the \textit{d}-separation rules.

We present two initial conceptual graphical representations of the theory underlying the processes which relate to these variables in Figure~\ref{fig:practical_ex_theory}. The structure is initially organized according to a supposed temporal ordering between groups of variables. In Figure~\ref{fig:practical_ex_theory}a we see demographic/`global'/trait level variables on the left, then retrospective reports of support and distress, followed by current coping constructs, and finally by the outcomes relationships satisfaction and depression. In Figure~\ref{fig:practical_ex_theory}b we also incorporate direct and indirect mediation between certain groups of variables, as well as an explicit specification of the relationship between our global/demographic variables and these groups. Indeed, in Figure~\ref{fig:practical_ex_theory}b we see an acknowledgement for possible links between all global variables and all other variables.

In some cases, any further mathematical specification of a phenomenon may be too difficult to do without high likelihood of misspecification, particularly if there exist known problems with data collection/methodology/measurement, the complexity of the phenomenon is too high and it is not clear how to simplify it, or possibly because of a paucity of supporting evidence. Nonetheless, it may be possible use a loose or abstract specification of the theory to inform a predictive approach, and researchers may pursue such an approach after this Step 1, or indeed after any step during Stage A. A predictive approach can be useful in providing either an initial indication of the predictive validity of the theory, or as a tool for automating decision processes (such as automating assessments).

\subsection{Step 2 - Data Collection and Constraining Discovery}
\label{sec:stageAstep2}
Once we have the initial structural representation of our theory, such as the one in Figure~\ref{fig:practical_ex_theory}b we have two options. If we already have secondary data, we can proceed to causal discovery, and if not, we can begin to plan our data collection methodology. In case of the latter, it may be worth fully expanding the graph to include all plausible links in order to ensure that, in the worst case, our data collect methods are sufficient to capture the complexity of the underlying phenomenon. For example, we may `unroll' our graph over time, representing repeated measures as individual nodes in our graph. This may help us understand the necessary frequency at which we collect variables, which can be key for understanding time fluctuating phenomena \parencite{VowelsSpectralTutorial}. Doing so may also help us understand which scales to use - if it is necessary to collect data frequently it may be necessary to use scales which are sensitive to short term variation and which are short enough to mitigate problems with respondent fatigue \parencite{Lavrakas2008}. Most importantly, expanding the initial representation to capture `worst-case' complexity helps us understand which variables (particularly confounders, which we need to adjust for) are necessary to help us answer our research questions. 

On the other hand, assuming we are ready to undertake causal discovery, we can begin by constraining our discovery according to the missing links in our initial graphical representation. It is worth noting that the absence or omission of a link represents a much greater assumption than the inclusion of a link \parencite{Vowels2021, vanderLaan2011}. Indeed, in the case where a superfluous link is specified, the effect can still be zero, thus no harm is done and no misspecification occurs. Conversely, it is clear that if a necessary link is omitted there is no way to compensate without introducing bias and the model will be misspecified. This is why, in Figure~\ref{fig:practical_ex_theory}b, we have specified the temporal dependence, but have not made any assumptions about the absence links between the grouped variables. 

\begin{tcolorbox}
\textbf{Application/Code Note :}
In the associated code, the constraints we have are specified in the graph `skeleton', which is a square adjacency matrix representing the discoverable cause-effect pairs.
\end{tcolorbox}

\subsection{Step 3 - Discovery}
\label{sec:stageAstep3}
Assuming we have some data with which to perform causal discovery, the next task involves the selection of a relevant algorithm. For a review of some well known options, readers are encouraged to reference \textcite{Glymour2019}, and for an extended review, see work by \textcite{Vowels2021DAGs}. The use of \textit{multiple} techniques/approaches to causal discovery is encouraged so as to provide a means to establish which discovered structures are consistent across algorithms. The approach we demonstrate in this work is the Structural Agnostic Model (SAM) \parencite{Kalainathan2020}, which makes no assumptions about the functional or parametric form (\textit{i.e.}, potentially non-linear and non-Gaussian, mixed-type variables). The approach is a neural network based \parencite{goodfellow} approach incorporating a number of constraints. Neural network based methods operate on the basis of gradient updates. A putative structure is proposed, and the gradient of some measure of performance is used to update the parameters of the neural network in such a way that this measure of performance improves. The approach includes a number of constraints, and these constraints prevent the network from, for example, simply proposing the fully-connected graph such that the likelihood is maximised. The constraints penalise the network for proposing graphs for which the conditional independencies implied by the putative structure contradict the conditional independencies measurable from the data, and also so that distributional asymmetries associated with cause-effect directions can be leveraged. Neural network methods a often have many thousands of parameters, and an optimal solution is rarely guaranteed. In order to mitigate problems with convergence to locally (rather than globally) optimal solutions, we perform 50 runs of SAM and take the average over these runs.

\begin{figure}[!]
\centering
\includegraphics[width=1\linewidth]{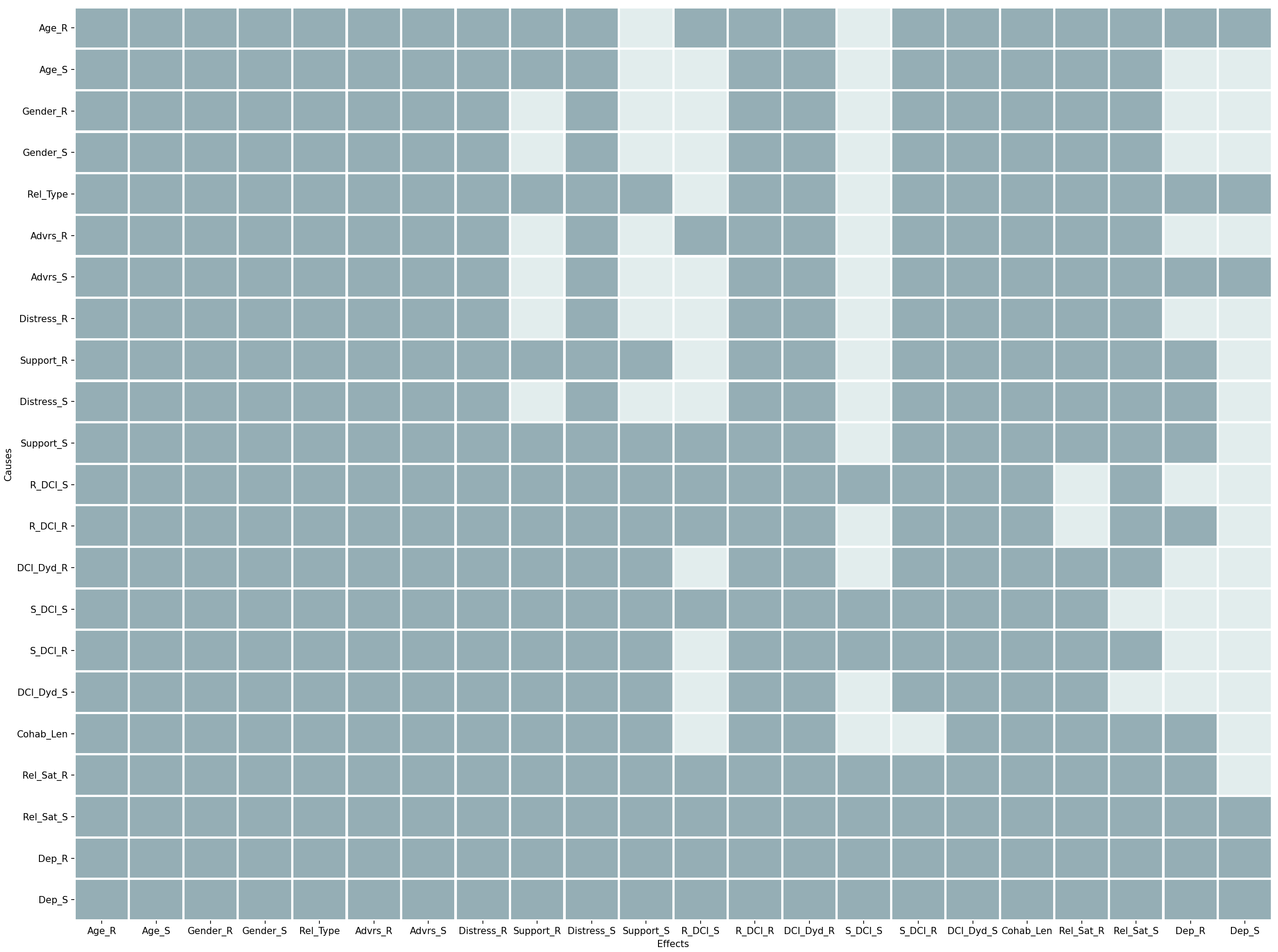} 
\caption{The assumed data generating process for the current study.}
\label{fig:adj_full}
\end{figure}

\begin{figure}[!]
\centering
\includegraphics[width=1\linewidth]{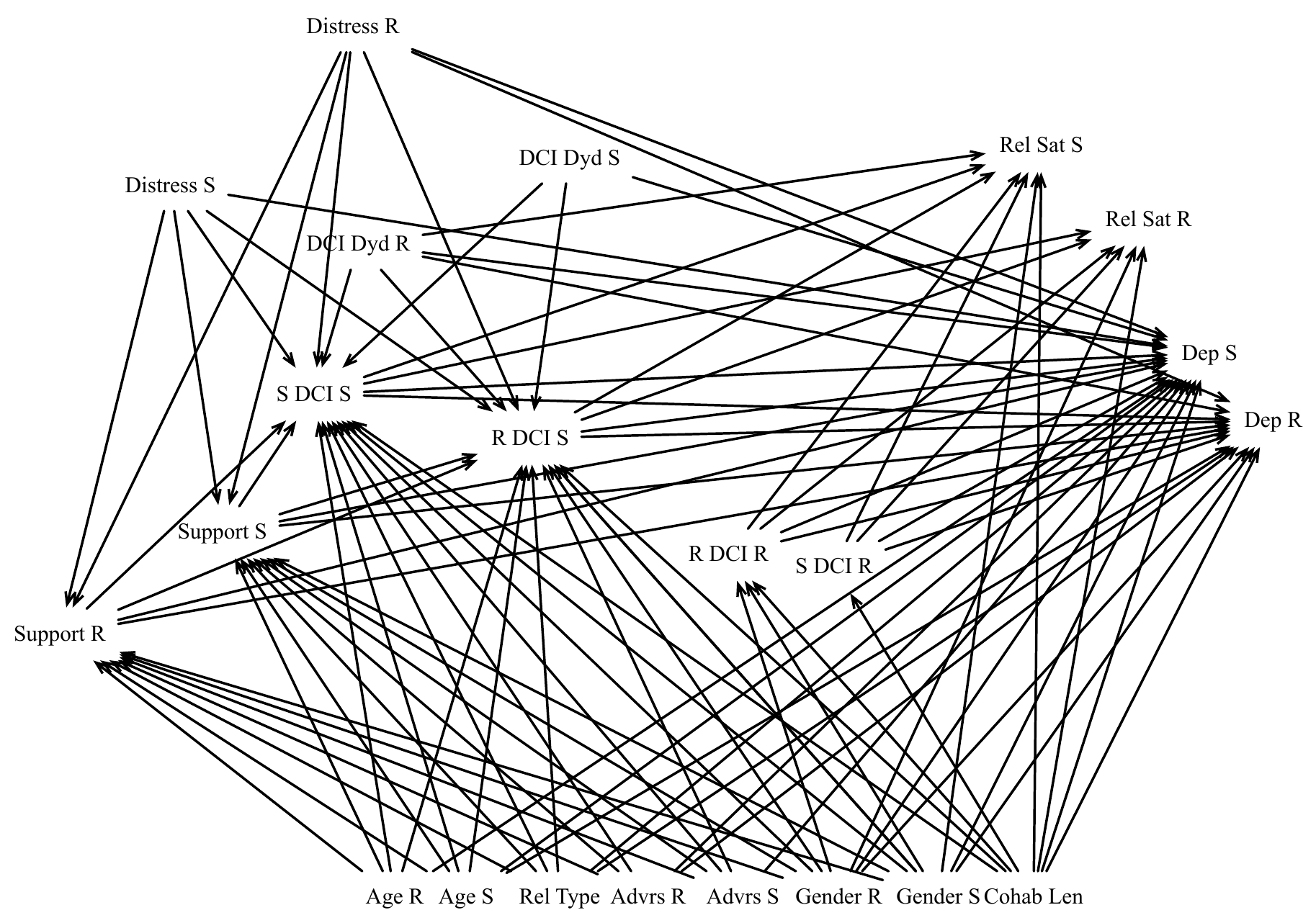} 
\caption{The assumed data generating process for the current study.}
\label{fig:adj_full_graphical}
\end{figure}

The output of SAM (following the averaging over 50 runs) can be represented as an adjacency matrix, with continuous weights or confidences between 0 and 1 representing the presence of each edge (1 indicates confident edge presence, 0 indicates confident edge absence). One can threshold these weights to derive a binarized adjacency matrix, and an example is shown in Figure~\ref{fig:adj_full} (threshold of 0.5). The y-axis represents the causes, and the x-axis represents the effects. Lighter shades indicate the presence of a putative cause-effect pair. Such a representation is useful when the number of variables starts to increase. For instance, the equivalent graphical representation of this discovered structure is shown in Figure~\ref{fig:adj_full_graphical}, which is starting to get visually cluttered despite the comparatively sparse adjacency matrix representation.

\subsection{Step 4 - Theory Specification and Unobserved Confounders}
\label{sec:stageAstep4}
\begin{figure}[!]
\centering
\includegraphics[width=0.7\linewidth]{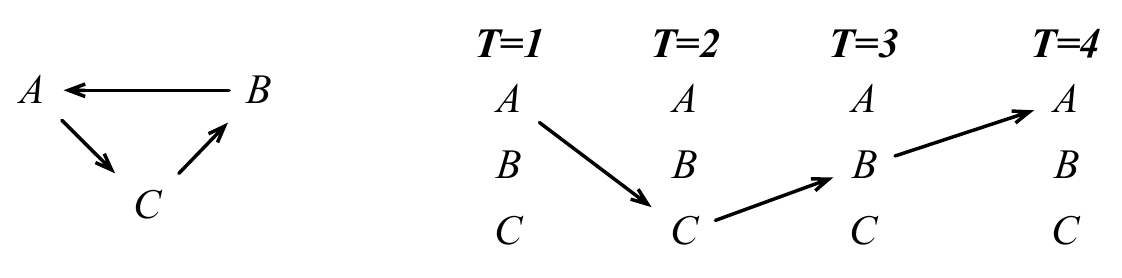} 
\caption{A summary graph including a cycle (left) and one possible representation of this graph unrolled over time (right).}
\label{fig:unroll}
\end{figure}

Owing to limitations associated with causal discovery from observational data, once a researcher has a putative structure it is essential that they check for discrepancies with their existing expertise and theories. The orientation of edges, in particular, is challenging and impossible in certain circumstances (see above for linear Gaussian case and Markov equivalence class). Faced with such discrepancies, researchers can attempt to confirm the robustness of the discovered structure by utilizing other causal discovery algorithms, checking the results on an alternative sample, adjusting the constraints to prevent such links being proposed to begin with, or experimenting with different questionnaires/scales. Of course, any such decisions should be made transparently, and be well recorded. We should, on the whole, expect the proposed structure (or at least its Markov Equivalence Class) to reflect and support the prior theory, but some iteration may be required (hence the pipeline presented in Figure~\ref{fig:pipeline} includes a loop in Stage A). Furthermore, links between variables which are not consistently included as part of a theory may be incorporated formally (particularly demographics, which may induce certain confounding relationships between key constructs). In the absence of any discrepancies between the graph and the theory, one can use the discovered graph to specify and formalize their theory. Note that, depending on the causal discovery method used, the result may involve bidirected edges (which may indicate unobserved confounders) and cycles. One should (as with all other elements of the proposed graph) check the validity of such proposals, particularly in light of the data being used to discover them. For example, a cycle can be unrolled over time, as in Figure~\ref{fig:unroll}.

\begin{figure}[!]
\centering
\includegraphics[width=0.75\linewidth]{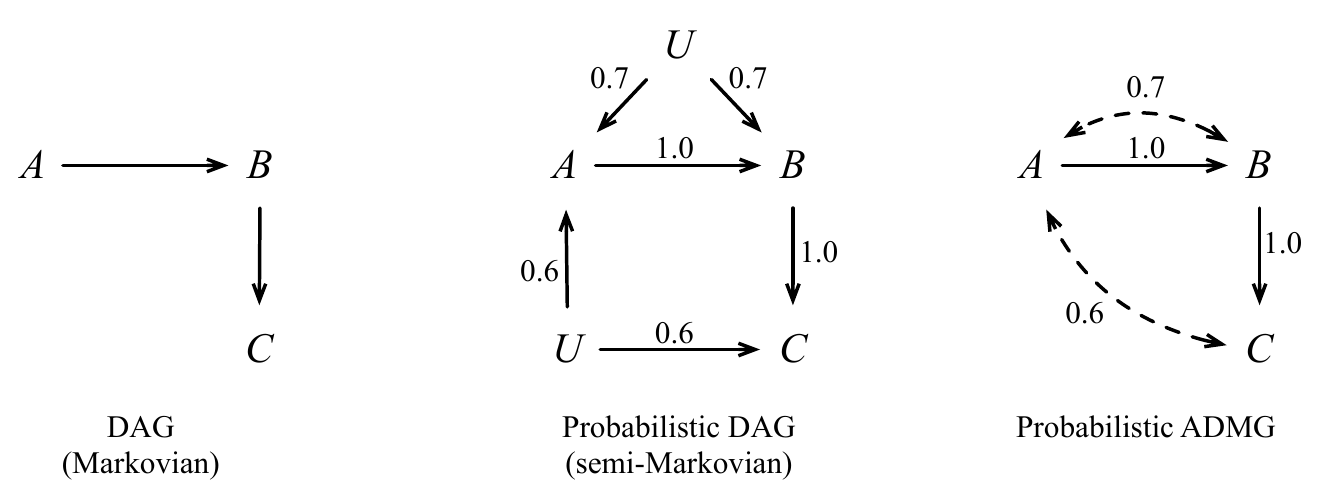} 
\caption{Demonstrating the extension of a DAG (left) to a semi-Markovian probabilistic DAG with edge existent probabilities / plausibility scores (middle) to a probabilistic Acyclic Directed Mixed Graph (ADMG) (right), following \textcite{AkbariProbADMG2022}.}
\label{fig:probdag}
\end{figure}

\begin{figure}[!]
\centering
\includegraphics[width=0.7\linewidth]{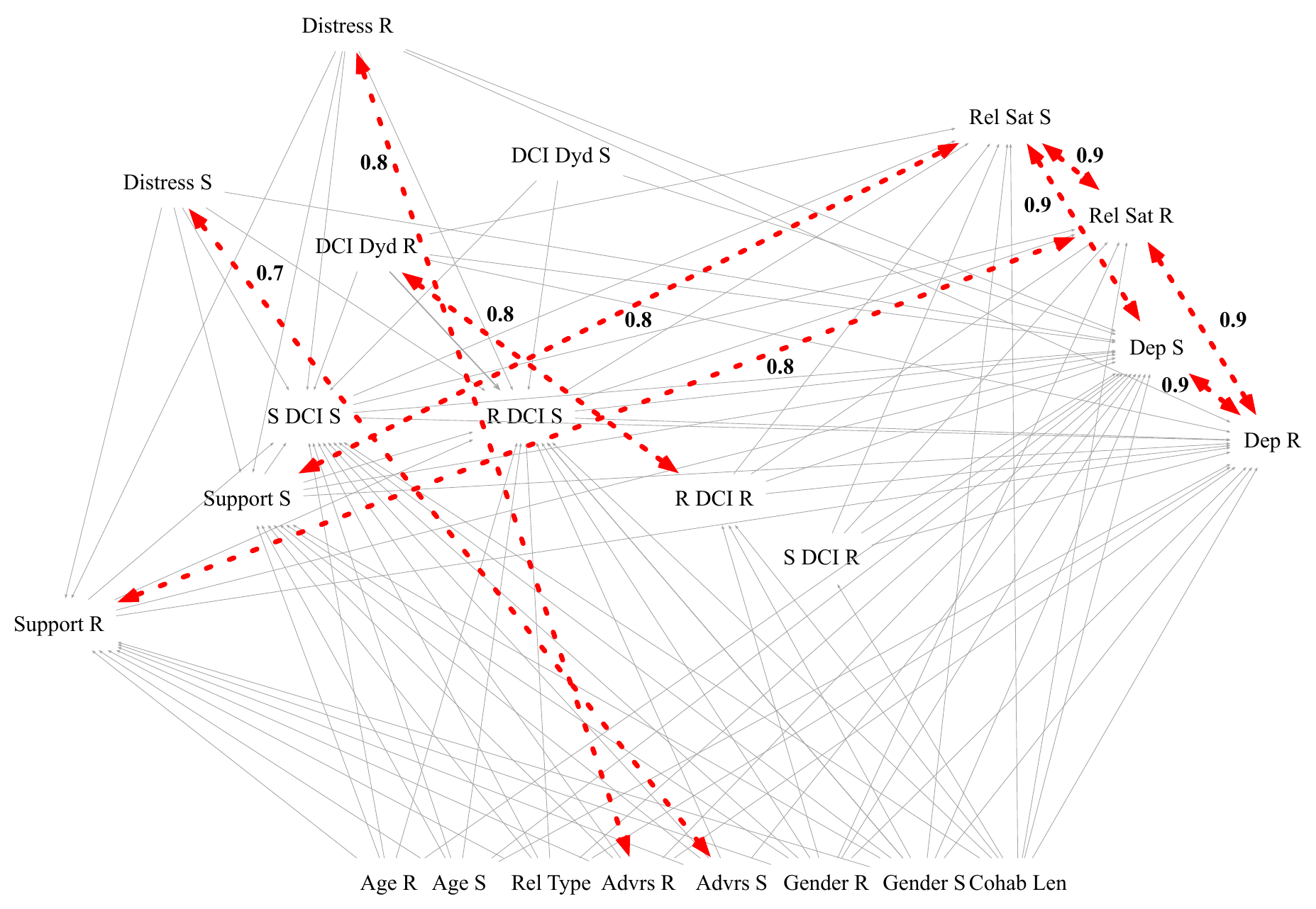} 
\caption{Augmenting the discovered structure shown in Figure~\ref{fig:adj_full_graphical} to include bidirected edges /unobserved confounders (red dashed lines) and corresponding plausibility scores. Directed arrows greyed-out for clarity.}
\label{fig:adj_full_graphicalADMG}
\end{figure}

Recent developments in the domain of structure learning and causal discovery \parencite{AkbariProbADMG2022} also provides researchers with a means to specify unobserved confounders. The idea is that, because causal discovery algorithms cannot reliably discover unobserved confounders (except in restricted cases \parencite{Spirtes2000}) researchers can instead specify the presence of such unobserved confounders and assign to them a plausibility score between 0 and 1 (although note that scores less than 0.5 are essentially `more unlikely' to exist than to exist, and can be omitted). An example is shown in Figure~\ref{fig:probdag}. The graph on the left may derived from a causal discovery algorithm, which is then augmented by an expert to yield the graph in the middle, which includes unobserved confounders as well as edge plausibility ratings. This graph, in turn, can be equivalently represented by transforming all unobserved confounder structures into bidirected edges, and this is depicted on the right in the form of an Acyclic Directed Mixed Graph (ADMG). This graph is `mixed' because it contains both directed and bidirected edges. The specification of such unobserved confounders is important in the representation of a theory which would, without any unobserved confounders, be potentially unrealistic, particularly in view of the complexity of most social and psychological phenomena. The assignment of edge probabilities/plausibilities is not necessary, but will be useful to us later when we need to identify a substructure for which the causal effect of interest is identifiable, and also helps to quantify the certainty of our assumed knowledge. In the example in Figure~\ref{fig:probdag}, all directed edges are assigned a weight of 1.0, and whilst this is not necessary for the specification stage, it will facilitate the use of a graph reduction algorithm in Stage B (which only functions when directed edge plausibilities are 1.0). In Figure~\ref{fig:adj_full_graphicalADMG} we have augmented the previous graph in Figure~\ref{fig:adj_full_graphical} to include a number of bidirected edges in red (reflecting unobserved confounders) and corresponding edge weights. We assume all edge weights on directed edges are 1.0. As this graph concerns dyads, we have included some bidirected edges between equivalent variables for each partner. For example, there exists a bidirected edge between `Rel Sat R' and `Rel Sat S' with a certainty of 0.9 (the relationship satisfaction ratings for each partner).

\section{Stage B: Model Reduction and Causal Inference}
\label{sec:stageB}
Stage B is concerned with the estimation of a particular effect in order to answer a particular research question or test a particular hypothesis. In order to engage seriously with Stage B one must first have access to a formalized representation of their theory (in the form of \textit{e.g.} a Directed Acyclic Graph, or an Acyclic Directed Mixed Graph), such that it is clear what exactly these research questions and hypotheses refer to. Without a direct link between the hypothesis and a formal representation of the theory it relates to, there exist too many degrees of freedom when it comes to model specification to be able to reliably analyse the data. Indeed, without a formal model, one has great leeway in deciding which variables to include in their statistical models, and this choice can have a huge impact on the size and direction of effects \parencite{Vowels2021}. However, assuming we have a mathematical object which represents our theory (in our case, we have a DAG or ADMG) and a particular research question or hypothesis, we can progress through Stage B.

The process involves a reduction of the graph (which hitherto captured the full scope of our theory) into the key elements necessary for answering our research question or testing our hypothesis. Sometimes no reduction is possible. Using the resulting graph, we can express the hypothesis in terms of attributes belonging to that graph. As such, the hypothesis is clearly tied to a formalised and unambiguous object, thus aiding in transparency, reproducibility, and helping us reduce bias by leveraging the rules of \textit{do}-calculus and \textit{d}-separation to specify the desired causal quantity as a function of the observed distribution. We then need to estimate this desired quantity, and here we recommend that researchers use non- or semi-parametric approaches. Even though it remains common to use linear, parametric methods, the validity of such approaches rest on the assumption that the processes being modeled are themselves linear and parametric - an assumption which we know is unlikely to hold. Once the quantity is estimated, we need to derive confidence intervals and test statistics in order to perform statistical inference. In turn, we can quantify the degree to which our results depend on the correct specification of the model, by undertaking sensitivity analysis (for example, we can test the impact of failing to account for unobserved confounders). Finally, we can interpret these results, relating them back to our theory and drawing qualified conclusions in response to our original research questions and hypotheses.

\subsection{Step 1 - Specifying the Hypothesis and Establishing Identification}
\label{sec:stageBstep1}
Firstly, if we assume we are using the DAG/ADMG framework, we must ensure that no cycles exist in the graph. Although it is possible to handle cycles with Structural Causal Models (see, for example, \cite[pp.28]{Pearl2009}) this lies beyond the scope of the present work. In order to remove cycles one can unroll the graph over time, as shown in Figure~\ref{fig:unroll}, although note that reliably discovering such longitudinal structures is associated with increased data requirements. See work by \textcite[pp.197-211]{Peters2017} for more information on on time series causal discovery. Otherwise, the removal of cycles should be done with care, and may require more iterations of Stage A.

Once we have a DAG, we can use it to design a hypothesis which can be directly associated with our theory. For instance, given the DAG in Figure~\ref{fig:adj_full_graphicalADMG}, we might be interested in the Average Causal Effect (ACE) of the respondent's distress `Distress R'  on the respondent's depression `Dep R'. Our hypothesis may be that the ACE is non-zero, enabling us to establish typical null $h_0$ and alternative $h_1$ hypotheses. The challenges lies in expressing the ACE in an unambiguous manner, and in terms of the graphical object. Indeed, there are many ways we could ostensibly test our hypothesis, but almost all of them would be misspecified with respect to the graph. Thus, we should start by expressing our target estimand in the general sense. The ACE can be expressed as:

\begin{equation}
    \Psi = \mathbb{E}_{y \sim P(Y|do(T=t))}[y] - \mathbb{E}_{y \sim P(Y|do(T=t'))}[y]
    \label{eq:backdoorcont1}
\end{equation}.

Here, we note that $Y$ is Dep R, $T$ is Distress, and $t$ and $t'$ are specific contrasts we are interested in estimating. In this working example, there are five discrete, ordinal levels of distress measured, ranging from 0-4 inclusive. We may be interested in understanding the effect on depression when distress is increased from group 0 to group 2 (first contrast), and from group 0 to group 4 (second contrast), thereby resulting in two estimands, $\Psi_{0-2}$ and $\Psi_{0-4}$, respectively. As described in the background section, we need to express this estimand in terms of the observed data (rather than hypothetical $do$ operations). If this is not possible, the estimands are referred to as \textit{unidentifiable}. One way to achieve identification is via an application of the backdoor adjustment process. Checking identification is possible by eye using the \textit{d}-separation rules or \textit{do}-calculus, or by identifying a set of variables which we can condition on in order to block all backdoor paths.

\begin{tcolorbox}
\textbf{Application/Code Note :}
In the associated code, we include a means to check for identification using a package by \textcite{Pedemonte2021causaleffect} (code included in \texttt{auto\_IF.py}). The algorithm indicates whether the effect of interest is identifiable (by providing the identification formula) or not (by throwing an error). Note that in the absence of unobserved confounders, this step is not necessary, because identification is always possible in the absence of unobserved confounders. We also provide our own method in \texttt{bd\_SMDAG.py} which checks for identification via the backdoor criterion. If the backdoor criterion is fulfilled, the code provides one possible backdoor adjustment set. Finally, in \texttt{reducer.py}, we provide an algorithm which takes a graph for which the backdoor criterion holds (which can be checked with \texttt{bd\_SMDAG.py}), and which also yields a sufficient set of backdoor adjustment variables, as well as a list of precision variables. Precision variables which may help in estimation but which are not necessary for identification.
\end{tcolorbox}

In the worked example shown in Figure~\ref{fig:adj_full_graphicalADMG}, we have both directed edges, and bidirected edges corresponding with fork structures from unobserved confounders. Furthermore, we also have edge probabilities - the directed edges are assumed to have probabilities of 1.0, whereas the unobserved confounders (which cannot reliably be discovered from observational data) have been given ratings between 0.5 and 1. These unobserved confounders may prevent backdoor adjustment from being possible. Indeed, they may prevent \textit{any} kind of identification. Thus the question arises - if one had to assume any of those confounders with edge probabilities $<1$ did \textit{not} exist in order to facilitate backdoor adjustment, which unobserved confounders would they be? For this, we modified the work of \textcite{AkbariProbADMG2022} (which provides algorithms for general estimands and general identification) to provide a way to identify the most plausible subgraph in which the backdoor criterion holds. The algorithm considers the cost of removing all possible combinations of unobserved confounders (with some search-space pruning), and removes the combination with the lowest cost. It also provides a rating of plausibility, which is given as $P(\hat{\mathcal{G}}^*)/P(\mathcal{G})$, where $P(\hat{\mathcal{G}}^*)$ is the plausibility of the graph $\mathcal{G}$ \textit{after} removal of unobserved confounders (and for which the backdoor criterion holds), and $/P(\mathcal{G})$ is the plausibility of the graph before any modification. If backdoor adjustment is possible in the original graph, then this ratio will be equal to one. The algorithm therefore provides researchers with a quantification of the strength of whatever assumptions were necessary to facilitate backdoor identifiability.

\begin{tcolorbox}
\textbf{Application/Code Note :}
In \texttt{bd\_SMDAG.py} we provide an exact, brute-force algorithm which finds the most plausible subgraph in a provided graph for which the backdoor criterion holds. The subgraph is found by identifying the set of unobserved confounders (or, equivalently, bidrected edges) which can be removed with the lowest cost whilst nonetheless yielding an identifiable causal effect according to backdoor adjustment. 
\end{tcolorbox}

In general, researchers may feel encouraged to include as many covariates as possible with the goal of satisfying the unconfoundness assumption \parencite{Zhang2020, Vowels2020b}. This, however, can easily lead to problems with the unintentional inclusion e.g. of mediators or colliders which open backdoor paths, and can thereby significantly bias the estimates \parencite{Cinelli2020}. This is why a careful consideration for the structure of the problem is so important, and also highlights the utility of a graphical representation for identification of suitable adjustment sets.

Once one has a set of adjustment variables, one can use the following expression to represent the desired estimand in terms of observed quantities:

\begin{equation}
    \Psi = \mathbb{E}_{\mathbf{c}\sim P(\mathbf{C})}[\mathbb{E}_{y \sim P(Y|T=t, \mathbf{C}=\mathbf{c})}[y] - \mathbb{E}_{y \sim P(Y|T=t', \mathbf{C}=\mathbf{c})}[y]]
    \label{eq:estimandex}
\end{equation}.

In the case of multiple contrasts / multiple treatments, we may have a number of estimands. In the worked example, we can contrast, for instance, people who reported experiences of low levels of distress (level 0), with those in groups with medium levels of distress (level 2) and high levels (level 4):

\begin{equation}
\begin{split}
    \Psi_{2-0} = \mathbb{E}_{\mathbf{c}\sim P(\mathbf{C})}[\mathbb{E}_{y \sim P(Y|T=2, \mathbf{C}=\mathbf{c})}[y] - \mathbb{E}_{y \sim P(Y|T=0, \mathbf{C}=\mathbf{c})}[y]],\\
        \Psi_{4-0} = \mathbb{E}_{\mathbf{c}\sim P(\mathbf{C})}[\mathbb{E}_{y \sim P(Y|T=4, \mathbf{C}=\mathbf{c})}[y] - \mathbb{E}_{y \sim P(Y|T=0, \mathbf{C}=\mathbf{c})}[y]].
        \end{split}
    \label{eq:estimandes}
\end{equation}.

The target quantities $\Psi$ can further be integrated into a clear statements concerning the null and alternative hypothesis. For example: $h_{0, 2-0}: \Psi_{2-0} = 0$ and  $h_{1, 2-0}: \Psi_{2-0} \neq 0$. In Eq.~\ref{eq:estimandes} we use $\mathbf{C}$ to represent the set of confounders \textit{and} precision variables. In our working example, the set $\mathbf{C}$ actually comprises no confounders (there are no backdoor paths from Distress R to outcome Dep R), but a considerable number of precision variables including, for instance, Age and Gender.\footnote{We provide an algorithm in the supplementary which provides a set of confounders \textit{and} precision variables for a given graph and a given specified cause and effect.}  Given that an absence of confounders between Distress R and Dep R is unlikely, we might review whether (a) we have collected enough pertinent variables (and missed key confounders), or (b) whether the causal discovery algorithm has provided us with a plausible solution that sufficiently aligns with our expectations as domain experts. Indeed, one way to diagnose problems in this regard (and we will explore this below) is to compute the naive group mean differences (which assumes no confounders) and compare it with the estimates including precision variables. If there is a substantial difference between the naive group mean differences and our estimates with a set of precision variables but no confounders, we know that the set of precision variables likely contains either confounders, mediators, colliders, or a combination thereof.

\subsection{Step 2 - Estimation}
\label{sec:stageBstep2}
Now that we have specified the estimand in terms of attributes in our formalized theory (see Eq.~\ref{eq:estimandex}), we can start thinking about estimating this quantity. We will use plug-in estimation approach, which means that we can express the approximation to $\Psi$ in Eq.~\ref{eq:estimandex} for the two contrasts as:

\begin{equation}
\begin{split}
 \hat{\Psi}_{2-0} = N^{-1} \sum_i^N \left[\hat Q(T=2, \mathbf{C}_i) - \hat Q(T=0, \mathbf{C}_i) \right],\\
  \hat{\Psi}_{4-0} = N^{-1} \sum_i^N \left[\hat Q(T=4, \mathbf{C}_i) - \hat Q(T=0, \mathbf{C}_i) \right].
  \end{split}
 \label{eq:estimatorex}
\end{equation}

In words, for the first contrast the approximation can be found by taking the average over the differences in predictions from estimator $Q$ for each row in the evaluation dataset whilst setting $T$ (which in the working example is Distress R) to group 2 and setting $T$ to group 0. The second contrast can be found the same way but instead of setting $T$ to group 2, we set it to group 4. For these $\hat{Q}$ estimators we use Super Learners \parencite{Polley2007}, which are ensembles of diverse learners (some linear, some highly flexible and non-linear) from which a set of predictions are combined via a weighted linear sum. The learners themselves are fit via a k-fold cross-validation scheme, and the weights are estimated based on the combined test fold predictions, stored during the k-fold process. The advantage of using Super Learners is to avoid any unnecessary and unreasonable assumptions about the functional form relating the quantities of interest. Indeed, we are already making strong assumptions about the structure of the data generating process, and so it is crucial to avoid any further opportunities for misspecification to be introduced to the model.

\begin{tcolorbox}
\textbf{Application/Code Note :}
For recommendations on the choice and training of the Super Learner, readers are directed towards the accessible overview by \textcite{PhillipsSuperLearner}, and related evaluations by \textcite{VowelsFreeLunch2022}.
\end{tcolorbox}

\subsection{Step 3 - Debiasing and Inference}
\label{sec:stageBstep3}
Using Eq.~\ref{eq:estimatorex} we can derive an initial estimate for the ATE. However, the estimates are non-parametric as a result of the use of Super Learners, and therefore do not readily facilitate statistical inference (\textit{p}-values and confidence intervals). Furthermore, it is well known from semiparametric theory \parencite{Hines2021, Hampel1974, Fisher2019, VowelsFreeLunch2022} that when we are required to estimate quantities which are non-central to our target estimand, which are typically known as nuisance parameters, our effect of interest is often biased. Consider a simple case where we fit a linear regression for estimating $\Psi$, and let us assume the model is correctly specified both in terms of its functional and implied structural form. In this case, the parameter coefficient $\beta^T$ on the treatment variable $T$ will approximate $\Psi$. However, this model is simultaneously being used to estimate $\beta^{X_1},... \beta^{X_k}, \beta^{R_1},... \beta^{R_d} $, which represent nuisance parameters because they are necessary for the specification of the model but are not what we are fundamentally interested in estimating. Indeed, the Ordinal Least Squares objective for the regression is only tangentially related to the estimation of the target quantity of interest. The resulting estimate $\hat\beta^T$ will therefore exhibit residual bias, associated with finite sample deviations and interference from the other quantities in the model.

In order to achieve inference, and to `target' the model to deal with the bias which results from the nuisance parameters, we can look to the tools of semiparametric statistics. Specifically, we use the targeted learning approach \parencite{vanderLaan2011, Coyle2020}. This approach yields doubly robust estimates, so-called because they require specification of both an outcome model, and a treatment model, and the results are consistent even if one of the models is misspecified.\footnote{A consistent estimator is one which converges to the true value as the sample size increases.} Although the theoretical details of the approach are beyond the scope of this paper, readers are encouraged to consult introductions to the general topics by \textcite{Hines2021, Kennedy2017, Fisher2019, Coyle2020, VowelsFreeLunch2022}. The key steps (following those already taken above) for deriving a targeted estimate for the ATE are outlined below. For these steps we split our set of adjustment variables into confounders $\mathbf{C}$ and precision variables $\mathbf{R}$. As mentioned earlier, our set of confounders in this worked example is actually empty, and we describe how to handle this situation below. For now, let us continue as if we have a non-empty set of necessary adjustment variables. Note that the following procedure works for either a binary outcome, or a bounded continuous outcome (\textit{i.e.}, an outcome $Y$ which is continuous but limited according to its range) which has been rescaled to the range zero to one. Interested readers are directed to  \parencite[pp.121:132]{vanderLaan2011} for proof of equivalence.

\begin{enumerate}
    \item Fit \textit{initial} plug-in Super Learner estimator $\hat{Q}_0$ to the datasets (via the k-fold cross-validation process) and derive initial estimates for the interventional outcome distributions of each of the desired contrasts across all data points  $\hat{Q}_0(T=4, \mathbf{C},\mathbf{R})$, $\hat{Q}_0(T=2, \mathbf{C},\mathbf{R})$, and $\hat{Q}_0(T=0, \mathbf{C},\mathbf{R})$. \textit{i.e.} we generate predictions from our Super Learner for all data points.
    \item Fit a nuisance parameter model for the propensity score\footnote{Propensity scores were proposed by \textcite{Rosenbaum1983}, and describe the probability of receiving treatment or not, and the Generalised Propensity Score (GPS) proposed by \textcite{ImbensGPS2000} extends the propensity score to multiple treatment groups. For an example application see \textcite{Siddique2019}.} $G_{T}(\mathbf{C}) = P(T|\mathbf{C})$, using a separate Super Learner, and generate predictions from the model $\hat G_{T}(\mathbf{C})$ for all data points. \textit{i.e.}, we use the confounders $\mathbf{C}$ to predict the treatment group. 
    \item Compute what are known as the `clever covariates', this will help us model the residual bias caused by the link between the confounders and our treatment. For only one contrast, the clever covariates are computed as follows:
    \begin{itemize}
        \item $\hat{H}_{2-0}(T, \mathbf{C}) = \frac{\mathbbm{1}_{2}(T)}{\hat{G}_{T=1}(\mathbf{C})} - \frac{\mathbbm{1}_{0}(T)}{\hat{G}_{T=0}(\mathbf{C})}$,
        \item $\hat{H}_2(T=2, \mathbf{C}) = \frac{\mathbbm{1}_{2}(T)}{\hat{G}_{T=2}(\mathbf{C})}$,
        \item $\hat{H}_0(T=0, \mathbf{C}) = \frac{-\mathbbm{1}_{0}(T)}{\hat{G}_{T=0}(\mathbf{C})}$.
    \end{itemize}
    Here, $\mathbbm{1}$ is the indicator function, providing an output which is equal to one when the input value is equal to the value indicated in the subscript, and zero otherwise. 
    \item Fit a linear model representing the degree to which the clever covariates (which contain information about the propensity scores) bias our predictions from the true values. In this linear model, the outcome are the true outcome $Y$, and the `predictors' are the predictions from our initial plug-in estimator $\hat{Q}_0(T, \mathbf{X}, \mathbf{R})$ and the clever covariates $\hat{H}(T, \mathbf{C})$. Importantly, note that $\hat{Q}_0(T, \mathbf{C}, \mathbf{R})$ is treated as a fixed \textit{offset} in the linear model, and no coefficient is estimated for this term.
    \begin{equation}
    \begin{split}
        Y = \sigma(\mbox{logit}(\hat{Q}_0(T,\mathbf{C},\mathbf{R})) + \epsilon_{2-0} \hat{H}_{2-0}(T, \mathbf{C})),\\
        Y = \sigma(\mbox{logit}(\hat{Q}_0(T,\mathbf{C},\mathbf{R})) + \epsilon_{4-0} \hat{H}_{4-0}(T, \mathbf{C})). 
        \end{split}
    \end{equation}
    Here, $\epsilon$ is the parameter coefficient in this linear model which represents, on average, the amount by which our predictions differ from the true values according to the magnitude of the clever covariates. The $\sigma$ is the sigmoid or `expit' function, where $\sigma(x) = 1 / (1 + \exp^{-x})$, and $\mbox{logit}(x) = \log ( x / (1-x))$ is the logistic quantile function. Thus, the sigmoid function constrains the output of this regression to fall within the range of 0 and 1, even for continuous outcomes (assuming they are bounded and have been rescaled for the analysis) which helps with stability. 
    \item Fitting the linear model above yields $\hat{\epsilon}$. This can then be used, in turn, to update our initial estimates for the different contrasts:
    \begin{itemize}
        \item $\hat{Q}_{*,4-0}(T=4,\mathbf{C},\mathbf{R}) = \sigma(\mbox{logit}(\hat{Q}_{0}(T=4,\mathbf{C},\mathbf{R})) + \hat{\epsilon}_{4-0}\hat{H}_{4-0}(T=4,\mathbf{C}))$
        \item $\hat{Q}_{*,2-0}(T=2,\mathbf{C},\mathbf{R}) = \sigma(\mbox{logit}(\hat{Q}_{0}(T=2,\mathbf{C},\mathbf{R})) + \hat{\epsilon}_{2-0}\hat{H}_{2-0}(T=2,\mathbf{C}))$
        \item $\hat{Q}_{*,4-0}(T=0,\mathbf{C},\mathbf{R}) = \sigma(\mbox{logit}(\hat{Q}_{0}(T=0,\mathbf{C},\mathbf{R})) + \hat{\epsilon}_{4-0}\hat{H}_{4-0}(T=0,\mathbf{C}))$
        \item $\hat{Q}_{*,2-0}(T=0,\mathbf{C},\mathbf{R}) = \sigma(\mbox{logit}(\hat{Q}_{0}(T=0,\mathbf{C},\mathbf{R})) + \hat{\epsilon}_{2-0}\hat{H}_{2-0}(T=0,\mathbf{C}))$
    \end{itemize}
\item Finally, the updated estimates above can be plugged into Eq.~\ref{eq:estimatorex} to yield:
\begin{equation}
    \begin{split}
     \hat\Psi_{*, 2-0} = n^{-1} \sum_i^N \left[ \hat{Q}_{*,2-0}(T_i=2, \mathbf{C} = \mathbf{c}_i, \mathbf{R}=\mathbf{r}_i) \right. \\ \left.- \hat{Q}_{*,2-0}(T_i=0, \mathbf{C}=\mathbf{c}_i, \mathbf{R}=\mathbf{r}_i) \right],\\
         \hat\Psi_{*,4-0} = n^{-1} \sum_i^N \left[ \hat{Q}_{*,4-0}(T_i=4, \mathbf{C} = \mathbf{c}_i, \mathbf{R}=\mathbf{r}_i) \right. \\ \left.- \hat{Q}_{*,4-0}(T_i=0, \mathbf{C}=\mathbf{c}_i, \mathbf{R}=\mathbf{r}_i) \right]
    \label{eq:psihat}
    \end{split}
\end{equation}
\item From here we can derive the Influence Function (IF), which can be used to derive standard errors, confidence intervals, and \textit{p}-values:
\begin{itemize}
    \item First compute the IFs for each data and for each contrast according to:
    \begin{equation}
        \begin{split}
IF_{i, 2-0}= \hat{H}(T=t_i,\mathbf{C}=\mathbf{c}_i)\left[y_i - \hat{Q}_{*,2-0}(T=t_i, \mathbf{C}=\mathbf{c}_i, \mathbf{R}=\mathbf{r}_i))  \right] \\ + \left( \hat{Q}_{*,2-0}(T=2, \mathbf{C}=\mathbf{c}_i, \mathbf{R}=\mathbf{r}_i)) - \hat{Q}_{*,2-0}(T=0, \mathbf{C}=\mathbf{c}_i, \mathbf{R}=\mathbf{r}_i))\right) - \hat\Psi_{*},\\
IF_{i, 4-0}= \hat{H}(T=t_i,\mathbf{C}=\mathbf{x}_i)\left[y_i - \hat{Q}_{*,4-0}(T=t_i, \mathbf{C}=\mathbf{c}_i, \mathbf{R}=\mathbf{r}_i))  \right] \\ + \left( \hat{Q}_{*,4-0}(T=4, \mathbf{C}=\mathbf{c}_i, \mathbf{R}=\mathbf{r}_i)) - \hat{Q}_{*,4-0}(T=0, \mathbf{C}=\mathbf{c}_i, \mathbf{R}=\mathbf{r}_i)) \right) - \hat\Psi_{*}.
    \end{split}
    \end{equation}
    
    \item Then compute the variance of the IFs:
    \begin{equation}
    \begin{split}
        \widehat{\mbox{Var}}(IF_{2-0}) = n^{-1} \sum_i^N\left(IF_{i,2-0} -  \bar{IF}_{2-0}\right)^2,\\
        \widehat{\mbox{Var}}(IF_{4-0}) = n^{-1} \sum_i^N\left(IF_{i,4-0} -  \bar{IF}_{4-0}\right)^2,
    \end{split}
    \end{equation}
    where $\bar{IF}$ is the empirical average of the IF evaluated over the sample data points.
    \item The standard errors for each contrast are derived from the variance of the influence function according to the standard definitions:
    \begin{equation}
        \hat{\sigma}_{se} = \sqrt{\frac{\mbox{Var}(IF)}{N}}.
    \end{equation}
    \item From here we can derive the upper and lower bounds for the 95\% confidence interval:
    \begin{equation}
        \hat\Psi_* \pm 1.96 \hat{\sigma}_{se}
    \end{equation}
    \item And \textit{p}-values, in turn, may be derived using standard definitions also:
    \begin{equation}
        2\left(1 - \mbox{CDF}\left(\frac{\hat\Psi_*}{\hat{\sigma}_{se}}\right)\right),
    \end{equation}
    where CDF represent the cumulative distribution function for a normal distribution.
    \end{itemize}
\end{enumerate}

Note that in addition to the resulting valid inference, the key advantages of this procedure is that the estimates are \textit{doubly robust}, meaning that they are consistent if \textit{either} the outcome model or the propensity score model are misspecified. Note that this double robustness does not immediately extend to the inference itself (\textit{i.e.}, the standard error and confidence intervals). For the inference to also be doubly robust, an augmented procedure is required which is described in \textcite{BenkeserTMLE2017} and implemented in the code in the supplementary material. 

\begin{figure}[!]
\centering
\includegraphics[width=0.6\linewidth]{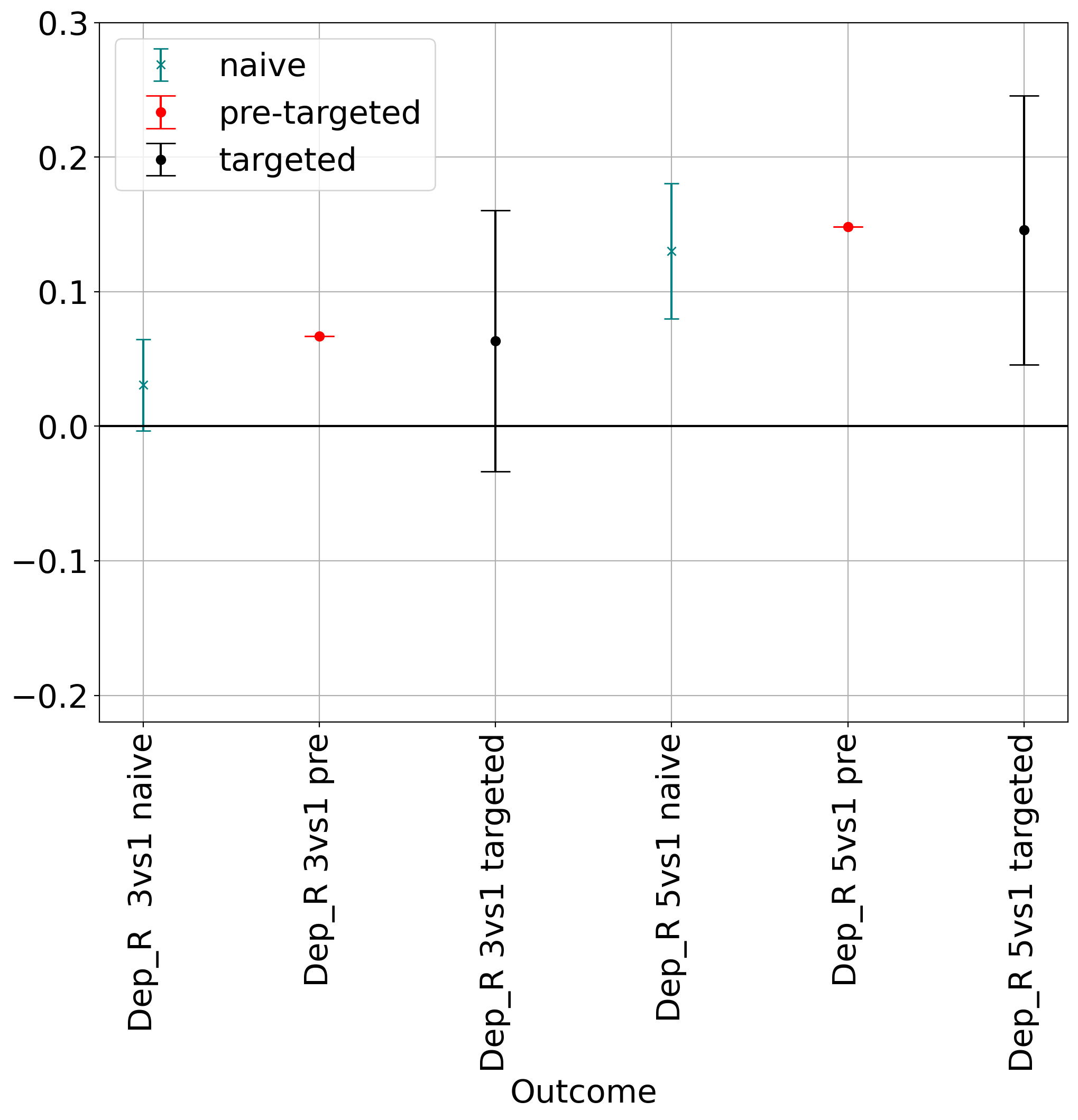}
\caption{Results from the targeted learning analysis for the average causal effect of respondent's distress on their depression - displaying the naive group differences estimates, the pre-targeted `plug-in' estimates from step 1 of the targeted learning procedure, and the updated, targeted estimates.}
\label{fig:tlresults}
\end{figure}

 The results for our working example are shown in Figure~\ref{fig:tlresults}. Note that these effects concern changes in a bounded continuous outcome which has been rescaled to the range 0-1. The naive estimates are derived by simply computing the group mean differences under the different contrasts, and the pre-targeted effects are derived from undertaking step 1 in the procedure above. It can be seen that, in spite of the fact that no confounders were identified according to the structure derived using the causal discovery process, the differences between the naive estimates and the estimates from the pre- and targeted process are quite substantial. However, precision variables should not significantly affect the estimate and these results therefore suggest that the set of precision variables contains confounders, mediators, or colliders (or a combination thereof). When faced with such results, researchers may wish to revisit the specification of their theory,  or check for positivity violations which can make the process unstable and lead to spurious estimation (we discuss positivity violations below). For the sake of the working example (in particular, the sensitivity analysis below), we will assume that the set of precision variables actually contained a number of key confounders: dyadic coping variables `DCI Dyd S' and `DCI Dyd R'; partner's distress and depression variables `Distress S' and `Dep S'; and the couple's cohabitation length `Cohab Len'. The differences between the naive group mean differences and the targeted estimates can therefore be attributed to the backdoor adjustment (although note, again, that this structure was not indicated by the causal discovery algorithm itself and is only assumed for the purposes of the following sensitivity analysis.).

\begin{tcolorbox}
\textbf{Application/Code Note :}
In the accompanying code, we include functions for both the targeted learning procedure described above in the python class \texttt{TLP}, which uses our implementation of the Super Learner specified in \texttt{super\_learner.py}.  \texttt{TLP} also includes a function which includes the extension proposed by \textcite{BenkeserTMLE2017} to facilitate doubly robust inference (in addition to double robust estimation). 
\end{tcolorbox}

Based on the results so far, and considering our earlier hypotheses $h_{0, 2-0}: \Psi_{2-0} = 0$ and  $h_{1, 2-0}: \Psi_{2-0} \neq 0$, one can say that 
the effect on depression of increasing levels of distress from level 0 to level 2 is not statistically significant, whereas the effect on depression of increasing levels of distress from level 0 to level 4 would be significant. Of course, in practice one would be advised to present a table of effect sizes, \textit{p}-values, demographics, and number of participants in each treatment group. One may also wish to specify a minimum meaningful effect size of interest (rather than assuming the nill null), but in any case, the focus of this work is not so much reporting standards but the methods and approaches to research, and we therefore leave such an exercise to motivated readers.

Before continuing, it is worth noting that the procedure for deriving the results for our worked example is slightly different to the one described above, owing to the absence of confounders for our choice of cause and effect. However, and as one might expect, the absence of confounders actually makes the targeting process simpler. In the absence of confounders, the propensity score is now simply the likelihood of receiving treatment (\textit{i.e.}, conditional on the empty set) - instead of the propensity score being equal to $P(T|\mathbf{C})$ it is $P(T)$. The corresponding expression for the influence function for the 2-0 contrast is:

\begin{equation}
    \begin{split}
    IF_{i, 2-0}= \left(\frac{\mathbbm{1}_{2}(t_i)}{\hat{G}_{T=2}}- \frac{\mathbbm{1}_{0}(t_i)}{\hat{G}_{T=0}}\right)\left[y_i - \hat{Q}_{*}(T=t_i, \mathbf{R}=\mathbf{r}_i))  \right] \\ + \left( \hat{Q}_{*,2-0}(T=2,  \mathbf{R}=\mathbf{r}_i)) - \hat{Q}_{*,2-0}(T=0,\mathbf{R}=\mathbf{r}_i))\right) - \hat\Psi_{*,2-0},
    \end{split}
    \label{eq:emptyIF}
\end{equation}

where $\hat{G}_{T}$ is an empirical estimate for $P(T)$, and can be derived simply by counting the number of participants in each treatment group, and normalizing by the total number of participants. The expressions for the influence functions of the estimands for the other contrasts follow straightforwardly.

\begin{tcolorbox}
\textbf{Application/Code Note :}
In the accompanying code, we automatically handle cases where the confounder / adjustment set is empty, and compute the influence function according to the expression given in Eq.~\ref{eq:emptyIF}.
\end{tcolorbox}

\subsection{Step 4 - Sensitivity Analysis}
\label{sec:stageBstep4}
Once one has derived these targeted estimates and corresponding standard errors, one can undertake statistics inference in the usual way, as we did above.  However, we would argue that at least one more important step remains, and  we should first try to capture the robustness of these results to changes in model specification. Indeed, the validity of the results we have so far derived rests on the correct specification of the graph, which was constructed using a combination of domain knowledge and causal discovery - this implicates ignorability (\textit{i.e.}, that there are no unobserved confounders). This is a particularly strong assumption in psychology, where the complexity of phenomena may implicate a complex web of interacting components - alluding, in turn, to Meehl's infamous `crud' factor \parencite{Orben2020, Meehl1990}.

Sensitivity analysis provides us with a way to understand how much our estimates would vary if our model was subject to some degree of specification. Whilst numerous approaches exist for sensitivity analysis \parencite{Rotnitzky1998, Rotnitzky2001, Gabriel2021, Imbens2015, Pearl2009}, we follow an adaptation of the one described by \textcite{Diaz2013} and \textcite{Gruber2022}. Rather than specifying a single estimand for each contrast \textit{e.g} $\Psi_{2-0}$ for the contrast between those in the low (level 0) and medium (level 2) levels of distress groups, we specify a family of estimands for each contrast. The family is indexed by a parameter $\delta$ which characterizes the degree to which our assumption about ignorability (and therefore our assumption about the identifiability of the estimand) holds. Specifically, if $\bar{\Psi}_{2-0}$ is the true parameter expressed as a function of the full distribution, and $\Psi_{2-0}$ is the estimand in terms of the \textit{observed} distribution (and the one given earlier in Eq.~\ref{eq:estimandex}), in the case where identifiability does not hold, perhaps because the observed distribution is missing certain key confounders, we can express the difference between these quantities as $\Psi_{2-0} - \bar{\Psi}_{2-0} \leq \delta$. If the ignorability assumptions hold, $\delta = 0$ and there is no resulting bias. Note that we have not dealt with $\hat \Psi_{2-0}$, which represents an empirical estimate for $\Psi_{2-0}$, because identification concerns the specification of the estimand itself, \textit{before} estimation. If we can propose a plausible way in which ignorability does not hold, we can quantify the impact of such a violation and degree to which our estimates (such as the ones shown in Figure~\ref{fig:tlresults}) become biased.

Continuing our worked example, we can begin by assuming that the results in Figure~\ref{fig:tlresults} correspond with the estimates for when $\delta =0$ (\textit{i.e.}, ignorability holds). Next, we construct a scenario which represents a conservative degree of non-identifiability, and observe the correspond impact on the estimates. Specifically, we assume that violations of our ignorability assumption could not be worse than if we made no adjustment at all, and that such a scenario is given be $\delta = 1$. We can then quantify the impact on our estimands by re-running the targeted analysis with the empty adjustment set, and produce a set of results for multiples of $\delta$, such as $0.5\delta$, $\delta$, $1.5\delta$, $2\delta$, etc. so that we can understand to what degree any unobserved confounders could still be affecting our inference, expressed as multiples of a \textit{worst case} scenario where we have no adjustment at all.

\begin{figure}[!]
\centering
\includegraphics[width=0.6\linewidth]{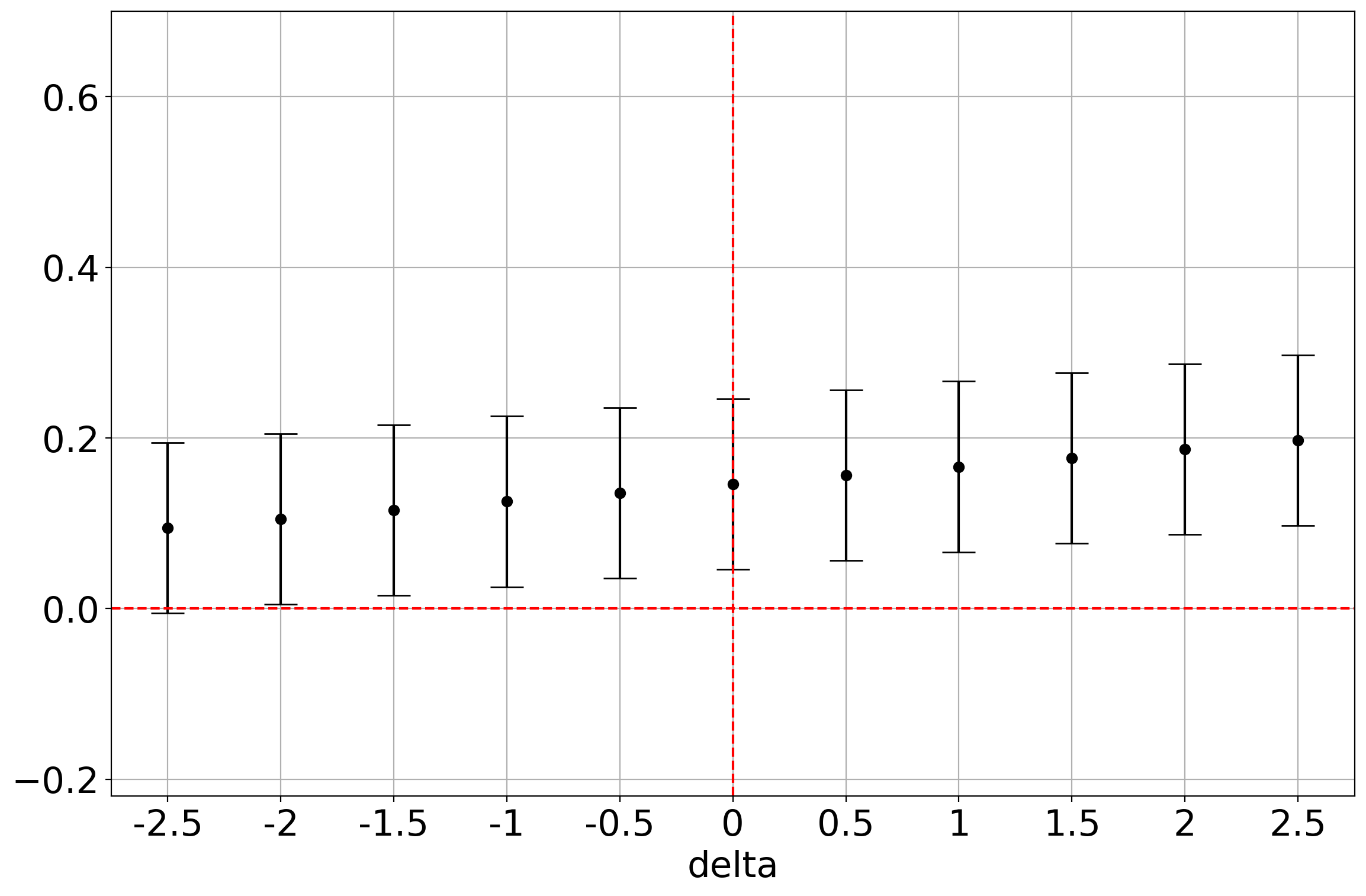}
\caption{Sensitivity analysis for the contrast 4-0 in the worked example.}
\label{fig:sensitivityresults}
\end{figure}

In Figure~\ref{fig:sensitivityresults} we show results for $\hat{\Psi}_{4-0} \pm \delta$ for different values of $\delta = \{0.0, 0.5, 1.0, 1.5, 2.0, 2.5\}$. When $\delta=0$ we have the same estimate for this contrast as in Figure~\ref{fig:tlresults}, and we use the same confidence interval from these earlier results. It can be seen from the plot that we would need the unobserved confounding to be 2 times more severe than the confounders we already adjusted for, for the result to be non-significant, assuming $h_{0, 4-0}: \Psi_{4-0} = 0$. One can also interpret the impact of misspecification in the other direction (\textit{i.e.}, our estimate may \textit{underestimate} the effect if it is in fact misspecified). Thus we are able to get an understanding for the degree to which modeling assumptions affect our results.

\section{Prediction/Exploration}
\label{sec:prediction}
Prediction and exploration is not the primary focus of this work, and so we keep this section brief.

Although it is possible to engage with a predictive/exploratory approach at any stage of the research pipeline, such an approach may be particularly useful during the early stages of the theory development in providing either an initial indication of the predictive validity of the theory, as a tool for automating decision processes (such as automating assessments), or for identifying clusters and identifying additional putative theoretically relevant components. As described already, in some cases, researchers may find their phenomenon of interest be too difficult to specify mathematically (\textit{e.g.}, as a DAG) without high likelihood of misspecification, particularly if there exist known problems with data collection/methodology/measurement. In cases where there is not (yet) a clear route for developing an appropriately simplified representation of an otherwise complex model, one may nonetheless have enough information to construct a loose theory which allows a researcher to identify relevant variables and constructs. Even if one cannot undertake causal discovery and causal inference without severe risk of misspecification, one can instead collect data to undertake a purely associational, predictive or exploratory approach. Under the assumption that causally/theoretically relevant variables should also be predictive, the results from a predictive analysis may provide indication of predictive validity of the initial theory \parencite{Yarkoni2017}. Clustering and dimensionality reduction may similarly provide useful insights into substructures within the data. Furthermore, researchers may wish to pursue a predictive approach above a causal approach, possibly for the purpose of developing decision tools for automating behavioural coding processes \parencite{Biggiogera2021}, or automating assessment \parencite{Cohn2018}.

Following our worked example, we may decide, for instance, that there are too many unobserved confounding variables to be able to reliably estimate the effects we are interested in once we arrive at Stage B, and that instead it may be more productive to pursue the predictive/exploratory approach. Of course, the predictive approach will not yield interpretable causal results, but then neither will a misspecified representation of our theory. The approach itself involves the specification of a set of theoretically relevant or predictive variables. These variables can be used to train predictive algorithms \parencite{Rosenbusch2021} such as deep learning models \parencite{goodfellow} or Super Learners \parencite{Polley2007}, or for exploration and identification of clusters in the data \parencite{luxburg2007}.

\section{Assumptions}
\label{sec:assumptions}
When undertaking causal inference, particularly with observational data, it is normal and encouraged to state clearly the assumptions upon which the validity of the inference rests \parencite{Yao2020, Guo2019, Rubin2005, Imbens2015, Grosz2020, Petersen2012positivity, Rubin1980SUTVA, Hernan2018}. Specifically, the validity of causal inference rests on three key assumptions: (1) Ignorability/Unconfoundedness/Conditional Exchangeability: There are no unobserved confounders, such that the likelihoods of treatment for two individuals with the same covariates are equal, and the potential outcomes for two individuals with the same latent covariates are also equal s.t. $T \indep  (Y(1),Y(0))|\mathbf{X}$. (2) Positivity: the assignment of treatment probabilities are non-zero and non-deterministic $P(T=t_i | \mathbf{X}=\mathbf{x}_i) > 0, \: \forall \; t,\mathbf{x}$. In other words, for each possible substratum in the data, the probability of being assigned a particular treatment is non-zero. (3) Stable Unit Treatment Value Assumption (SUTVA)  \parencite{Rubin1980SUTVA, Laffers2020, Schwartz2012}: this assumptions holds that the potential outcomes for each individual or data unit are independent of the treatments assigned to all other individuals, such that there are no interactions between individuals, and that the effect of treatment on an individual is consistent regardless of the way the participant came to be treated \parencite{Schwartz2012}. In this section we discuss each of these assumptions in turn, and finally discuss how to deal with missing data.

\subsection{Ignorability}
\label{sec:ignorability}
In practice, we may never know whether we have truly collected and adjusted for all possible confounders in order to isolate the effect we care about. Randomizing the assignment of people to treatment and control groups (\textit{i.e.}, in experimental settings) helps considerably in this regard, but again, it may not always be practical or ethical to do so. Of course, and as described in Section~\ref{sec:stageBstep4}, one can undertake sensitivity analysis to understand to what extent violations of this assumption might affect the estimate.

\subsection{Positivity}
\label{sec:positivity}
In terms of the second assumption, violations of the positivity assumption can have an impact on the stability of the estimates, in particular in the derivation of the clever covariates which involves the inverse propensity score $\hat{G}(\mathbf{C})^{-1}$. If any of the propensity scores are close to $0$, meaning that some participants are very unlikely to be in a certain `treatment' group, this reciprocal can take on arbitrarily large values. This results in misspecification of the propensity score model and therefore threatens the double-robustness of the targeted estimator (making it sensitive to misspecification of the outcome model) \parencite{Petersen2012positivity}. It also indicates that certain participants have covariates for which there is little support across contrast groups in the data. This can occur either because there are simply certain types of people who do not fall into these `treatment' groups in the population, or because our particular empirical sample does not contain adequate coverage over covariates and treatment groups. In the former case, collecting more data will not help, and in both cases, the estimation is clearly hindered. There are a number of mitigation strategies. The most straightforward involves simply clipping the estimated propensity scores between the ranges of, say, 0.025 and 0.975 \parencite{Cole2008,Petersen2012positivity}, and this is the approach we used in the analysis above and one which has been shown to function adequately for the targeted approach. Alternatively, one can trim the dataset to consider only participants with a non-zero probability of being in any of the groups \parencite{Petersen2012positivity}. This is, perhaps, not as drastic as it sounds, because, as we mentioned above, some participants may simply be unlikely to fall in certain groups. Trimming results in what is referred to as estimation of the effect for the `intention to treat' regimes \parencite{Petersen2012positivity}. One can alternatively restrict the covariate adjustment set to exclude variables which contribute to positivity violations. Of course, owing to the possible complex interaction between covariates and identification, this may yield other adverse consequences (and threaten identification). We leave a detailed explication of positivity checks and remedies in psychology to future work, but point interested readers to the excellent overview by \textcite{Petersen2012positivity}.

\subsection{Stable Unit Treatment Value Assumption (SUTVA)}
\label{sec:sutva}
 Violations of the SUTVA occur when there exist multiple possible counterfactual / potential outcomes for a participant and a given treatment \parencite{Schwartz2012, Laffers2020}. These violations can occur, for instance, when participants interfere with the effect of treatment of other participants (peer effects), and as such, such violations are quite plausible depending on the data collection methodology. In our worked example, a violation of SUTVA could mean that participants from one couple interfere with the treatment effects of participants in a different couple. Alternatively, SUTVA violations can occur when there exist multiple `hidden' versions of the same treatment. In particular, \textcite[pp.4]{Laffers2020} note that the SUTVA only holds if `neither the treatment nor the observed outcomes are measured with error', although they note that this condition is rarely discussed or considered. This has potential implications for psychology because of the inherent challenges associated with the measurement of certain psychological constructs \parencite{Flake2020}. The proposal by \textcite{Laffers2020} involves the derivation of bounds on the estimand (similar to the family of intervals derived for the sensitivity analysis). Indeed, with SUTVA we assume that if, for participant $i$, $T=t_i$, then $Y_i(t_i) = Y_i$. In other words, the measured outcome for a participant receiving treatment $t_i$ has a potential outcome under this treatment which is equivalent to the outcome measured. This sounds somewhat trivial, but it formalizes the assumption that the effect of treatment itself is a consistent and stable quantity. We leave an implementation and demonstration of SUTVA checks and bounds to future work.

\subsection{Missingness}
\label{sec:missingness}
Missing data is straightforward to handle in the targeted learning framework \parencite{Petersen2014longitudinal, Schomaker2019, vanderLaan2014, vanderlaan2018longitudinal}. In short, for any variables with missing data, each of the missing values are replaced by a single arbitrary value (\textit{e.g}, the average of that variable over participants), and a new `censoring' variable $Q$ is created which indicates missingness. This censoring variable is set to one whenever a value in the original variable is missing, and zero when it is not missing. The censoring variable can assume the same causal input/output causes/effects in the structural representation as the original variable, or, alternatively (particularly in cases where missingness is not completely at random), have a causal structure guided by theory and/or empirical data. During Stage B, the censoring variable $Q$ is treated as another variable upon which we intervene (in addition to the treatment), setting it to one during inference. For instance, the average causal effect for treatment $T=1$ vs. $T=0$ can be reformulated as: $\mathbb{E}_{\mathbf{c}\sim P(\mathbf{C})}[\mathbb{E}_{y\sim P(Y|T=1, \mathbf{C}=\mathbf{c}, Q=1)}[y] - \mathbb{E}_{y\sim P(Y|T=0, \mathbf{C}=\mathbf{c}, Q=1)}[y] ]$. In this expression it can be seen that $Q=1$ in for both $T=1$ and $T=0$, ensuring that we estimate the difference between potential outcomes in the case where missingness does not occur.

\section{Limitations and Discussion}
\label{sec:discussion}
 Even though we agree with \textcite{Hernan2018} and others \parencite{Grosz2020, Vowels2021} insofar as we should start transparently engaging with causality, we also acknowledge the serious challenges which psychologists face in doing so. It is well known that results in psychology are difficult to interpret, and this comes as a consequence of the crud factor \parencite{Meehl1990}, inevitably fat-handed interventions \parencite{Eronen2021}, and complex dynamic phenomena \parencite{Hilpert2019, Vowels2021}. These concerns are also empirically justified: In a large scale evaluation of causal methods for online advertisement involving 500 million observations and 1.6 billion advert impressions, \textcite{Gordon2022} found that even recent methods in causal methods were not able to yield accurate estimates of the true causal effects (some of the data were collected experimentally, and so ground truth existed for comparison). This is particularly notable given that psychological data is likely to be of lower quality (certainly in terms of sample size, if not in terms of measurement error and system uncertainty), and therefore the chances of deriving accurate estimates is even lower. When faced with such challenges, researchers may wish to follow the predictive/exploratory path. We have discussed how predictive and/or exploratory models can be used to establish the predictive validity of a theory and to derive potentially useful tools for automating decision processes (such as behavioural coding or assessment). Indeed, researchers may take this route in parallel to the causal route, and do so at any stage of the research pipeline (although some initial knowledge/theory may be important in identifying important variables and constructs). A predictive approach may well be advised, for instance, for the example data used to demonstrate the methods in this paper (and again, we stress that these data were used \textit{only} for such an illustration).

If one chooses the causal route, one has to take extra care to be transparent about assumptions, and think carefully about the possibility implications when these assumptions fail. It is our view that this approach falls also in line with the goals of most psychologists: We wish to \textit{understand} phenomena so that we may design interventions and manipulate our environment for the better. We should therefore once again argue that taking a transparent and structured approach to the derivation of causal estimates has to be better than the typical but comparatively opaque and unfocused approach to theory specification and analysis commonly adopted in psychology and social science \parencite{Grosz2020, Shmueli2010, Hernan2018}. 

Some researchers may argue that they already employ a similar approach, by virtue of their use of mediation or the Actor Partner Interdependence Model \parencite{Kenny2010APIM}, but in our view these models are (besides the assumption of linearity) too simple to be likely to yield meaningful estimates \parencite{Vowels2021}. Consideration should, at least, be given to the causally proximal structures which influence the mediation subgraph (\textit{e.g.}, possible collider and confounding structures). Indeed, these models are overly simple in spite of the relatively rich descriptions of the corresponding theories they are intended to reflect, highlighting the significant specification gap between theory and analysis. In our working example, the discovered graph presented in Figure~\ref{fig:adj_full_graphical} may also represent a significant simplification, omitting many relevant variables with key structures affecting the estimates we care about. Recall that there were no identified confounders between distress and depression - a situation which seems highly unlikely, and which urges us to reconsider whether we have a sufficient data collection methodology. Of course, this example was used only to demonstrate the techniques, and in practice we would encourage researchers to think more carefully about whether they have collected what they need to continue from Stage A to Stage B in the pipeline.

In terms of the limitations associated with specific algorithms, it is our experience that, in general, non-parametric approaches to causal discovery work poorly and are computationally experience. Indeed, our brief demonstration in Figure~\ref{fig:cdresults} illustrated the computational cost associated with nonparametric approaches to constraint-based causal discovery. Our choice of the Structural Agnostic Model \parencite{Kalainathan2020} is equally subject to variations over hyperparameters and random initalizations. We provided results across 50 runs in order to mitigate such issues, but still one cannot use the output blindly, and one should attempt to validate and modify the putative structure where necessary (though, for example, qualitative research, domain expertise, or by comparing the results from multiple causal discovery algorithms).

In addition to our specific recommendations for the use of DAGs and targeted learning, researchers may wish to also explore extensions to time series data \parencite{Runge2019pcmci,Runge2020pcmci+, Runge2018chaos}, dynamic structural equation models \parencite{Mooij2013}, and complex systems, for which causal discovery methods also exist \parencite{Sugihara2012, Vowels2021NSM, Clark2014, Ye2015, rEDM}. Interested readers are directed to a useful survey of causal discovery methods for time series by \textcite{Assaad2022}. In the case where the resulting structure is still a DAG (\textit{e.g.}, one which has been unrolled over time), one can still continue with Stage B of our proposed pipeline by using, for example, the longitudinal extension of targeted learning \parencite{Schomaker2019, vanderlaan2018longitudinal}, which we plan to integrate into the codebase in the near future. 

We note that many real-time processes may be too dynamic, stochastic, or complex to represent with a DAG and test with the tools of causal inference. For instance, if researchers are interested in the the unfolding of real-time interactions between romantic partners or patients and therapists, pre- and post-session reports will be insufficient. Furthermore, in order to understand how emotions unfold during interactions, we might use human behavioural coding, but such approaches are expensive and time consuming, requiring raters to watch through hours of video manually annotating interactions according to a prescribed codebook \parencite{Hilpert2019}. Even such codes are limited according to the cues the coder has been trained to identify, and there may exist many rich features in video data which are otherwise wasted. In such cases, researchers might turn to modern, real-time, multi-modal techniques which allow for the extraction of facial keypoints \parencite{openface2}, motion \parencite{openpose}, audio \parencite{Eben2010opensmile} and text features \parencite{BERT}. Researchers can then use such features for predicting various outcomes to test, for example, the predictive validity of certain verbal or para-verbal features \parencite{Biggiogera2021, Yarkoni2017}. Such an approach may not help one derive a causal structure, but may be useful in automating coding processes and assessments, as well as identifying associations between behaviors and outcomes. However, in order to know whether such a real-time approach is really necessary, it may still be an important exercise for a researcher to attempt to sketch a possible DAG for the phenomenon of interest. The (potentially in-) tractability of the problem may quickly become apparent, thereby influencing the methodological and analytical direction that they take.

Finally, we note that the abundance of observational data is growing, and some describe the current time as the era of Big Data \parencite{barnes, raghavan, struijs, Harlow2016, Cheung2019}. However, it is important to note that more data does not necessarily mean better data. The same challenges with measurement and the same challenges associated with identifying causal structures exist. In some ways, they are worse. Indeed, and as  \textcite{Canali2016} explains, ``the problem is not finding correlations but rather finding too many of them''. Even though larger samples sizes will always help us with more precise estimation (whether that be for identifying putative causal links or estimating effect sizes), having access to more variables only puts more emphasis on the importance of correctly specifying a theory.

\section{Conclusion}

In this work, we have argued that without a framework to interrogate data for causal quantities, psychologists will not have the tools needed to adequately specify and/or test their theories. Accordingly, we presented a new research pipeline, presented in Figure~\ref{fig:pipeline}. By incorporating tools developed as part of recent progress in the intersection of the machine learning, semiparametric statistics, causal inference, and causal discovery fields, psychologists can move their approach to research to fall more in line with their (causal) goals as scientists. In spite of numerous potential pitfalls and limitations associated with the causal endeavour, it is only through trying that progress is made, and we otherwise resign ourselves to ambiguous pseudo-causal interpretations of arbitrary estimates of correlations and associations.

\printbibliography

@article{Borsboom2021,
	author = {Borsboom, D. and {van der Maas}, H.L.J. and Dalege, J. and Kievit, R.A. and Haig, B.D.},
	doi = {doi:10.1177/1745691620969647},
	journal = {Perspectives on Psychological Science},
	number = {4},
	pages = {756-766},
	title = {Theory construction methodology: a practical framework for building theories in psychology},
	volume = {16},
	year = {2021}}

@article{luxburg2007,
	author = {{von Luxburg}, U.},
	doi = {10.1007/s11222-007-9033-z},
	journal = {Statistics and Computing},
	pages = {395-416},
	title = {A tutorial on spectral clustering},
	volume = {17},
	year = {2007}}

@article{Rosenbusch2021,
	author = {Rosenbusch, H. and Soldner, F. and Evans, A.M. and Zeelenberg, M.},
	doi = {10.1111/spc3.12579},
	journal = {Social and Personality Psychology Compass},
	number = {2},
	title = {Supervised machine learning methods in psychology: A practical introduction with annotated {R} code},
	volume = {15},
	year = {2021}}

@article{Himabindu2021compvision,
	author = {Himabindu, D.D. and Kumar, S.P.},
	doi = {10.14569/IJACSA.2021.0121013},
	journal = {Journal of Advanced Computer Science and Applications},
	number = {10},
	title = {A survey on computer vision architectures for large scale image classification using deep learning},
	volume = {12},
	year = {2021}}

@article{Qiu2020NLPsurvey,
	author = {Qiu, X.P. and Sun, T.X. and Xu, Y.G. and Shao, Y.F. and Dai, N. and Huang, X.J.},
	doi = {10.1007/s11431-020-1647-3},
	journal = {Science China Technological Sciences},
	pages = {1872-1897},
	title = {Pre-trained models for natural language processing: a survey},
	volume = {63},
	year = {2020}}

@article{Petersen2014longitudinal,
	author = {Petersen, M. and Schwab, J. and Gruber, S. and Blaser, N. and Schomaker, M. and {van der Laan}, M.J.},
	doi = {10.1515/jci-2013-0007},
	journal = {Journal of Causal Inference},
	number = {2},
	pages = {147-185},
	title = {Targeted maximum likelihood estimation for dynamic and static longitudinal marginal structural working modles},
	volume = {2},
	year = {2014}}

@book{Glymour2014,
	author = {Glymour, C. and Scheines, R. and Spirtes, P.},
	publisher = {Academic Press},
	title = {Discovering causal structure: Artificial intelligence, philosophy of science, and statistical modeling},
	year = {2014}}

@book{Mackie1974causalcement,
	address = {Oxford},
	author = {Mackie, J.L},
	publisher = {Clarendon Press},
	title = {The cement of the universe: A study of causation},
	year = {1974}}

@article{Machamer2000causalphil,
	author = {Machamer, P. and Darden, L. and Craver, C.F.},
	doi = {jstor.org/stable/188611},
	journal = {Philosophy of Science},
	number = {1},
	pages = {1-25},
	title = {Thinking about mechanisms},
	volume = {67},
	year = {2000}}

@article{Fisher1970sem,
	author = {Fisher, F.M.},
	doi = {10.2307/1909242},
	journal = {Econometrica: Journal of the Econometric Society},
	number = {1},
	pages = {73-92},
	title = {A correspondence principle for simultaneous equation models},
	volume = {38},
	year = {1970}}

@article{Forney2022causaledu,
	author = {Forney, A. and Mueller, S.},
	journal = {Journal of Causal Inference},
	title = {Causal inference in {AI} education: a primer},
	year = {2022}}

@book{vanderlaan2018longitudinal,
	address = {Cham, Switzerland},
	author = {{van der Laan}, M.J. and Rose, S.},
	doi = {10.1007/978-3-319-65304-4},
	publisher = {Springer Series in Statistics},
	title = {Targeted Learning in Data Science: Causal inference for complex longitudinal studies},
	year = {2018}}

@article{Assaad2022,
	author = {Assaad, C.K. and Devijver, E. and Gaussier, E.},
	doi = {10.1613/jair.1.13428},
	journal = {Journal of Artificial Intelligence Research},
	title = {Survey and evaluation of causal discovery methods for time series},
	volume = {73},
	year = {2022}}

@article{Biggiogera2021,
	author = {Biggiogera, J. and Boateng, G. and Hilpert, P. and Vowels, M. and Bodenmann, G. and Neysari, M. and Nussbeck, F. and Kowatsch, T.},
	doi = {10.1145/3461615.3485423},
	journal = {Companion Publication of the 2021 International Conference on Multimodal Interaction},
	number = {385-389},
	title = {{BERT} meets {LIWC}: Exploring state-of-the-art language models for predicting communication behavior in couples' conflict interactions},
	year = {2021}}

@article{Eben2010opensmile,
	author = {Eyben, F. and Wollmer, M. and Schuller, B.},
	doi = {10.1145/1873951.1874246},
	journal = {Proceedings 18th ACM international conference on multimedia},
	number = {1459-1642},
	title = {Opensmile: the munich versatile and fast open-source audio feature extractor},
	year = {2010}}

@article{Shepard1987,
	author = {Shepard, R.N.},
	doi = {10.1126/science.3629243},
	journal = {Science},
	number = {4820},
	pages = {1317-1323},
	title = {Toward a universal law of generalization for psychological science},
	volume = {237},
	year = {1987}}

@article{Kenny2010APIM,
	author = {Kenny, D.A. and Lederman, T.},
	doi = {10.1037/a0019651},
	journal = {Journal of Family Psychology},
	pages = {359-366},
	title = {Detecting, measuring, and testing dyadic patters in the actor-partner interdependence model},
	volume = {24},
	year = {2010}}

@article{Navarro2021perspective,
	author = {Navarro, D.J.},
	doi = {10.1177/1745691620974769},
	journal = {Perspectives on Psychological Science},
	number = {4},
	title = {If mathematical psychology did not exist we might need to invent it: A comment on theory building in psychology},
	volume = {16},
	year = {2021}}

@article{Runge2018chaos,
	author = {Runge, J.},
	doi = {10.1063/1.5025050},
	journal = {Chaos: An interdisciplinary journal of nonlinear science},
	number = {7},
	title = {Causal network reconstruction from time series: from theoretical assumptions to practical estimation},
	volume = {28},
	year = {2018}}

@article{Runge2020pcmci+,
	author = {Runge, J.},
	journal = {Proceedings of the 36th Conference on Uncertainty in Artificial Intelligence},
	pages = {1388-1397},
	title = {Discovering contemporaneous and lagged causal relations in autocorrelated nonlinear time series data},
	volume = {124},
	year = {2020}}

@article{Runge2019pcmci,
	author = {Runge, J. and Nowack, M. and Kretschmer, M. and Flaxman, S. and Sejdinovic, D.},
	doi = {10.1126/sciadv.aau4996},
	journal = {Science Advances},
	number = {11},
	title = {Detecting and quantifying causal associations in large nonlinear time series datasets},
	volume = {5},
	year = {2019}}

@article{Schwartz2012,
	author = {Schwartz, S. and Gatto, N.M. and Campbell, U.B.},
	doi = {10.1186/1742-5573-9-3},
	journal = {Epidemiologic Perspectives and Innovations},
	number = {3},
	title = {Extending the sufficient component cause model to describe the stable unit treatment value assumption ({SUTVA})},
	volume = {9},
	year = {2012}}

@article{Venderweele2006,
	author = {{VanderWeele}, T.J. and Hernan, M.A.},
	doi = {10.1007/s10654-006-9075-0.},
	journal = {European Journal of Epidemiology},
	pages = {855-858},
	title = {From counterfactuals to sufficient component causes and vice versa},
	volume = {21},
	year = {2006}}

@article{Cole2008,
	author = {Cole, S.R. and Hernan, M.A.},
	doi = {10.1093/aje/kwn164},
	journal = {American Journal of Epidemiology},
	number = {6},
	title = {Constructing inverse probability weights for marginal structural models},
	volume = {168},
	year = {2008}}

@article{Gruber2022,
	author = {Gruber, S. and Phillips, R.V. and Lee, H. and Ho, M. and Concato, J. and {van der Laan}, M.J.},
	journal = {arXiv preprint},
	title = {Targeted learning: towards a future informed by real-world evidence},
	volume = {arXiv:2205.08643},
	year = {2022}}

@article{Rotnitzky1998,
	author = {Rotnitzky, A. and Robins, J. and Scharfstein, D.},
	doi = {jstor.org/stable/2670049},
	journal = {Journal of the American Statistical Association},
	number = {444},
	pages = {1321-1339},
	title = {Semiparametric regression for repeated outcomes with nonignorable nonresponse},
	volume = {93},
	year = {1998}}

@article{Rotnitzky2001,
	author = {Rotnitzky, A. and Scharfstein, D. and Su, {T.-L.} and Robins, J.},
	doi = {jstor.org/stable/2676847},
	journal = {International Biometric Society},
	number = {1},
	pages = {103-113},
	title = {Methods for conducting sensitivity analysis of trials with potentially nonignorable competing causes of censoring},
	volume = {57},
	year = {2001}}

@article{Gabriel2021,
	author = {Gabriel, E.E. and Sjolander, A. and Sachs, M.C.},
	doi = {10.1080/01621459.2021.1950734},
	journal = {Journal of the American Statistical Association},
	title = {Nonparametric bounds for causal effects in imperfect randomized experiments},
	year = {2021}}

@article{Laffers2020,
	author = {Laffers, L. and Mellace, G.},
	journal = {Discussion Papers on Business and Economics},
	title = {Identification of the average treatment effect when {SUTVA} is violated},
	volume = {3},
	year = {2020}}

@article{Rubin1980SUTVA,
	author = {Rubin, D.B.},
	journal = {Journal of the American Statistical Association},
	pages = {591-593},
	title = {Comment on `Randomization analysis of experimental data in the {Fisher} randomization test' by {D. Basu}},
	volume = {75},
	year = {1980}}

@article{Petersen2012positivity,
	author = {Petersen, M.L. and Porter, K.E. and Gruber, S. and Wang, Y. and {van der Laan}, M.J.},
	doi = {10.1177/0962280210386207},
	journal = {Statistical Methods in Medical Research},
	number = {1},
	pages = {31-54},
	title = {Diagnosing and responding to violations in the positivity assumption},
	volume = {21},
	year = {2012}}

@article{AkbariProbADMG2022,
	author = {Akbari, S. and Jamshidi, M. and Mokhtarian, E. and Vowels, M.J. and Etesami, J. and Kiyavash, N.},
	journal = {arXiv preprint},
	title = {Causal Discovery in Probabilistic Networks with an Identifiable Causal Effect},
	volume = {tbc},
	year = {2022}}

@article{Umberson2015,
	author = {Umberson, D.},
	doi = {10.3886/ICPSR37404.v2},
	journal = {Inter-university consortium for political and social research},
	title = {Health and relationships project},
	year = {2014-2015}}

@article{Siddique2019,
	author = {Siddique, A.A. and Schnitzer, M.E. and Bahamyirou, A. and Wang, G. and Holtz, T.H. and Migliori, G.B. and {et al.}},
	doi = {10.1177/0962280218808817},
	journal = {Statistical Methods in Medical Research},
	number = {12},
	pages = {3534-3549},
	title = {Causal inference with multiple concurrent medications: A comparison of methods and an application in multidrig-resistance tuberculosis},
	volume = {28},
	year = {2019}}

@article{Gordon2022,
	author = {Gordon, B.R. and Moakler, R. and Zettelmeyer, F.},
	journal = {arXiv preprint},
	title = {Close enough? A large-scale exploration of non-experimental approaches to advertising measurement},
	volume = {arXiv:2201.07055v1},
	year = {2022}}

@article{Pedemonte2021causaleffect,
	author = {Pedemonte, M. and Vitria, J. and Parafita, A.},
	journal = {arXiv preprint},
	title = {Algorithmic causal effect identification with causaleffect},
	volume = {arXiv:2107.04632v1},
	year = {2021}}

@article{Eronen2020,
	author = {Eronen, M.I.},
	doi = {doi:10.1016/j.newideapsych.2020.100785},
	journal = {New Ideas in Psychology},
	title = {Causal discovery and the problem of psychological interventions},
	volume = {59},
	year = {2020}}

@article{Eronen2021,
	author = {Eronen, M.I. and Bringmann, L.F.},
	doi = {doi:10.1177/1745691620970586},
	journal = {Perspectives on Psychological Science},
	number = {4},
	title = {The theory crisis in psychology: how to move forward},
	volume = {16},
	year = {2021}}

@article{Greenland1999,
	author = {Greenland, S. and Pearl, J. and Robins, J.M.},
	journal = {Epidemiology},
	number = {1},
	pages = {37-48},
	title = {Causal diagrams for epidemiological research},
	volume = {10},
	year = {1999}}

@book{Rothman2008,
	address = {Philadelphia},
	author = {Rothman, K.J. and Greenland, S. and Lash, T.L.},
	edition = {3rd},
	publisher = {Lippincott Williams and Wilkins},
	title = {Modern Epidemiology},
	year = {2008}}

@article{Pearl1999,
	author = {Pearl, J.},
	journal = {Synthese},
	number = {93-149},
	title = {Probabilities of causation: Three counterfactual interpretations and their identification},
	volume = {1-2},
	year = {123}}

@article{Robinaugh2020,
	author = {Robinaugh, D.J. and Hoekstra, R.H. and Toner, E.R. and Borsboom, D.},
	doi = {10.1017/S0033291719003404},
	journal = {Psychological Medicine},
	number = {3},
	pages = {353-366},
	title = {The network approach to psychopathology: a review of the literature 2008-2018 and an agenda for future research},
	volume = {50},
	year = {2020}}

@article{Epskamp2018PMRF,
	author = {Epskamp, S. and Waldorp, L.J. and Mottus, R. and Borsboom, D.},
	doi = {10.1080/00273171.2018.1454823},
	journal = {Multivariate Behavioral Research},
	number = {4},
	pages = {453-480},
	title = {The {Gaussian} graphical model in cross-sectional and time series data},
	volume = {53},
	year = {2018}}

@article{borkulo2014PMRF,
	author = {{van Borkulo}, C.D. and Borsboom, D. and Epskamp, S. and {et al.}},
	doi = {10.1038/srep05918},
	journal = {Nature Scientific Reports},
	number = {5918},
	title = {A new method for constructing networks from binary data},
	volume = {4},
	year = {2014}}

@article{Zhou2021sens,
	author = {Zhou, M. and Yao, W.},
	doi = {10.1080/02664763.2021.1999398},
	journal = {Jouranl of Applied Statistics},
	title = {Sensitivity analysis of unmeasured confounding in causal inference based on exponential tilting and super learner},
	year = {2021}}

@article{Ryan2022,
	author = {Ryan, O. and Bringmann, L.F. and Schuurman, N.K.},
	journal = {PsyArXiv preprint},
	title = {The challenge of generating causal hypotheses using network models},
	volume = {10.31234/osf.io/ryg69},
	year = {2022}}

@article{BenkeserTMLE2017,
	author = {Benkeser, D. and Carone, M. and {van der Laan}, M.J. and {et al.}},
	doi = {10.1093/biomet/asx053},
	journal = {Biometrika},
	number = {4},
	pages = {863-880},
	title = {Doubly robust nonparametric inference on the average treatment effect},
	volume = {104},
	year = {2017}}

@article{Bonvini2020,
	author = {Bonvini, M. and Kennedy, E.H.},
	journal = {arXiv preprint},
	title = {Sensitivity analysis via the proportion of unmeasured confounding},
	volume = {arXiv:1912.02793v2},
	year = {2020}}

@article{Diaz2013,
	author = {Diaz, I. and {van der Laan}, M.J.},
	journal = {The International Journal of Biostatistics},
	number = {2},
	pages = {149-160},
	title = {Sensitivity analysis for causal inference under unmeasured confounding and measurement error problems},
	volume = {9},
	year = {2013}}

@article{Thabane2013Sensitivity,
	author = {Thabane, L. and Mbuagbaw, L. and Zhang, S. and Samaan, Z. and Marcucci, M. and et al.},
	doi = {10.1186/1471-2288-13-92},
	journal = {BMC Medical Research Methodology},
	number = {92},
	title = {A tutorial on sensitivity analyses in clinical trials: the what, why, when and how},
	volume = {13},
	year = {2013}}

@article{PhillipsSuperLearner,
	author = {Phillips, R.V. and {van der Laan}, M.J. and Lee, H. and Gruber, S.},
	journal = {arXiv preprint},
	title = {Practical considerations for specifying a super learner},
	volume = {arXiv:2204.06139v2},
	year = {2022}}

@article{ImbensGPS2000,
	author = {Imbens, G.W.},
	journal = {Biometrika},
	number = {3},
	pages = {706-710},
	title = {The role of the propensity score in estimating dose-response functions},
	volume = {87},
	year = {2000}}

@article{Vowels2022minsem,
	author = {Vowels, M.J.},
	journal = {arXiv preprint},
	title = {Prespecification of structure for increasing research transparency and for the optimization of data collection},
	volume = {arXiv:2203.13331},
	year = {2022}}

@article{Efron1981,
	author = {Efron, B.},
	doi = {10.1093/biomet/68.3.589},
	journal = {Biometrika},
	number = {3},
	pages = {589-599},
	title = {Nonparametric estimates of standard error: the jackknife, the bootstrap and other methods},
	volume = {68},
	year = {1981}}

@article{Mueller2021ite,
	author = {Mueller, S. and Li, A. and Pearl, J.},
	journal = {arXiv preprint},
	title = {Causes of effects: Learning individual responses from population data},
	volume = {arXiv:2104.13730v2},
	year = {2021}}

@article{Pearl1993backdoor,
	author = {Pearl, J.},
	journal = {Proceedings of the 49th Session of the International Statistical Institute},
	pages = {399-401},
	title = {Aspects of graphical models connceted with causality},
	year = {1993}}

@article{Tennant2021,
	author = {Tennant, P.W.G. and Murray, E.J. and Arnold, K.F. and Berrie, L. et al.},
	doi = {10.1093/ije/dyaa213},
	journal = {International Journal of Epidemiology},
	title = {Use of directed acyclic graphs {DAGs} to identify confounders in applied health research: review and recommendations},
	year = {2021}}

@article{Yazdani2015,
	author = {Yazdani, A. and Boerwinkle, E.},
	doi = {10.4172/2153-0602.1000163},
	journal = {Journal of Data Mining Genomics Proteomics},
	number = {1},
	title = {Causal inference in the age of decision medicine},
	volume = {6},
	year = {2015}}

@article{Szucs2017,
	author = {Szucs, D. and Ioannidis, J.},
	journal = {Frontiers in Human Neuroscience},
	title = {When null hypothesis significance testing is unsuitable for research: A reassessment},
	volume = {11},
	year = {2017}}

@article{Henrich2010,
	author = {Henrich, J. and Heine, S.J. and Norenzayan, A.},
	journal = {Behavioral and Brain Sciences},
	pages = {61-83},
	title = {The weirdest people in the world?},
	volume = {33},
	year = {2010}}

@article{Barry2014,
	author = {Barry, A.E. and Chaney, B. and {Piazza-Gardner}, A.K. and Chavarria, E.A.},
	doi = {10.1177/1090198113483139},
	journal = {Health Education and Behavior},
	pages = {12-18},
	title = {Validity and reliability reporting practices in the field of health education and behavior: A review of seven journals},
	volume = {41},
	year = {2014}}

@inbook{Lavrakas2008,
	address = {Thousand Oaks, CA},
	author = {Lavrakas, P.J.},
	chapter = {Respondent fatigue},
	doi = {10.4135/9781412963947},
	publisher = {SAGE Publications},
	title = {Encyclopedia of survey research methods},
	volume = {1},
	year = {2008}}

@article{Sedlmeier1989,
	author = {Sedlmeier, P. and Gigerenzer, G.},
	doi = {10.1037/0033-2909.105.2.309},
	journal = {Psychological Bulletin},
	number = {2},
	pages = {309-316},
	title = {Do studies of statistical power have an effect on the power of studies},
	volume = {105},
	year = {1989}}

@article{Scheel2022,
	author = {Scheel, A.M.},
	doi = {10.1002/icd.2295},
	journal = {Infant and Child Development},
	number = {1},
	title = {Why most psychological research findings are not even wrong},
	volume = {31},
	year = {2022}}

@article{Crutzen2017,
	author = {Crutzen, R. and Peters, {G.J.Y}},
	doi = {10.3389/fpsyg.2017.01184},
	journal = {Frontiers in Psychology},
	title = {Targeting next generations to change the common practice of underpowered research},
	volume = {8},
	year = {2017}}

@article{Vankov2014,
	author = {Vankov, I. and Bowers, S. and Munafo, M.R.},
	doi = {10.1080/17470218.2014.885986},
	journal = {The Quarterly Journal of Experimental Psychology},
	number = {5},
	pages = {1037-1040},
	title = {On the persistence of low power in psychological science},
	volume = {67},
	year = {2014}}

@article{Maxwell2004,
	author = {Maxwell, S.E.},
	doi = {10.1037/1082-989X.9.2.147},
	journal = {Psychological Methods},
	number = {2},
	pages = {147-163},
	title = {The persistence of underpowered studies in psychological research: Causes, consequences, and remedies},
	volume = {9},
	year = {2004}}

@article{Wagenmakers2012,
	author = {Wagenmakers, E-J. and Wetzels, R. and Borsboom, D. and {van der Maas}, H.L.J. and Kievit, R.A.},
	doi = {10.1177/1745691612463078},
	journal = {Perspectives on Psychological Science},
	number = {6},
	pages = {632-638},
	title = {An agenda for purely confirmatory research},
	volume = {7},
	year = {2012}}

@article{Hullman2022,
	author = {Hullman, J. and Kapoor, S. and Nanayakkara, P. and Gelman, A. and Narayanan, A.},
	journal = {arXiv preprint},
	title = {The worst of both worlds: A comparative analysis of errors in learning from data in psychology and machine learning},
	volume = {arXiv:2203.06498},
	year = {2022}}

@article{Maclaren2020,
	author = {Maclaren, O.J. and Nicholson, R.},
	journal = {arXiv preprint},
	title = {What can be estimated? Identifiability, estimability, causal inference and ill-posed inverse problems},
	volume = {arXiv:1904.02826v4},
	year = {2020}}

@article{pgmpy,
	author = {Ankan, A. and Panda, A.},
	journal = {Proc. 14th Python in Science Conference (SPIPY)},
	title = {pgmpy: {Probabilistic} graphical models using python},
	year = {2015}}

@article{Runge2018,
	author = {Runge, J.},
	journal = {ICML},
	title = {Conditional indpendence testing based on a nearest-neighbor estimate of conditional mutual information},
	year = {2018}}

@article{VowelsFreeLunch2022,
	author = {Vowels, M.J. and Akbari, S. and Etesami, J. and Camgoz, N.C. and Bowden, R.},
	journal = {arXiv preprint},
	title = {A Free Lunch with Influence Functions? Improving Neural Network Estimates with Concepts from Semiparametric Statistics},
	year = {2022}}

@article{Hunermund2021,
	author = {H{\"u}nermund, P. and Bareinboim, E.},
	journal = {arXiv preprint},
	title = {Causal inference and data fusion in econometrics},
	volume = {arXiv:1912.09104v3},
	year = {2021}}

@article{Canali2016,
	author = {Canali, S.},
	doi = {10.1177/2053951716669530},
	journal = {Big Data and Society},
	title = {{Big Data}, epistemology and causality: {K}nowledge in and knowledge out in {EXPOsOMICS}},
	volume = {1},
	year = {2015}}

@article{Wright1923,
	author = {Wright, S.},
	journal = {Genetics},
	pages = {239-255},
	title = {The theory of path coefficients: a reply to {N}iles' criticism},
	volume = {8},
	year = {1923}}

@article{Wright1921,
	author = {Wright, S.},
	journal = {Journal of Agriculture Research},
	pages = {557-585},
	title = {Correlation and causation},
	volume = {20},
	year = {1921}}

@article{huang2012pearls,
	author = {Huang, Y. and Valtorta, M.},
	doi = {10.5555/3020419.3020446},
	journal = {Proceedings of the Twenty-Second Conference on Uncertainty in Artificial Intelligence},
	pages = {217--224},
	title = {Pearl's calculus of intervention is complete},
	volume = {arXiv:1206.6831},
	year = {2006}}

@article{Richardson2017,
	author = {Richardson, T.S. and Evans, R.J. and Robins, J.M. and Shpitser, I.},
	journal = {arXiv preprint},
	title = {Nested {Markov} properties for {Acyclic Directed Mixed Graphs}},
	volume = {arXiv:1701.06686v2},
	year = {2017}}

@article{Hines2021,
	author = {Hines, O. and Dukes, O. and {Diaz-Oraz}, K. and Vansteelandt, S.},
	journal = {arXiv preprint},
	title = {Demystifying statistical learning based on efficient influence functions},
	volume = {arXiv:2107.00681},
	year = {2021}}

@article{Polley2007,
	author = {{van der Laan}, M.J. and Polley, E.C. and Hubbard, A.E.},
	doi = {10.2202/1544-6115.1309},
	journal = {Statistical Applications of Genetics and Molecular Biology},
	number = {25},
	title = {{Super Learner}},
	volume = {6},
	year = {2007}}

@article{Reisach2021,
	author = {Reisach, A.G. and Seiler, C. and Weichwald, S.},
	journal = {arXiv preprint},
	title = {Beware of the simulated {DAG}! Varsortability in additive noise models},
	volume = {arXiv:2102.13647},
	year = {2021}}

@article{Schomaker2019,
	author = {Schomaker, M. and {Luque-Fernandez}, M.A. and Leroy, V. and Davies, M.A.},
	doi = {10.1002/sim.8340},
	journal = {Statistics in Medicine},
	number = {24},
	pages = {4888-4911},
	title = {Using longitudinal targeted maximum likelihood estimation in complex settings with dynamic interventions},
	volume = {38},
	year = {2019}}

@article{BERT,
	author = {Devlin, J. and Chang, {M.-W.} and Lee, K. and Toutanova, K.},
	journal = {arXiv preprint},
	title = {{BERT}: Pre-training of deep bidirectional transformers for language understanding},
	volume = {arXiv:1810.04805v2},
	year = {2019}}

@article{Kennedy2017,
	author = {Kennedy, E.H.},
	journal = {arXiv:1709.06418v1},
	title = {Semiparametric Theory},
	year = {2017}}

@article{Fisher2019,
	author = {Fisher, A. and Kennedy, E.H.},
	journal = {arXiv:1810.03260v3},
	title = {Visually communicating and teaching intuition for influence functions},
	year = {2019}}

@article{Cheung2019,
	author = {Cheung, M. W. -L. and Jak, S.},
	doi = {10.1027/2151-2604/a000348},
	journal = {Zeitschrift fur Psychologie},
	number = {4},
	pages = {209-211},
	title = {Challenges of big data analyses and applications in psychology},
	volume = {226},
	year = {2018}}

@article{Harlow2016,
	author = {Harlow, L.L. and Oswald, F.L.},
	doi = {10.1037/met0000120},
	journal = {Psychological Methods},
	number = {4},
	title = {Big data in psychology: introduction to the special issue},
	volume = {21},
	year = {2016}}

@article{Vowels2021NSM,
	author = {Vowels, M.J. and Camgoz, N.C. and Bowden, R.},
	journal = {IEEE Conference on Computer Vision and Pattern Recognition Causality in Vision Workshop},
	title = {Shadow-mapping for unsupervised neural causal discovery},
	year = {2021}}

@article{Kaiser2021,
	author = {Kaiser, M. and Sipos, M.},
	journal = {arXiv:2104.05441},
	title = {Unsuitability of {NOTEARS} for causal graph discovery},
	year = {2021}}

@article{Clark2014,
	author = {Clark, A.T. and Ye, H. and Isbell, F. and Deyle, E.R. and Cowles, J. and Tilman, G.D. and Sugihara, G.},
	journal = {Ecology},
	number = {5},
	title = {Spatial convergent cross mapping to detect causal relationships from short time series},
	volume = {96},
	year = {2015}}

@article{Vowels2021DAGs,
	author = {Vowels, M.J. and Camgoz, N.C. and Bowden, R.},
	doi = {10.1145/3527154},
	journal = {ACM Comput. Surv.},
	title = {D'ya like {DAGs}? {A} survey on structure learning and causal discovery},
	year = {2022}}

@article{Ye2015,
	author = {Ye, H. and Deyle, E. and Gilarranz, L.J. and Sugihara, G.},
	journal = {Scientific Reports},
	number = {14750},
	title = {Distinguishing time-delayed causal interactions using convergent cross mapping},
	volume = {5},
	year = {2015}}

@article{rEDM,
	author = {Park, J. and Smith, C. and Sugihara, G. and Deyle, E. and Saberski, E. and Ye, H.},
	title = {{rEDM}: Empirical Dynamic Modeling},
	url = {https://CRAN.R-project.org/package=rEDM},
	year = {2021}}

@article{Sugihara2012,
	author = {Sugihara, G. and May, R. and Hsieh, {C.-h.} and Deyle, E. and Fogarty, M. and Munch, S.},
	journal = {Science},
	title = {Detecting causality in complex ecosystems},
	volume = {338},
	year = {2012}}

@article{Mooij2013,
	author = {Mooij, J.M. and Janzing, D. and Sch\"{o}lkopf, B.},
	journal = {UAI},
	title = {From ordinary differential equations to structural causal models: the deterministic case},
	year = {2013}}

@article{n}

@article{Korb1997,
	author = {Korb, K.B. and Wallace, C.S.},
	journal = {British Journal for the Philosophy of Science},
	pages = {543-553},
	title = {In search of the philosopher's stone: Remarks on {Humphreys and Freedman's} critique of causal discovery},
	year = {1997}}

@article{Verma1990,
	author = {Verma, T. and Pearl, J.},
	journal = {Proc. 6th Conf. on Uncertainty in Artificial Intelligence},
	title = {Equivalence and synthesis of causal models},
	year = {1990}}

@article{LewisCausation1973,
	author = {Lewis, D.},
	journal = {The Journal of Philosophy},
	number = {17},
	pages = {556-567},
	title = {Causation},
	volume = {70},
	year = {1973}}

@book{Spirtes2000,
	address = {Cambridge, Massachusetts},
	author = {Spirtes, P. and Glymour, C. and Scheines, R.},
	edition = {2nd},
	publisher = {MIT Press},
	title = {Causation, prediction, and search},
	year = {2000}}

@article{Triantafillou2010,
	author = {Triantafillou, S. and Tsamardinos, I. and Tollis. I.},
	journal = {Proceedings of the 13th International Conference on Artificial Intelligence and Statistics},
	title = {Learning causal structure from overlapping variable sets},
	year = {2010}}

@article{Hoyer2008b,
	author = {Hoyer, P.O. and Janzing, D. and Mooij, J.M. and Peters, J.},
	journal = {Advances in Neural Information Processing Systems},
	title = {Nonlinear causal discovery with additive noise models},
	year = {2008}}

@article{Shah2020,
	author = {Shah, R. D. and Peters, J.},
	journal = {The Annals of Statistics},
	number = {3},
	title = {The hardness of conditional independence testing and the generalised covariance measure},
	volume = {48},
	year = {2020}}

@article{Dawid2008,
	author = {Dawid, A.P.},
	journal = {{NeurIPS} Workshop on Causality},
	title = {Beware of the {DAG!}},
	year = {2008}}

@article{Kalainathan2020,
	author = {Kalainathan, D. and Goudet, O. and Guyon, I. and {Lopez-Paz}, D. and Sebag, M.},
	journal = {arXiv:1803.04929v3},
	title = {Structural agnostic modeling: Adversarial learning of causal graphs},
	year = {2020}}

@article{Glymour2019,
	author = {Glymour, C. and Zhang, K. and Spirtes, P.},
	journal = {Frontiers in Genetics},
	title = {Review of causal discovery methods based on graphical models},
	volume = {10},
	year = {2019}}

@article{Heinze2018,
	author = {{Heinze-Deml}, C. and Maathuis, M.H. and Meinshausen, N.},
	journal = {Annual Review of Statistics and Its Application},
	title = {Causal structure learning},
	volume = {5},
	year = {2018}}

@article{VowelsSpectralTutorial,
	author = {Vowels, M. J. and Vowels, L. M. and Wood, N.},
	journal = {Psychological Methods},
	title = {Spectral and Cross-Spectral Analysis-a Tutorial for Psychologists and Social Scientists},
	year = {2021}}

@article{Vowels2021,
	author = {Vowels, M. J.},
	doi = {10.1037/met0000429},
	journal = {Psychological Methods},
	title = {Misspecification and unreliable interpretations in psychology and social science},
	year = {2021}}

@article{Guo2020,
	author = {Guo, R. and Cheng, L. and Li, J. and Hahn, P.R. and Liu, H.},
	journal = {ACM Comput. Surv.},
	number = {1},
	title = {A survey of learning causality with data: Problems and methods},
	volume = {1},
	year = {2020}}

@article{Flake2020,
	author = {Flake, J. and Fried, E.},
	journal = {Advances in Methods and Practices in Psychological Science},
	title = {Measurement schmeasurement: questionable measurement practices and how to avoid them},
	year = {in press}}

@article{Spellman2015,
	author = {Spellman, B.A.},
	doi = {10.1177/1745691615609918},
	journal = {Perspectives on Psychological Science},
	number = {6},
	pages = {886-899},
	title = {A short (personal) future history of revolution 2.0},
	volume = {10},
	year = {2015}}

@article{Claesen2019,
	author = {Claesen, A. and Gomes, S.L.B.T. and Tuerkinckx, F. and Vanpaemel, W.},
	doi = {10.31234/osf.io/d8wex},
	journal = {PsyArXiv},
	title = {Preregistration: Comparing dream to reality},
	year = {2019}}

@article{Scheel2020,
	author = {Scheel, A.M. and Tiokhin, L. and Isager, P.M. and Lakens, D.},
	journal = {Perspectives on Psychological Science},
	title = {Why hypothesis testers should spend less time testing hypotheses},
	year = {in press}}

@article{Orben2020,
	author = {Orben, A. and Lakens, D.},
	doi = {10.1177/2515245920917961},
	journal = {Advances in Methods and Practices in Psychological Science},
	number = {2},
	pages = {238-247},
	title = {Crud (re)defined},
	volume = {3},
	year = {2020}}

@book{Loehlin2017,
	address = {New York},
	author = {Loehlin, J.C. and Beaujean, A.A.},
	publisher = {Routledge Taylor and Francis},
	title = {Latent Variable Models: An introduction to factor, path, and structural equation analysis},
	year = {2017}}

@book{Sayama2015,
	address = {Geneseo, New York},
	author = {Sayama, H.},
	publisher = {Open {SUNY} Textbooks},
	title = {Introduction to the modeling and analysis of complex systems},
	year = {2015}}

@article{Meehl1990,
	author = {Meehl, P.E.},
	doi = {10.2466/pr0.1990.66.1.195},
	journal = {Psychological Reports},
	pages = {195-244},
	title = {Why summaries of research on psychological theories are often uninterpretable},
	volume = {66},
	year = {1990}}

@article{Imbens2020,
	author = {Imbens, G.W.},
	journal = {arXiv:1907.07271v2},
	title = {Potential outcome and directed acyclic graph approaches to causalty: Relevance for empirical practice in economics},
	year = {2020}}

@article{Shmueli2010,
	author = {Shmueli, G.},
	doi = {doi:10.1214/10-STS330},
	journal = {Statistical Science},
	number = {3},
	pages = {289-310},
	title = {To explain or to predict?},
	volume = {25},
	year = {2010}}

@article{Hernan2018,
	author = {Hernan, M.},
	doi = {10.2105/AJPH.2018.304337},
	journal = {American Journal of Public Health},
	number = {5},
	pages = {625-626},
	title = {The c-word: scientific euphemisms do not improve causal inference from observational data},
	volume = {108},
	year = {2018}}

@book{Glymour2001,
	author = {Glymour, C.},
	publisher = {MIT Press},
	title = {The mind's arrows: {B}ayes nets and graphical causal models in psychology},
	year = {2001}}

@article{Cassidy2019,
	author = {Cassidy, S.A. and Dimova, R. and Giguere, B. and Spence, J.R. and Stanley, D.J.},
	doi = {10.1177/2515245919858072},
	journal = {Advances in Methods and Practices in Psychological Science},
	number = {3},
	title = {Failing grade: 89 percent of introduction-to-psychology textbooks that define or explain statistical significance do so incorrectly},
	volume = {2},
	year = {2019}}

@article{Mooij2016,
	author = {Mooij, J.M. and Peters, J. and Janzing, D. and Zscheischler, J. and Scholkopf, B.},
	journal = {Journal of Machine Learning Research},
	number = {32},
	pages = {1-102},
	title = {Distinguishing cause from effect using observational data: methods and benchmarks},
	volume = {17},
	year = {2016}}

@article{Vowels2020b,
	author = {Vowels, M. J. and Camgoz, N.C. and Bowden, R.},
	journal = {IEEE SMDS},
	title = {Targeted {VAE}: Structured inference and targeted learning for causal parameter estimation},
	year = {2021}}

@article{Yarkoni2019,
	author = {Yarkoni, T.},
	doi = {10.31234/osf.io/jqw35},
	journal = {PsyArXiv},
	title = {The generalizability crisis},
	year = {2019}}

@book{Koller2009,
	address = {Cambridge, Massachusetts},
	author = {Koller, D. and Friedman, N.},
	publisher = {MIT Press},
	title = {Probabilistic Graphical Models: Principles and Techniques},
	year = {2009}}

@article{Yarkoni2017,
	author = {Yarkoni, T. and Westfall, J.},
	doi = {10.1177/1745691617693393},
	journal = {Perspectives on Psychological Science},
	title = {Choosing prediction over explanation in psychology: lessons from machine learning},
	year = {2017}}

@article{Blanca2018,
	author = {Blanca, M.J. and Alarcon, R. and Bono, R.},
	doi = {10.3389/fpsyg.2018.02558},
	journal = {Frontiers in Psychology},
	title = {Current practices in data analysis procedures in psychology: what has changed?},
	year = {2018}}

@article{Fiedler2017,
	author = {Fiedler, K.},
	doi = {10.1177/1745691616654458},
	journal = {Perspectives on Psychological Science},
	number = {1},
	title = {What constitutes strong psychological science? The (neglected) role of diagnosticity and a priori theorizing},
	volume = {12},
	year = {2017}}

@article{Grosz2020,
	author = {Grosz, M.P. and Rohrer, J.M. and Thoemmes, F.},
	doi = {10.1177/1745691620921521},
	journal = {Perspectives on Psychological Science},
	pages = {1-13},
	title = {The taboo against explicit causal inference in nonexperimental psychology},
	year = {2020}}

@article{Oberauer2019,
	author = {Oberauer, K. and Lewandowsky, S.},
	doi = {10.3758/s13423-019-01645-2},
	journal = {Psychonomic Bulletin and Review},
	pages = {1596-1618},
	title = {Addressing the theory crisis in psychology},
	volume = {26},
	year = {2019}}

@book{Kline2005,
	author = {Kline, R.B.},
	publisher = {Guilford Press},
	title = {Principles and practice of structural equation modeling},
	year = {2005}}

@article{Coyle2020,
	author = {Coyle, J.R. and Hejazi, N.S. and Malenica, I. and Phillips, R.V. et al.},
	journal = {arXiv2006.07333},
	title = {Targeted learning: Robust statistics for reproducible research},
	year = {2020}}

@article{Rohrer2018,
	author = {Rohrer, J.M.},
	doi = {https://doi.org/10.1177/2515245917745629},
	journal = {Advances in Methods and Practices in Psychological Science},
	title = {Thinking clearly about correlations and causation: Graphical causal models for observational data},
	year = {2018}}

@article{Button2013,
	author = {Button, K.S. and Ioannidis, J.P.A. and Mokrysz, C. and Nosek, B.A. and Flint, J. and Robinson, E.S.J. and Munafo, M.R.},
	journal = {Nature Reviews: Neuroscience},
	title = {Power failure: why small sample size undermines the reliability of neuroscience},
	year = {2013}}

@article{Chernozhukov2017,
	author = {Chernozhukov, V. and Chetverikov, D. and Demirer, M. and Duflo, E. and Hansen, C. and Newey, W.},
	journal = {American Economic Review},
	title = {Double/debiased/{N}eyman machine learning of treatment effects},
	volume = {5},
	year = {2017}}

@article{Deaton2018,
	author = {Deaton, A. and Cartwright, N.},
	doi = {10.1016/j.socscimed.2017.12.005},
	journal = {Social Science and Medicine},
	pages = {2-21},
	title = {Understanding and misundertstanding randomized controlled trials},
	volume = {210},
	year = {2018}}

@book{Imbens2015,
	address = {New York},
	author = {Imbens, G.W. and Rubin, D.B.},
	publisher = {Cambridge University Press},
	title = {Causal inference for statistics, social, and biomedical sciences. An Introduction.},
	year = {2015}}

@article{Hampel1974,
	author = {Hampel, F. R.},
	journal = {Journal of the American Statistical Association},
	number = {346},
	pages = {383-393},
	title = {The influence curve and its role in robust estimation},
	volume = {69},
	year = {1974}}

@book{vanderLaan2011,
	address = {New York},
	author = {{van der Laan}, M. J. and Rose, S.},
	publisher = {Springer International},
	title = {Targeted Learning - Causal Inference for Observational and Experimental Data},
	year = {2011}}

@article{Cinelli2020,
	author = {Cinelli, C. and Forney, A. and Pearl, J.},
	journal = {Technical Report R-493},
	title = {A crash course in good and bad controls},
	year = {2020}}

@article{Rosenbaum1983,
	author = {Rosenbaum, P. R. and Rubin, D. B.},
	journal = {Biometrika},
	number = {1},
	pages = {41-55},
	title = {The central role of the propensity score in observational studies for causal effects},
	volume = {70},
	year = {1983}}

@book{Peters2017,
	address = {Cambridge, Massachusetts},
	author = {Peters, J. and Janzing, D. and Scholkopf, B.},
	publisher = {MIT Press},
	title = {Elements of Causal Inference},
	year = {2017}}

@article{Rubin2005,
	author = {Rubin, D. B.},
	doi = {10.1198/016214504000001880},
	journal = {Journal of the American Statistical Association},
	number = {469},
	pages = {322-331},
	title = {Causal inference using potential outcomes: Design, modeling, decisions.},
	volume = {100},
	year = {2005}}

@book{Pearl2009,
	address = {Cambridge},
	author = {Pearl, J.},
	publisher = {Cambridge University Press},
	title = {Causality},
	year = {2009}}

@article{vanderLaan2014,
	author = {{van der Laan}, M. J. and Starmans, R. J. C. M.},
	journal = {Advances in Statistics},
	title = {Entering the era of data science: targeted learning and the integration of statistics and computational data analysis},
	year = {2014}}

@article{Bottou2013,
	author = {Bottou, L. and Peters, J. and {Quinonero-Candela} J. and Charles, D. X. and Chickering, D. M. and Portugaly, E. and Ray, D. and Simard, P. and Snelson, E.},
	journal = {Journal of Machine Learning Research},
	title = {Counterfactual reasoning and learning systems: the example of computational advertising},
	volume = {14},
	year = {2013}}

@article{Castro2019,
	author = {Castro, D. C. and Walker, I. and Glocker, B.},
	journal = {arXiv:1912.08142v1},
	title = {Causality matters in medical imaging},
	year = {2019}}

@article{Yao2020,
	author = {Yao, L. and Chu, Z. and Li, S. and Li, Y. and Gao, J. and Zhang, A.},
	doi = {10.1145/3444944},
	journal = {ACM Transactions on Knowledge Discovery from Data},
	number = {5},
	pages = {1-46},
	title = {A survey on causal inference},
	volume = {15},
	year = {2020}}

@article{Zhang2020,
	author = {Zhang, W. and Liu, L. and Li, J.},
	journal = {Proceedings of the Thirty-Fifth AAAI Conference on Artificial Intelligence},
	title = {Treatment effect estimation with disentangled latent factors},
	year = {2021}}

@article{Guo2019,
	author = {Guo, R. and Li, J. and Liu, H.},
	journal = {Association for Computing Machinery},
	title = {Learning individual causal effects from networked observational data},
	year = {2020}}

@article{Hilpert2019,
	author = {Hilpert, P. and Brick, T. R. and Flueckiger, C. and Vowels, M. J. and Ceuleman, E. and Kuppens, P. and Sels, L.},
	journal = {Journal of Counseling Psychology},
	title = {What Can Be Learned From Couple Research: Examining Emotional Co-Regulation Processes in Face-to-Face Interactions},
	year = {2019}}

@article{Cohn2018,
	author = {Cohn, J. F. and Cummins, N. and Epps, J. and Goecke, R. and Joshi, J. and Scherer, S.},
	journal = {Handbook of Multimodal-Multisensor Interfaces},
	pages = {375-417},
	title = {Multimodal assessment of depression from behavioral signals},
	volume = {2},
	year = {2018}}

@article{muthukrishna,
	author = {Muthukrishna, M. and Henrich, J.},
	doi = {10.1038/s41562-018-0522-1},
	journal = {Nature Human Behavior},
	pages = {221-229},
	title = {A problem in theory},
	volume = {3},
	year = {2019}}

@article{openpose,
	author = {Cao, Z. and Hidalgo, G. and Simon T. and Wei, S.-E. and Sheikh, Y.},
	journal = {arXiv:1812.08008v1},
	title = {OpenPose: Realtime multi-person 2D pose estimation using part affinity fields},
	year = {2018}}

@article{barnes,
	author = {Barnes, T. J. and Wilson, M. W.},
	journal = {Big Data and Society},
	title = {Big data, social physics, and spatial analysis: {T}he early years},
	volume = {1},
	year = {2014}}

@article{raghavan,
	author = {Raghavan, P.},
	journal = {Big Data and Society},
	pages = {1-4},
	title = {It's time to scale the science in the social sciences},
	volume = {1},
	year = {2014}}

@article{struijs,
	author = {Struijs, P. and Braaksma, B. and Daas, P. J. H.},
	journal = {Big Data and Society},
	title = {Official statistics and big data},
	volume = {1},
	year = {2014}}

@article{aarts,
	author = {Aarts, A. A. and others},
	doi = {10.1126/science.aac4716},
	journal = {Science},
	number = {6251},
	pages = {943-950},
	title = {Estimating the reproducibility of psychological science},
	volume = {349},
	year = {2015}}

@article{openface2,
	author = {Baltrusaitis, T. and Zadeh, A. and Lim, Y. C. and Morency, L-P.},
	journal = {13th IEEE International Conference on Automatic Face and Gesture Recognition},
	title = {{OpenFace} 2.0: Facial Behavior Analysis Toolkit},
	year = {2018}}

@article{mcinnes,
	author = {{McInnes}, L. and Healy, J.},
	journal = {arXiv:1802.03426v1},
	title = {{UMAP}: uniform manifold approximation and projection for dimension reduction},
	year = {2018}}

@book{strogatz,
	author = {Strogatz, S. H.},
	edition = {2nd},
	publisher = {Westview Press},
	title = {Nonlinear dynamics and chaos},
	year = {2015}}

@book{goodfellow,
	address = {Cambridge, Massachusetts},
	author = {Goodfellow, I. and Bengio, Y. and Courville, A.},
	publisher = {MIT Press},
	title = {Deep Learning},
	year = {2016}}
\end{document}